%% file: muse_cme.tex
\documentclass[twocolumn]{aastex631}
\usepackage[caption=false]{subfig}
\usepackage[shortlabels]{enumitem}

\newcommand{\musen}{{\em MUSE}}
\newcommand{\MUSE}{\musen}
\newcommand{\DKIST}{{\em DKIST}}
\newcommand{\EUVST}{{\em EUVST}}

\newcommand{\euvstn}{{\em EUVST}}
\newcommand{\dkistn}{{\em DKIST}}
\newcommand{\hic}{{\em Hi-C}}

\newcommand{\sst}{{\em SST}}
\newcommand{\gregor}{{\em GREGOR}}
\newcommand{\hinode}{{\em Hinode}}
\newcommand{\hinodee}{{\em Hinode/EIS}}

\newcommand{\iris}{{\em IRIS}}
\newcommand{\eui}{{\em EUI}}

\newcommand{\sdo}{{\em SDO}}
\newcommand{\trace}{{\em TRACE}}
\newcommand{\sot}{{\em SOT}}
\newcommand{\soho}{{\em SOHO}}
\newcommand{\sumer}{{\em SUMER}}

\newcommand{\goes}{{\em GOES}}
\newcommand{\suvi}{{\em SUVI}}
\newcommand{\aia}{{\em AIA}}
\newcommand{\solo}{{\em Solar Orbiter}}

\newcommand{\heiiw}{\ion{He}{2}~304~\AA}
\newcommand{\feix}{\ion{Fe}{9}}
\newcommand{\fexii}{\ion{Fe}{12}}
\newcommand{\fexv}{\ion{Fe}{15}}
\newcommand{\fexix}{\ion{Fe}{19}}
\newcommand{\fexxi}{\ion{Fe}{21}}
\newcommand{\feixw}{\ion{Fe}{9}~171~\AA}

\newcommand{\fexvw}{\ion{Fe}{15}~284~\AA}
\newcommand{\fexixw}{\ion{Fe}{19}~108~\AA}

\newcommand{\kms}{~km~s$^{-1}$}

\shorttitle{MUSE Diagnostics: II. Flares and Eruptions}
\shortauthors{MUSE Team}
\graphicspath{{./}{figures/}}

\begin{document}

\title{Probing the Physics of the Solar Atmosphere with the Multi-slit Solar Explorer (MUSE): \\ II. Flares and Eruptions}

\input{author_list}

\begin{abstract}
Current state-of-the-art spectrographs cannot resolve the fundamental spatial (sub-arcseconds) and temporal scales (less than a few tens of seconds) of the coronal dynamics of solar flares and eruptive phenomena. The highest resolution coronal data to date are based on imaging, which is blind to many of the processes that drive coronal energetics and dynamics. As shown by \iris\ for the low solar atmosphere, we need high-resolution spectroscopic measurements with simultaneous imaging to understand the dominant processes. In this paper: (1) we introduce the Multi-slit Solar Explorer (\musen ), a spaceborne observatory to fill this observational gap by providing high-cadence ($<$20\ s), sub-arcsecond resolution spectroscopic rasters over an active region size of the solar transition region and corona; (2) using advanced numerical models, we demonstrate the unique diagnostic capabilities of \musen\ for exploring solar coronal dynamics, and for constraining and discriminating models of solar flares and eruptions; (3) we discuss the key contributions \musen\ would make in addressing the science objectives of the Next Generation Solar Physics Mission (NGSPM), and how \musen, the high-throughput EUV Solar Telescope (\euvstn ) and the Daniel K Inouye Solar Telescope (and other ground-based observatories) can operate as a distributed implementation of the NGSPM. This is a companion paper to~\cite{DePontieu:MUSE_corheating}, which focuses on investigating coronal heating with \musen.
\end{abstract}

\keywords{Sun: corona -- Sun: coronal mass ejections (CMEs)}


\section{Introduction} \label{sec:intro}

We live in the Sun’s extended atmosphere, which is continually shaped by its changing magnetic landscape. The quiescent state of the Sun and its wind is punctuated by flares \citep{Fletcher:2005,2011SSRv..159..357Z,ToriumiWang:2019}, coronal mass ejections \citep[CMEs,][]{Forbes:1995}, and solar energetic particle (SEP) events \citep[][]{KleinDalla:2017}. While solar progenitors of space weather effects are recognized to pose a threat to our technological civilization, our ability to predict their occurrence and impact is hampered by an incomplete picture of underlying physical drivers, triggers and instabilities. 

Providing narrowband extreme ultraviolet (EUV) images at 12\ s cadence, the Atmospheric Imaging Assembly~\citep[\aia;][]{Lemen:2012} onboard NASA's Solar Dynamics Observatory~\citep[\sdo;][]{Pesnell:2012} has demonstrated the importance of high temporal cadence in order to `freeze' coronal dynamics, especially during eruptive events. Two shortcomings of \sdo/\aia\ are its spatial resolution ($\sim 1\arcsec$) and the lack of spectroscopic information necessary to understand the 3D flow fields in solar eruptive events. As demonstrated by the \hinode/Solar Optical Telescope~\citep[\sot;][]{Hinode:2019} and ground-based observatories~\citep[GBOs; e.g.,][]{Scullion:2014,Jess:2015,Jing:2016,Wang:2018,Yadav:2021} for the solar photosphere and by the Interface Region Imaging Spectrograph (\iris) for the chromosphere and transition region~\citep[see][and references therein]{DePontieu:IRISReview}, there is a plethora of sub-arcsecond structure in flaring and eruptive regions.

Our observational capacity to probe small-scale dynamic processes in the lower atmosphere will be enhanced as the Daniel K. Inouye Solar Telescope~\citep[\dkistn;][]{Rimmele:DKIST,DKISTCSP:2021} begins operations. Yet, we do not currently have regular sub-arcsecond coronal observations. Short observational campaigns lasting several minutes by the sounding rocket mission \hic~\citep[][]{Cirtain:2013,Rachmeler:2019,Tiwari:2019}, and by the Extreme Ultraviolet Imager~\citep[\eui;][]{Berghmans:2021} onboard the \solo\ mission~\citep{Mueller:2020} have demonstrated the discovery potential of sub-arcsecond resolutions. However, their chances of capturing solar eruptive events are hampered by the short suborbital flight time and deep space telemetry restrictions, respectively. In addition, for \solo, the restricted time interval of perihelion passage further reduces the chances of catching a flare/CME at the highest spatial resolution. Furthermore, like \sdo/\aia, these instruments provide narrowband imaging and lack spectroscopic information for measuring Doppler flows and non-thermal broadening of the targeted emission lines. 

Although primarily designed for studying the chromosphere and transition region, \iris\ has also provided important spectral diagnostics of the flaring corona. Thanks to its sub-arcsecond spatial resolution, IRIS has provided a new view of several aspects of flares \citep[see][for a review]{DePontieu:IRISReview}, ranging from finally resolving chromospheric evaporation sites \citep[fully blueshifted  \fexxi\ 1354\AA\ line profiles, e.g.,][]{Young:2015,Polito:2015,Tian:2015,GrahamCauzzi:2015,Li:2015,Brosius:2015,Tian:2018}--solving a long-standing problem  \citep[e.g.,][]{Doschek1986,Young2013}--, to observing \fexxi\ redshifts above the flare looptop  \citep[][]{Tian2014} which are a possible signature of downward-moving reconnection outflows or hot retracting loops, predicted by flare models. But with a single-slit spectrograph design, \iris\ has only allowed us a peek into the full richness of the waves and and nonlinear instabilities that pervade the solar corona. Without near-simultaneous spectral coverage of AR-scale fields of view, we will continue to be blindsided, unable to resolve fundamental questions of solar magnetic activity. 
As this paper demonstrates, spectroscopic observables at sub-arcsecond resolution and at cadences of $<20$s is crucial for understanding solar flaring and eruptive activity, including distinguishing physical triggers of eruptions, characterizing intermittency and turbulent structures in current sheets and reconnection outflows, and for constraining the initial conditions of CMEs.

This paper is structured as follows. In Section~\ref{sec:muse}, we introduce the Multi-slit Solar Explorer~\citep[\musen ,][]{BDP:MUSE,Cheung:SDC}, a mission undergoing a Phase A study as a NASA Heliophysics Medium-class Explorer which will fill the aforementioned observational gap. In Section~\ref{sec:sims}, we discuss the central role of advanced numerical modeling to the \MUSE\ science investigation. In Section~\ref{sec:ngspm}, we discuss the background and science objectives of the Next Generation Solar Physics Mission~\citep[NGSPM,][]{NGSPM}, and suggest how \musen, \euvstn, and \dkistn\ (and other GBOs) can coordinate as a distributed implementation of the NGSPM. In Section~\ref{sec:casestudies}, we use the numerical models to demonstrate how \musen\ is unique in its capability for exploring solar coronal dynamics, and for constraining models of solar flares and eruptions. A companion paper~\citep{DePontieu:MUSE_corheating} serves similar purposes but in the context of the coronal heating problem.

\section{The \musen\ Observatory and Science Investigation} \label{sec:muse}
\begin{figure*}[t]
\centering
\includegraphics[width=0.8\textwidth]{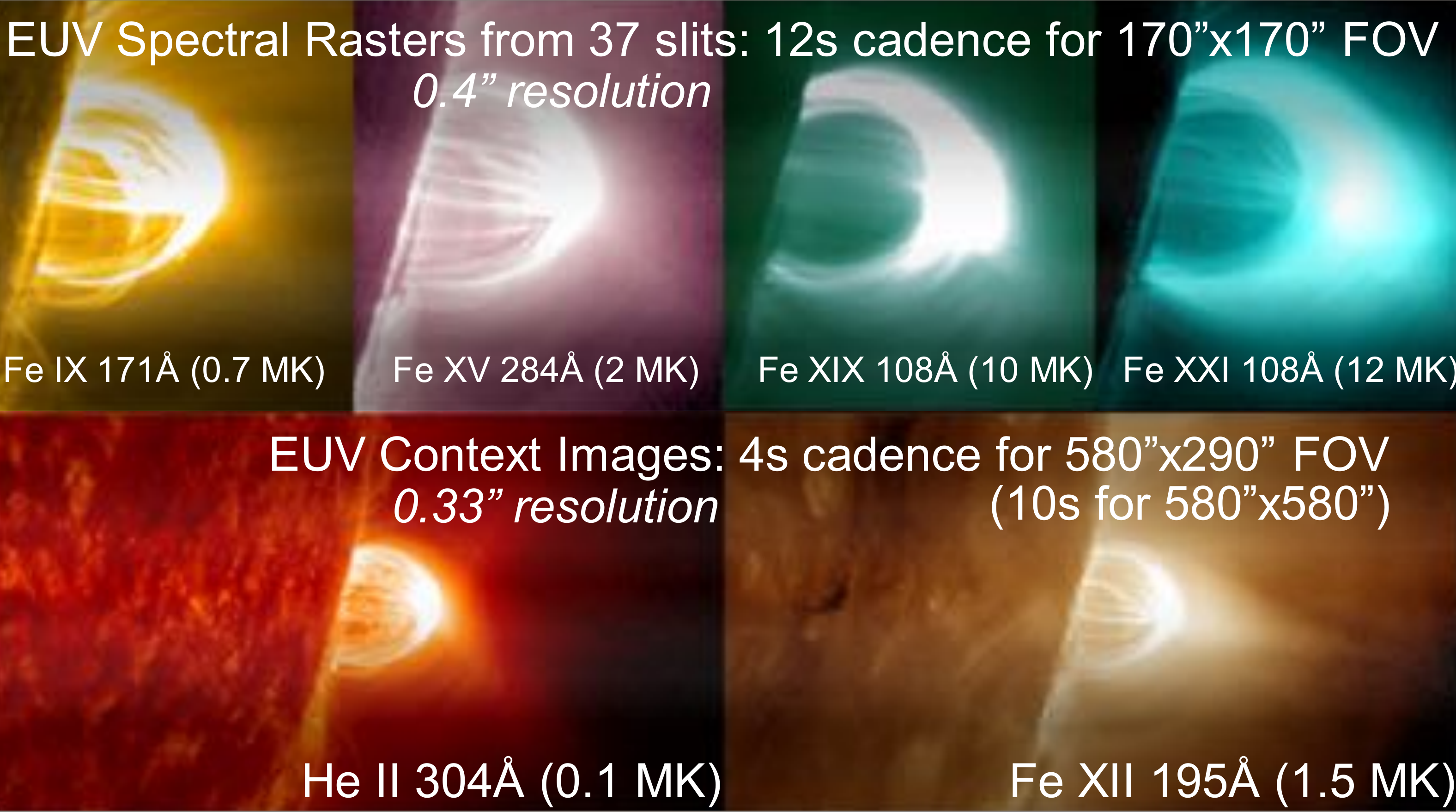}
\caption{Fields-of-view and cadences of \musen's multi-slit spectrograph (SG) and Context Imager (CI).}\label{fig:MUSE_FOV}
\end{figure*}

Currently undergoing Phase A study as a NASA Heliophysics Medium-class Explorer mission, \musen\ has the following science goals:
\begin{enumerate}
    \item Determine which mechanism(s) heat the corona and drive the solar wind,
    \item Understand the origin and evolution of the unstable solar atmosphere, and
    \item Investigate fundamental physical plasma processes.
\end{enumerate}
\musen's 37-slit extreme UV (EUV) spectrograph (\musen/SG) operating at three wavelength bands (108, 171, \& 284 \AA), offers active region (AR) scale (field-of-view, FOV, is 170\arcsec$\times$170\arcsec) spectral rasters at 0.4\arcsec\ along the slits and at 0.4\arcsec\ spatial sampling across the slits, all at a cadence as fast as of 12~s (see Fig.~\ref{fig:MUSE_FOV}). Each parallel slit (along the detector Y-direction) produces its own two-dimensional spectral image on the detector, but offset from each other on the detector in the X-direction. By choosing narrow bandpasses to target isolated spectral lines (\ion{Fe}{9} 171 \AA\ at 0.7 MK, \ion{Fe}{15} 284 \AA\ at  2 MK, \ion{Fe}{19} at 108 \AA\ at 10 MK, and \ion{Fe}{21} 108 \AA\ at 12 MK; see Fig.~\ref{fig:MUSE_FOV}), optimizing for the inter-slit spacing and by using a compressed sensing method, it has been shown that the detector signal can be processed to retrieve physical observables (e.g., emission measure, Doppler velocity) simultaneously sampled by the 37 slits \citep{Cheung:SDC,BDP:MUSE}.

This revolutionary multi-slit design allows \musen\ to capture AR-scale rasters at 100x the speed of existing or planned EUV spectrographs (e.g., \soho/\sumer, \hinodee\ and the upcoming \euvstn). This capability allows MUSE to effectively capture coronal/transition region (TR) dynamics, while delivering valuable spectroscopic information about the fundamental physical processes. In addition to the SG, the Context Imager (\musen/CI) provides 0.33\arcsec\ resolution narrowband images in an even larger FOV (580\arcsec$\times$290\arcsec) in the 304 and 195 \AA\ bands (4s cadence single channel, 8s cadence dual channel).

\section{Numerical simulations} \label{sec:sims}

\begin{figure*}
\centering
\includegraphics[width=0.9\textwidth]{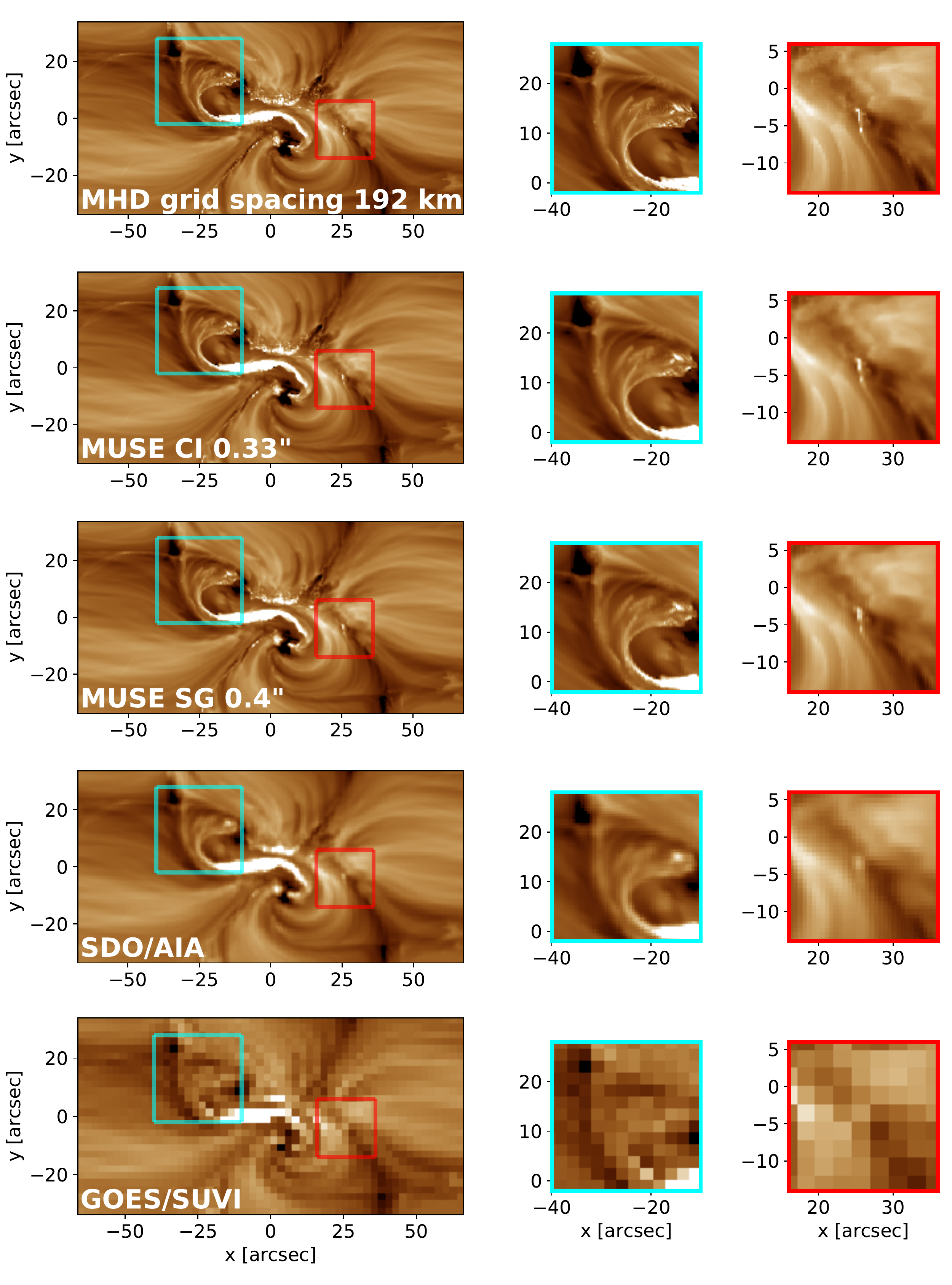}
\caption{High resolution coronal observations by \musen\ ($0.33 \arcsec$ for Context Imager, $0.40\arcsec$ for the multi-slit spectrograph), as well as the high sustained telemetry available will reveal small-scale structures (e.g., bright points and cross-loop striations) inaccessible to the current generation of solar instrumentation. This small-scale structure is important to constrain models of the coronal magnetic field and thermodynamic structure. The top row shows \ion{Fe}{15} 284~\AA~intensity images synthesized from the radiative MHD simulation \texttt{MURaM\_{collision}}, which has computational grid spacing of $192$ km. The blue and red boxes indicate the fields of view of the magnified regions (left two columns). The remaining rows show the synthetic images degraded to resolution and sampling of various instruments, including \musen/CI, \musen/SG, and \sdo/\aia\ and \goes/\suvi.}
\label{fig:multires}
\end{figure*}

A major component of the \musen\ investigation is the use of state-of-the-art numerical models, which are key to addressing \musen\ science goals, to demonstrate the need for high cadence, high-resolution imaging spectroscopy, and to illustrate how \musen\ observables will test current theory, and improve existing models. Such forward modeling exercises tell us how \musen\ can discriminate between models and physical mechanisms. Models used by the team to study the diagnostic potential of \musen\ in the context of flares and eruptions are listed in Table~\ref{table_sims}. They include 1D field-aligned radiation hydrodynamics (RHD), and 2D and 3D magnetohydrodynamics (MHD) models. For numerical models targeting the coronal heating question, refer to the companion paper~\citep{DePontieu:MUSE_corheating}. For further descriptions of the models, refer to Appendix~\ref{app:sim}. For a discussion of how synthetic observables are computed, the reader should consult Appendices~\ref{app:synthesis} and~\ref{app:moments}. 

As an example, consider Fig.~\ref{fig:multires}, which shows synthetic \ion{Fe}{15} 284 \AA~images computed from a radiative MHD simulation of an AR ({\tt MURaM\_collision}) in which two opposite polarity sunspots collide, eventually resulting in a flare and CME (see Figs.~\ref{fig:collision_zdir} and~\ref{fig:collision_xdir} for the eruptive phase of the model). The top row of Fig.~\ref{fig:multires} is the synthetic intensity image at the original simulation (horizontal) grid spacing of 192 km. The remaining rows show images degraded by (1) smoothing with a 2D Gaussian kernel of the form $p(r) \propto \exp\{-\frac{r^2}{2\sigma^2}\}$ ($r$ is distance from origin in arcsec), followed by sampling onto a plate-scale with pixel size $(\Delta_x,\Delta_y)$. The projected performance for \musen/CI is $\sigma=0.14\arcsec$ ($0.33$\arcsec\ full-width half max; FWHM). For \musen/SG $\sigma = 0.176$\arcsec\ ($0.4$\arcsec\ FWHM). The CI has pixel size $(\Delta_x,\Delta_y)=(0.143,0.143)$\arcsec. Note that the \musen\ CI has the \ion{Fe}{12} 195 \AA\ and \ion{He}{2} 304 \AA\ channels, but does not includes the \ion{Fe}{15} 284 \AA\ channel. \sdo/\aia\ lacks this channel. However, to illustrate the effect of instrumental resolution, we will use the same 284 \AA\ line. For a \musen\ dense SG raster with a step size of $0.4\arcsec$ and pixel size separation of $0.167\arcsec$ along the slit, $(\Delta_x,\Delta_y)=(0.4,0.167)$\arcsec. To mimic the resolution of \sdo/\aia\ (\goes/\suvi), we use $\sigma=0.49\arcsec$ ($0.49\arcsec$) and $\Delta_x=\Delta_y = 0.6\arcsec$ ($2.5\arcsec$).

As Fig.~\ref{fig:multires} shows, sub-arcsecond-scale bright points and brightness striations across neighboring loops will be effectively captured by both the \musen\ CI and SG, but are lost at \sdo/\aia\ and \goes/\suvi\ resolutions. This comparison illustrates only part of the benefit of \musen. For each pixel position in the \musen/SG raster, there will be spectroscopic information encoding the Doppler velocity and non-thermal broadening of coronal plasmas. In Section~\ref{sec:casestudies}, we consider how such spectroscopic information at high cadence and spatial resolution can be exploited for \musen's science goals, as well as to address the NGSPM Science Objectives, which are detailed in the following section.

\begin{deluxetable*}{llllcl}
\tablecaption{Numerical Simulations used to synthesize \musen\ observables}
\tablehead{
\colhead{Code}  &  \colhead{model} & \colhead{target} &\colhead{properties} & \colhead{NGSPM SO} & \colhead{Refs.\ \tablenotemark{a}}
}
\startdata
MURaM  & {\tt MURaM\_flare} \tablenotemark{b} & active region (AR), flares,  eruption & 3D MHD & II.1,II.2 & [1] \\ 
  & {\tt  MURaM\_circ\_rib} & circular ribbon flares,  eruption & 3D MHD & II.1,II.2 & [2] \\
  & {\tt  MURaM\_collision} & colliding sunspots,  eruption & 3D MHD & II.1,II.2,II.5 & [3] \\
  & {\tt MURaM\_emergence} &  plage, flare &3D MHD & II.1-II.2 & [4] \\ \cline{1-6}
Bifrost  & {\tt B\_npdns03} & coronal hole, bright point, jets & 2D MHD & II.1,II.2,II.4 & [5] \\ \cline{1-6} 
-  & {\tt Termination\_shocks} & flare, magnetic reconnection & 2D MHD & II.4 & [6] \\  \cline{1-6} 
PREFT  & {\tt Retracting\_tube} &  flare, magnetic reconnection & 1D MHD  & II.4 & [7]\\ \cline{1-6} 
RADYN  & {\tt RADYN\_1D} & flaring loops \& footpoints & 1D RHD, NTE & II.4  & [8,9] \\ 
  & {\tt RADYN\_Arcade} & flaring loops (LOS effects) & 1D RHD, NTE  + 3D AR loops & II.4 & [10]\\ \cline{1-6} 
\enddata
\label{table_sims}
\tablenotetext{}{RHD: radiative hydrodynamic; NTE: non-thermal electrons}
\tablenotetext{a}{References: [1] \cite{CheungRempel:2019}; [2] \citet{ChenRempelFan:2021}; [3] \cite{RempelChintzoglouCheung:2021}, [4]
 \cite{Sanja_AGU}; [5] \cite{Nobrega-Siverio:2021inprep}; [6] \cite{Takasao:2015,Takasao:2016}; [7] \cite{Longcope:2015,Longcope:2016}; [8] \cite{2015ApJ...809..104A}; [9] \cite{Polito2019}; [10] \cite{Kerr2020}.}
\tablenotetext{b}{Publicly available at https://purl.stanford.edu/dv883vb9686}

\end{deluxetable*}

\section{Next Generation \\ Solar Physics Mission}
\label{sec:ngspm}
The NGSPM~\citep{NGSPM} is a mission concept developed by a panel of solar physics experts designated by NASA, JAXA, and the European Space Agency (ESA) \footnote{https://hinode.nao.ac.jp/SOLAR-C/SOLAR-C/Documents/NGSPM\_report\_170731.pdf}. Following townhalls at international solar physics conferences and dozens of whitepaper submissions from the community, the Science Objectives Team (NGSPM SOT) developed a list of science objectives (SOs) based on the following criteria: (1) relevance to NASA/JAXA/ESA objectives, (2) scientific impact on solar physics,  (3) scientific impact on other disciplines and research fields, (4) inability of current/planned missions and ground-based facilities to accomplish the objective, (5) need for space observations, (6) maturity of technology, (7) maturity of methodology, and (8) widespread interest within the solar physics community. Based on these factors, the NGSPM prioritized SOs are:
\begin{enumerate}[I:]
    \item Formation mechanisms of the hot and dynamic outer solar atmosphere
    \item Mechanisms of large-scale solar eruptions and foundations for prediction
    \item Mechanisms driving the solar cycle and irradiance variation
\end{enumerate}
The companion paper~\cite{DePontieu:MUSE_corheating} focuses on how \musen\ and other observatories can coordinate to address NGSPM-I. This paper focuses on addressing NGSPM-II. Sub-objectives of NGSPM-II are listed in Table~\ref{table:NGSPM}. 

\begin{deluxetable*}{lccc}[]\label{table:NGSPM}
\tablecaption{\label{table_so} NGSPM Science Objectives\tablenotemark{a} and Corresponding Mission Science Objectives}
\tablehead{
\colhead{{\bf NGSPM Science Objectives}}  & \multicolumn{3}{l}{{\bf  Mission Science Objectives}} \\ \cline{2-4}
& \musen\ \tablenotemark{b}  & \euvstn\ \tablenotemark{c}  & \dkistn\ \tablenotemark{d}
}
\startdata
{\bf II. Mechanisms of large-scale solar eruptions}  & & & \\ 
{\bf \hspace{1cm} and foundations for prediction}  & & & \\ 
II-1 Measure the energy build-up processes in flaring and CME regions  & 2[a,b,d] & II-2-1 & 4.1, 5.6 \\
II-2 Identify the trigger mechanisms of solar flares and CMEs & 2[a,d],3c & II-2-2 & 4.1,5.7  \\
\hspace{1cm} and distinguish between the many CME models &  &  &  \\
II-3 Understand the evolution and propagation of CMEs and their effect & 2c & - & - \\
\hspace{1cm} on the surrounding corona &  & &  \\
II-4 Understand the processes of fast magnetic reconnection & 3c & II-1-[1,2,3] & 4.4, 5.3, 6.3 \\
II-5 Understand the formation mechanism of sunspots, & 1a,2b & II-[1,2]-1 & 3.4  \\
\hspace{1cm} in particular delta sunspots &  &  &  \\
\enddata
\tablenotetext{a}{https://hinode.nao.ac.jp/SOLAR-C/SOLAR-C/Documents/NGSPM\_report\_170731.pdf}
\tablenotetext{b}{\citet[][]{BDP:MUSE}}
\tablenotetext{c}{https://hinode.nao.ac.jp/SOLAR-C/SOLAR-C/Documents/2\_Concept\_study\_report\_part\_I.pdf}
\tablenotetext{d}{{\em DKIST} objective refers to section number in \citet{DKISTCSP:2021}.}
\end{deluxetable*}
The suite of instruments identified by the NGSPM report as the most suitable to address the prioritized SOs I and II are the following:
\begin{enumerate}[(a)]
    \item 0.3\arcsec\ resolution coronal / TR spectrograph,
    \item 0.2\arcsec-0.6\arcsec\ coronal imager, and
    \item 0.1\arcsec-0.3\arcsec\ resolution chromospheric/photospheric magnetograph/spectrograph
\end{enumerate}
This combined observational capability may be implemented on a single platform (e.g., the original proposed Solar-C mission), or on multiple platforms. We argue the latter option can be fulfilled by appropriate coordination between \musen\ and other observatories. The observational capability of (a) high-resolution coronal/TR spectrograph will be fulfilled by the high-throughput EUV Solar Telescope~\cite[\euvstn;][]{Shimizu:EUVST}, which has been selected for implementation by JAXA and NASA, with a planned launch date in 2026. \musen\ exceeds the requirements of (b), the high-resolution coronal/TR imager: it will provide context images continuously, at very high cadence, but its multi-slit spectrograph also provides spectroscopy, optimized to measure four spectral lines to facilitate high-cadence rasters.

As for (c), \dkistn\ \citep[currently in commissioning; ][]{Rimmele:DKIST,DKISTCSP:2021} and a number of other existing ground-based observatories (GBOs) have already achieved or exceeded 0.1\arcsec\ resolution observations. They include the Swedish 1-m Solar Telescope~\citep[\sst;][]{Scharmer:2003,Scharmer:2019}, \gregor\ \citep{Schmidt:2012,Kleint:2020} and the Goode Solar Telescope \citep[previously, New Solar Telescope;][]{Goode:2010,Cao:2010}. In particular, \sst\ observations have provided photospheric and chromospheric spectropolarimetric observation of flares at this spatial resolution~\citep{Yadav:2021}. The planned European Solar Telescope will add to this list of high-resolution GBOs for photospheric and chromospheric magnetometry. Coordination between \dkistn\ and other GBOs with \MUSE\ and \EUVST\ can then be considered a distributed implementation of the NGSPM concept. Table~\ref{table:NGSPM} shows the correspondence of NSGPM SOs and the science goals / objectives of \musen, \euvstn\ and \dkistn .

\section{Case studies addressing NGSPM Science Objective II} \label{sec:casestudies}
In the following, we present use cases of how \musen\ observations will address questions regarding the drivers and triggers of solar flares and eruptions, how these events impact the ambient corona, and the underlying physical processes (such as fast magnetic reconnection). We will also highlight the synergies available through coordination with \euvstn , and with GBOs like \dkistn . For this reason, the following sections are organized according to the sub-objectives of NGSPM SO II.

\subsection{II-1: Measure the energy build-up processes in flaring and CME regions}
\label{sec:ngspm_ii}
Eruptive events originate in the solar atmosphere and are powered by the release of energy stored in stressed magnetic fields. The energy stored in the corona of flare- and CME-productive ARs is carried by flux emergence~\citep[][]{Forbes:1995,Chen:2000,Archontis:2008,CheungIsobe:2014,ToriumiWang:2019}, shearing/twisting \citep[][]{Amari:1999,Wyper:2017}, perhaps driven by emergence of twisted field~\citep[][]{Manchester:2004,Okamoto:2010ApJ...719..583O,ToriumiHotta:2019}, and cancellation of flux at polarity inversion lines (PILs) to form flux ropes~\citep[e.g.,][]{VBMartns:1989,Savcheva:2012,CheungDeRosa:2012,Kazachenko:2014,Fisher:2015,Chintzoglou:2019}. Measuring energy build-up in flaring and CME regions can be done by modeling how the 3D magnetic field in the corona evolves~\citep[e.g., extrapolation methods;][]{Wiegelmann:2012,DeRosa:2015,Warren:2018}, or by estimating the amount of energy deposited in the corona via the Poynting flux through the photosphere \citep[e.g., see][]{Kazachenko:2014}. Both classes of methods rely on some combination of direct measurements of the field at the photosphere and chromosphere combined with measurements of field-aligned emission structures in the corona.

While photospheric vector magnetograms can be used as boundary conditions for non-linear force-free field (NLFFF) extrapolations, because of the non-force-free nature of the photospheric field (due to confinement by ambient plasma pressure, for instance), systematic errors in the extrapolation may occur. To alleviate this problem, there are methods to reconstruct the coronal magnetic field using a combination of line-of-sight (LOS, or radial component) magnetogram and EUV images of coronal loops~\citep{Aschwanden:2013,Malanushenko:2014,Plowman:2021}. Even when the reconstruction method does not directly use coronal imagery as input, the latter (e.g., by the locations of sigmoids and hot flux ropes) is needed for validation of the model~\citep[e.g.,][]{Savcheva:2012,Wiegelmann:2012,James:2018}. As indicated in Fig.~\ref{fig:multires}, existing instruments such as \aia\ and \goes/\suvi\ do not have sufficient spatial resolution to resolve small-scale loops (bright points) and cross-field striations between neighboring loops. Using loops traced at the coarse resolution available from existing instruments will lead to errors in the loop geometry, providing erroneous constraints on coronal magnetic field reconstructions.  This can have impacts on estimates of the free magnetic energy available to power solar flares. \euvstn\ can provide subarcsecond resolution raster imaging, but to keep a high cadence ($<20$s) rasters are limited to $\sim 5\arcsec$ in width (see Figs.~\ref{fig:hgcr_erupt_euvst_1} and~\ref{fig:hgcr_erupt_euvst_2}). AR-scale rasters requires several minutes (assuming 1s slit dwell time, 0.4\arcsec\ step sizes). Even at resolutions of $1$\arcsec\ or coarser, coronal loops evolve significantly over such timescales. This is especially true during the emerging phase of ARs. To provide observational constraints on the coronal magnetic field geometry, it is thus necessary to have both high resolution, and high cadence imaging from \musen. 

\begin{figure*}
\includegraphics[width=\textwidth]{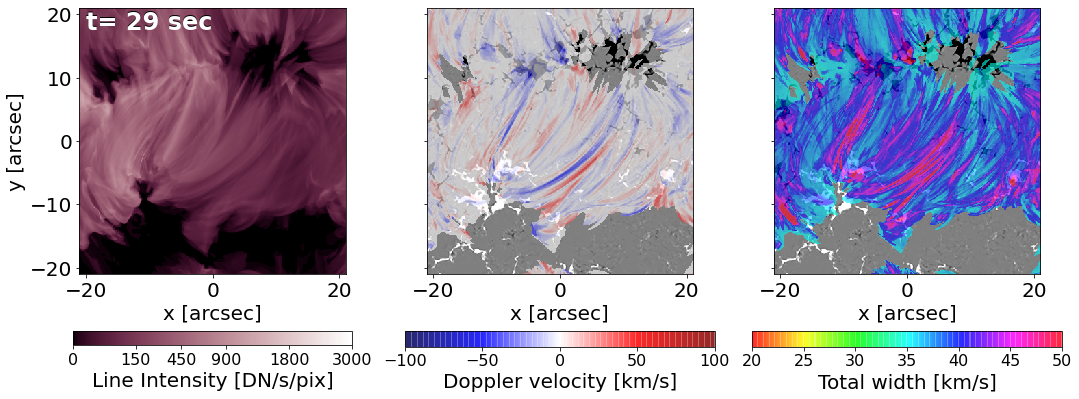}\\
\includegraphics[width=\textwidth]{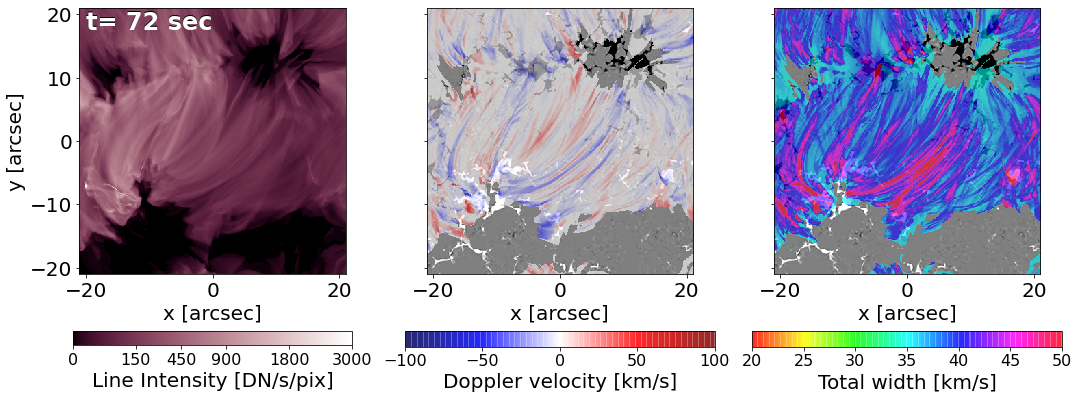}
\caption{\musen\ \ion{Fe}{15} 284 \AA\ maps synthesized from a radiative MHD simulation of an emerging flux region (EFR; model \texttt{MURaM\_emergence}, see Table~\ref{table_sims} and Appendix~\ref{app:sim}) reveals fine-scale coronal strands (subarcsecond widths) connecting opposite polarities of the emerging bipolar region, evolving on timescales of tens of seconds. \musen\ rasters with 12 s cadence meets the requirements to track the dynamics of loops in  EFRs. An animated version of this figure is available online.}\label{fig:emergence}
\end{figure*}

Figure~\ref{fig:emergence} shows \musen\ \ion{Fe}{15} 284 \AA\ maps synthesized from a flux emergence simulation ({\tt MURaM\_emergence}; see Table~\ref{table_sims} and Appendix~\ref{app:sim}). The spectroscopic rasters allows the retrieval of parameters such as total line intensity as count rates of data number~s$^{-1}$~pix$^{-1}$ (DN~s$^{-1}$~pix$^{-1}$), as well as the Doppler shift and total line width (given here in units of km~s$^{-1}$). The line intensity (integrated over wavelength), Doppler velocity and line width maps all show strands with sub-arcsecond widths. The loop structures vary significantly in the span of tens of seconds, while an \euvstn\ dense raster with 0.4\arcsec\ steps and slit dwell-time of 1s~step$^{-1}$ would require 100\ s to complete a single raster. In comparison, the multi-slit design of the \musen/SG allows for a dense raster to complete once every 12\ s, and \musen/CI will provide images in the $193$ \AA\ and $304$ \AA\ bands at 5\ s cadence.

\musen\ has a sufficiently large FOV to observe how nearby ARs interact (\musen/SG FOV is $170\arcsec\times170\arcsec$. The CI extends the FOV to $580\arcsec\times 290\arcsec$ (see Fig.~\ref{fig:MUSE_FOV}). This will enable studies of inter-AR interaction. \citet{Longcope:Heating:2020} analyzed coronal loops in the \aia\ 171~\AA~channel in and around two ARs. They suggest the coronal loops preferentially appear at the topological boundaries of magnetic subvolumes, and suggest these are preferential sites for plasma heating~\citep[see also][]{McCarthy:2019}. \aia\ and \hinodee\ observations of transequatorial coronal loops connecting ARs located at positive and negative latitudes suggest there are observable characteristics particular to topological features such as separators~\citep[][]{Ghosh:2020}.

\begin{figure*}
    \centering
    \includegraphics[width=1.\textwidth]{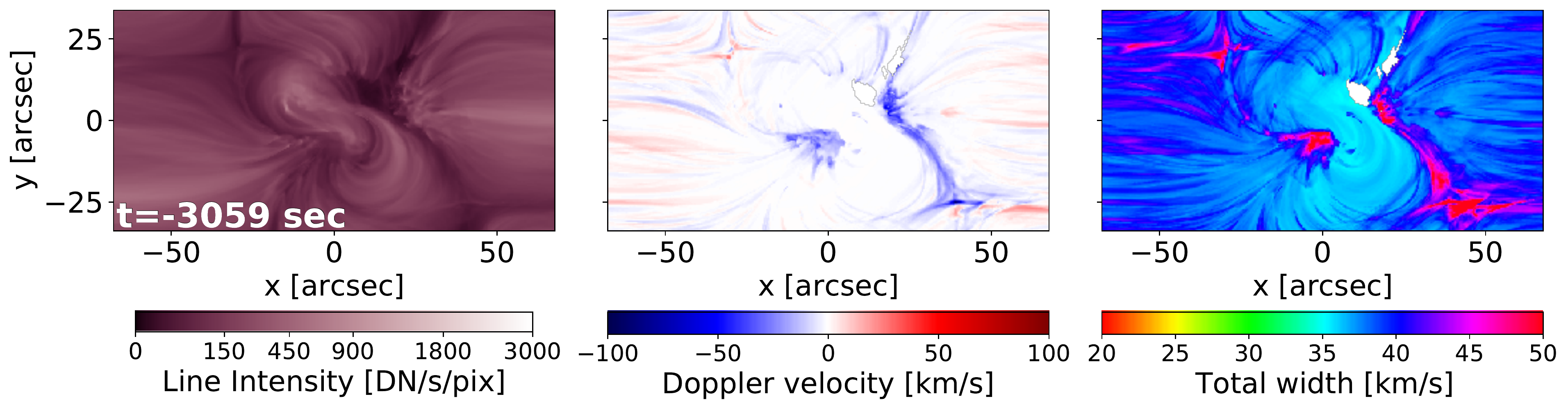}\\
    \caption{Line intensity, Doppler velocity and line width maps of the \ion{Fe}{15} 284 \AA\ line synthesized from a radiative MHD model of a pair of sunspots approaching each other ({\tt MURaM\_collision}, see Table \ref{table_sims}). The time stamp of this snapshot is relative to the time of the flare (see Fig.~\ref{fig:collision_zdir} for the eruption flare phase), so this frame is in the quiescent phase when the coronal field is quasi-steadily adjusting in response to photospheric flows advecting two sunspots. The two spots are linked by S-shaped coronal loops in the line intensity image. At the periphery of the closed loop system are fan loops with Doppler blueshifts and line widths of tens of \kms, signifying a change in connectivity. } 
    \label{fig:fan_loops}
\end{figure*}

Coronal field configurations have topological features which may leave imprints in spectroscopic observables. Synthetic maps of \ion{Fe}{15} emission, including line intensity, Doppler velocity and line width as would be observed with \musen\ as shown in Fig.~\ref{fig:fan_loops} from a quiescent AR (pre-flare phase of model {\tt MURaM\_collision}), corroborate the suggestion by previous work that AR fan loop outflows are located at quasi-separatrix layers (QSLs).~\citet{Baker:2009} used \hinodee\ rasters and coronal field extrapolations to show the fan loops have a different connectivity than the AR core loops. Therefore fan loop outflows (blueshifts and total line widths of tens of \kms) are tracers of QSLs of AR boundaries. \euvstn\ will be able to raster ARs in minutes (1s slit dwell time, 0.4\arcsec\ raster steps), which would be sufficient for tracking slow changes in connectivity. However, the higher cadence of \musen\ is needed to capture more dynamic changes, especially during flaring and eruptive scenarios.

\subsection{II-2: Identify the trigger mechanism of solar flares and CMEs and distinguish between the many CME models}
\label{sec:ngspm_ii2}
\begin{figure*}
    \centering
    \includegraphics[width=\textwidth]{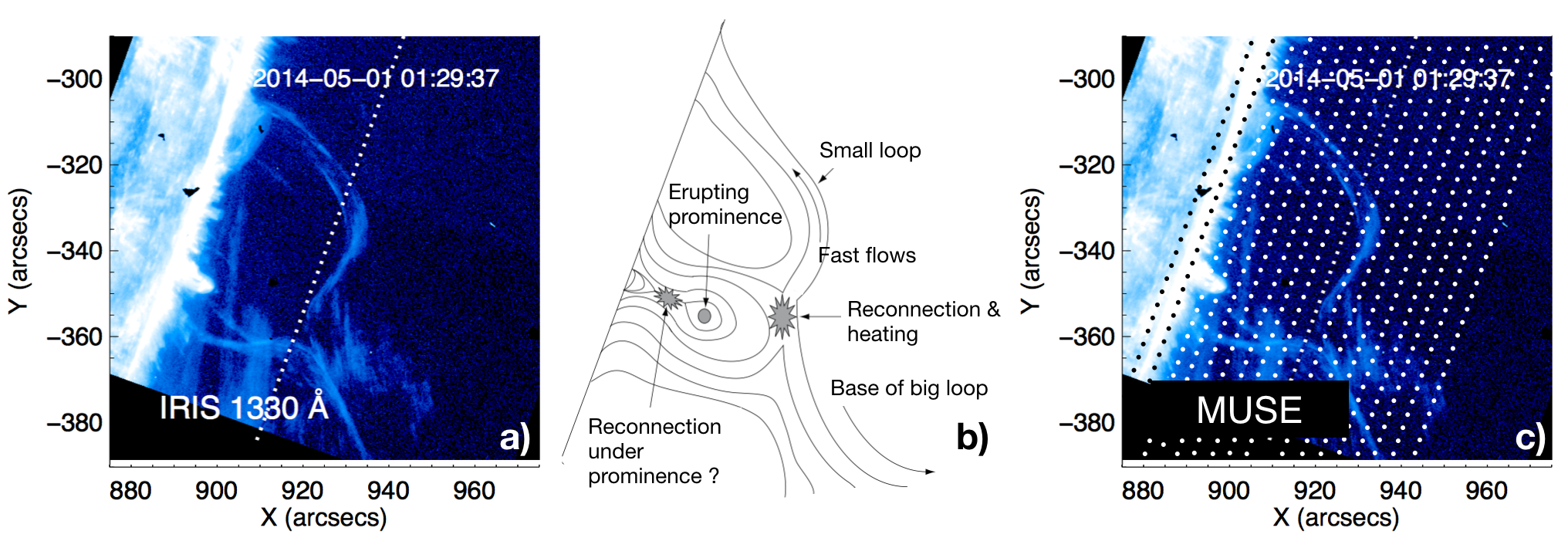}
    \caption{Panel a): A small eruption observed in the IRIS 1330 \AA\ SJI, from \citet{Reeves2015}.  The dotted line indicates the position of the single slit. Panel b) A cartoon indicating possible reconnection sites.  The timing of these reconnection events is critical for determining the triggering mechanisms for eruptions (adapted from \citealt{Reeves2015}).  Panel c) same as Panel a), except showing a subset of the 37 slits coverage of \MUSE. }
    \label{fig_kathy2015}
\end{figure*}

Different CME models postulate reconnection in different places and at different times. For example, breakout models~\citep{Antiochos:1999,Wyper:2017} hypothesize reconnection above the flux rope, but tether-cutting models~\citep{Moore:2001} say it is below. ~\citet{LongcopeForbes:2014} developed a unified 2D model showing either mechanism (or a combination of both) can evolve a multipolar system with a pre-existing flux rope to a loss of equilibrium, which explains CME acceleration as an ideal instability. In a series of 2.5D MHD experiments,~\citep{Karpen:2012} demonstrated - for a system with azimuthal symmetry (and with a coronal flux rope that lacks anchored footpoints at the photosphere) - that breakout reconnection is the first to occur. In contrast, the torus instability~\citep[][]{Kliem:2006} does not require reconnection to initiate acceleration of the coronal flux rope. Given the many proposed mechanisms for eruption triggers, it is necessary to capture the exact location and timing of reconnection (or lack thereof) to test which models are pertinent to solar eruptions.

\iris\ observations have found signatures of reconnection in various events that lend evidence to particular eruption initiation mechanisms, but \iris\ observations alone are often not enough to identify the trigger with certainty. For example, \citet{Kumar2019} identify breakout reconnection as the trigger for a small eruption at the limb (shown in Fig.~\ref{fig_kathy2015}) based on observations of bi-directional flows and small blobs in \iris\ slitjaw images. On the other hand, \citet{Reeves2015} attribute the triggering of this eruption to tether-cutting, based on the identification of brightenings below the flux rope that are observed in the slitjaw images that occur just as the fast rise phase of the eruption begins. A schematic drawing of this interpretation is shown in Fig.~\ref{fig_kathy2015}b.  These brightenings were not captured by the \iris\ slit, so it is not known if there were bi-directional outflows at this location, which would provide more solid evidence that reconnection is taking place there.

\begin{figure*}[!ht]
    \centering  
    \includegraphics[width=\textwidth]{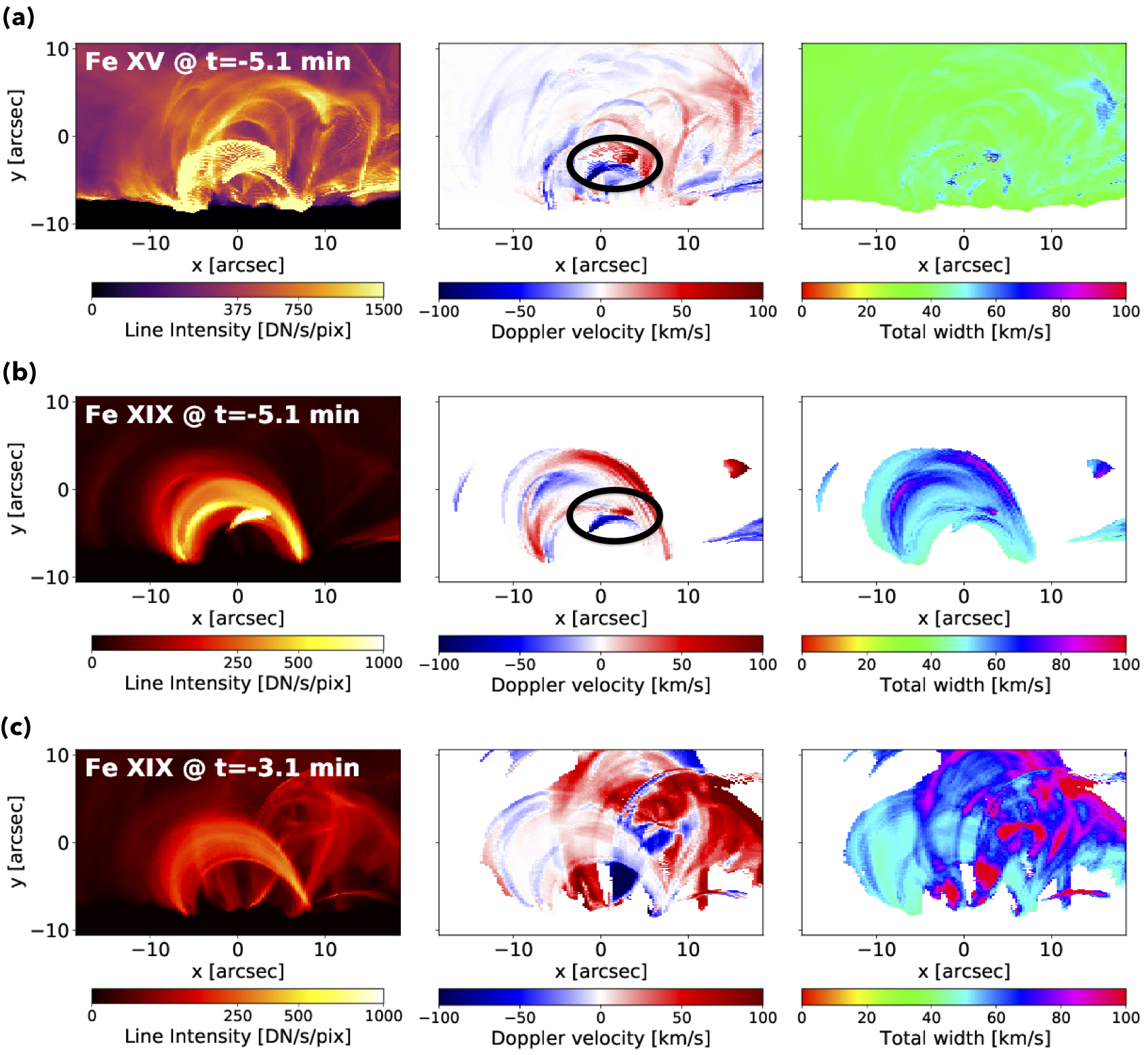}
\caption{High spatio-temporal resolution rasters possible with \musen\ are necessary for capturing the trigger(s) of flares and CMEs. For example, shown here are line intensity, Doppler velocity and total line width maps of \ion{Fe}{15} 284 \AA~and \ion{Fe}{19} 108 \AA~lines synthesized from a radiative MHD simulation of a C-class flare~\citep[{\tt MURaM\_flare}; see Table~\ref{table_sims} and Appendix~\ref{app:sim};][]{CheungRempel:2019}. Top \& middle rows: At $t=-5.1$ min, Doppler maps show bidirectional reconnection outflows at $(x,y)=(0,-5)\arcsec$ shown with black circle. In this simulated flare, this coronal magnetic reconnection event is the flare trigger that destabilized the pre-existing underlying magnetic flux rope. Bottom row: same as middle row but 2 min later.}\label{fig:hgcr_trigger}
\end{figure*}

\MUSE~will be ideal for identifying potential reconnection sites during a solar eruption because its multi-slit approach will simultaneously capture the plasma properties (e.g., Doppler velocity, plane-of-sky velocity, and non-thermal line broadening) over a wide area of the erupting site. For example, Fig.~\ref{fig_kathy2015}c shows the eruption observed by \citet{Reeves2015} and \citet{Kumar2019} as it might be observed by \musen.  As an additional illustration, Fig.~\ref{fig:hgcr_trigger} shows maps of the \ion{Fe}{19} 108 \AA\ line intensity, Doppler velocity and total width synthesized from a radiative MHD simulation of a solar flare (\texttt{MURaM\_flare}, see Table \ref{table_sims} and Appendix~\ref{app:sim}). As  discussed in detail by \citet{CheungRempel:2019}, the simulation was inspired by the observed evolution of NOAA AR 12017, in which a parasitic bipolar magnetic region emerged in the vicinity of a pre-existing sunspot  ($(x,y)=(0,-3)$\arcsec). In the MHD simulation mimicking this sequence of photospheric driving, the parasitic bipole undergoes flux cancellation at its internal PIL and creates a coronal flux rope. The pre-existing flux rope is then destabilized when overlying magnetic flux reconnects. The resulting bidirectional Doppler flows and localized brightening in the \ion{Fe}{19} 108 \AA\ map at $t=-5.1$ min~\citep[i.e., $5.1$ minutes before peak of the synthetic \goes\ soft x-ray light curve, see][]{CheungRempel:2019} are observable signatures with \musen\ (for a limb view). This observable signature is short-lived, and is no longer visible at $t=-3.1$ min (bottom row of the figure),  illustrating the importance of high cadence rasters by \musen.

\begin{figure*}
\centering
\includegraphics[width=\textwidth]{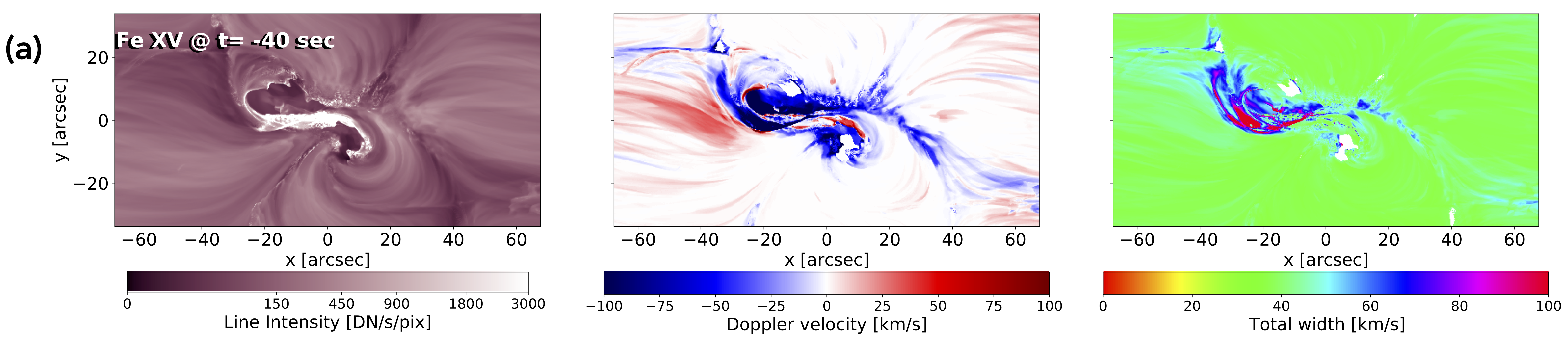} \\
\includegraphics[width=\textwidth]{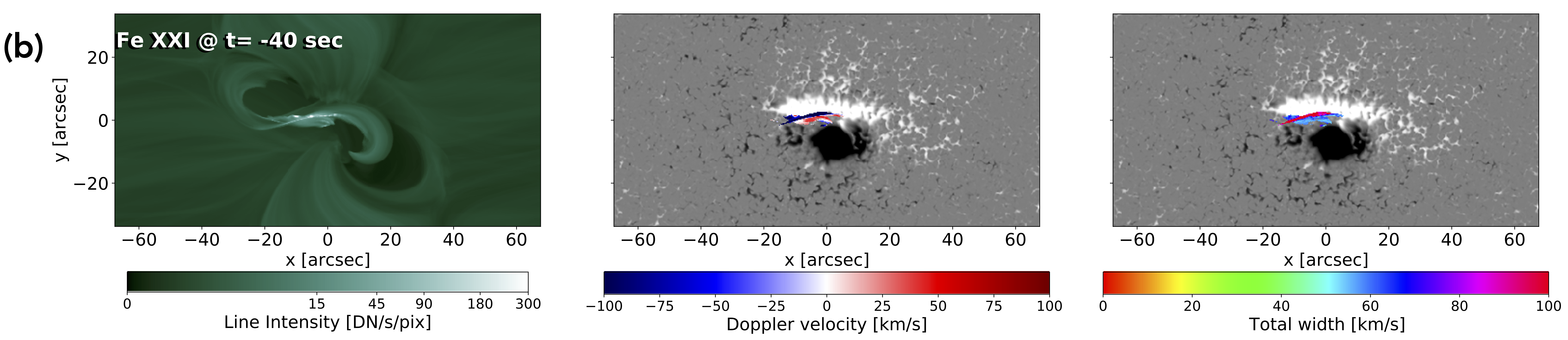} \\
\includegraphics[width=\textwidth]{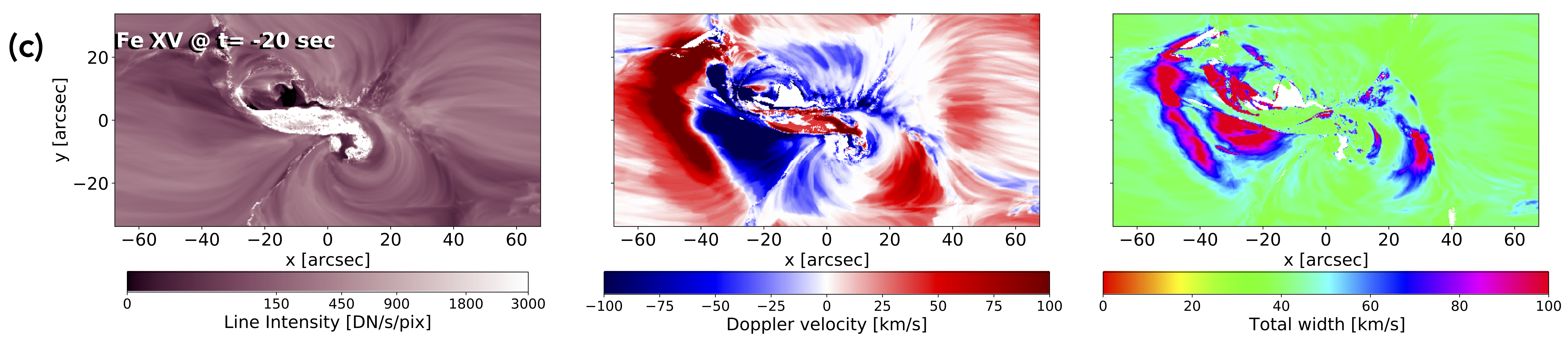} \\
\includegraphics[width=\textwidth]{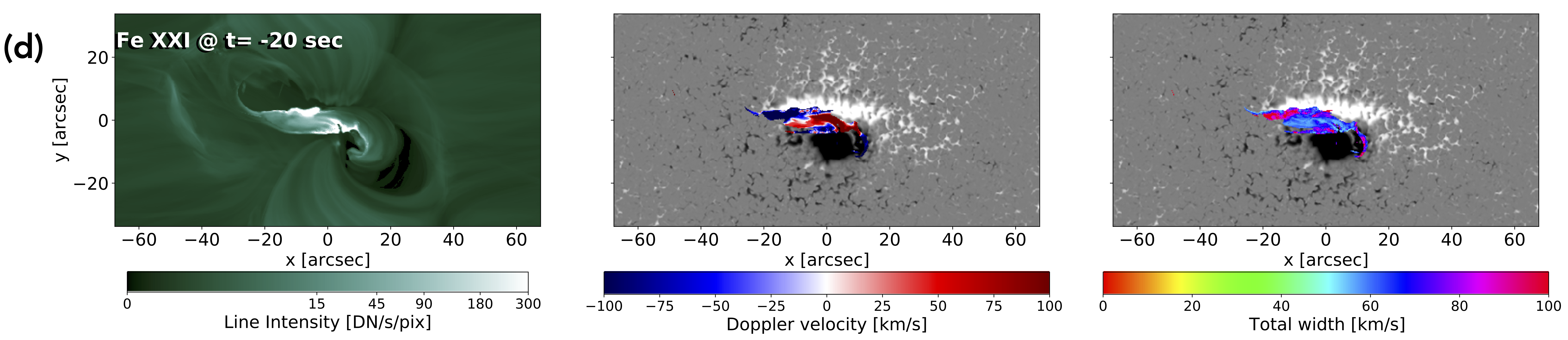} 
\caption{\musen\ will capture the triggering and evolution of solar eruptive events and their impact on the ambient corona. Rows (a) \& (b) show the synthetic \musen\ observables, namely the spectral line intensity (left column), Doppler velocity (middle) and line width (right column) for the \ion{Fe}{15} 284~\AA~and \ion{Fe}{21} 108 \AA~lines as synthesized from a radiative MHD simulation of a solar flare and CME (\texttt{MURaM\_collision}; see Table~\ref{table_sims} and Appendix~\ref{app:sim}). Rows (c) \& (d) show corresponding maps 20\ s later. The times indicated are relative to the peak of the \goes\ soft x-ray light curve (as synthesized from the model). The Doppler velocity and line width maps for the \ion{Fe}{21} line are overlaid over synthetic photospheric magnetograms. At $t=-20$s, the bidirectional Doppler flows centered at $(x,y)=(0,0)$ in the hot \ion{Fe}{21} line are detectable signatures of tether-cutting reconnection above the PIL, which triggers the flare and CME. The greyscale images in rows (b) and (c) show the vertical component of the photospheric magnetic field. See Fig.~\ref{fig:collision_xdir} for a limb-view of the same simulation.}\label{fig:collision_zdir}
\end{figure*}

\begin{figure*}
    \centering
    \includegraphics[width=\textwidth]{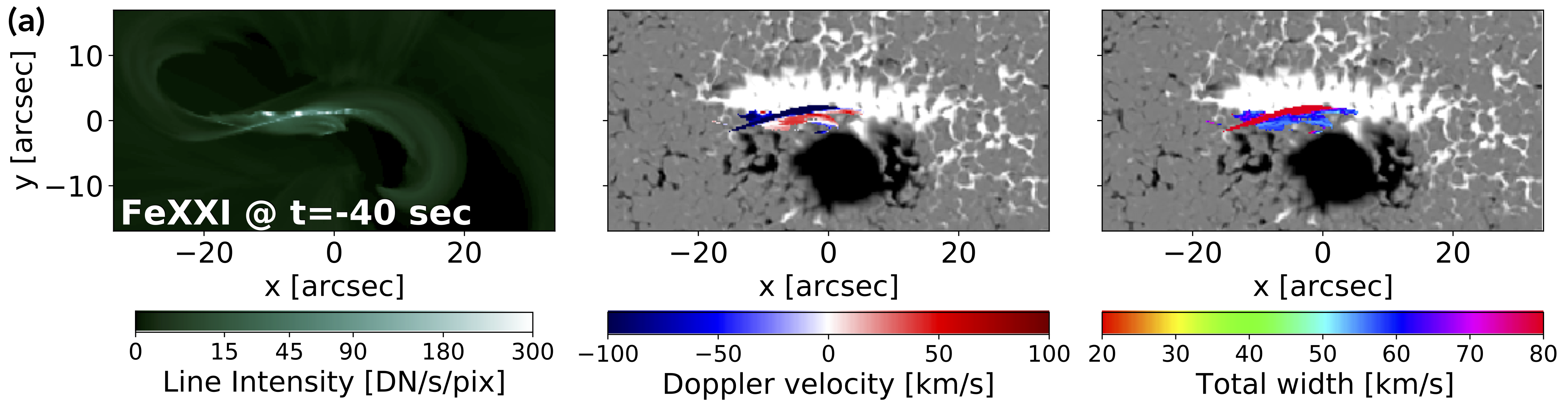} \\
    \includegraphics[width=\textwidth]{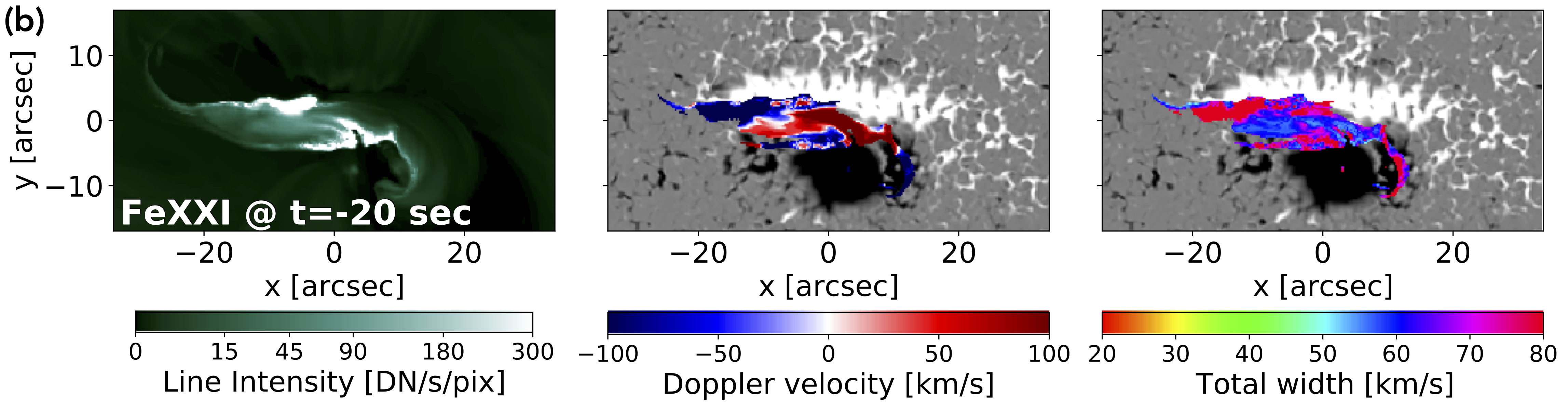}
    \caption{Zoomed-in \ion{Fe}{21} 108 \AA\ (12 MK) maps shown in Fig.~\ref{fig:collision_zdir} (from model \texttt{MURaM\_collision}; see Table~\ref{table_sims} and Appendix~\ref{app:sim}). Left: line intensity; middle: Doppler shift and right: line width. The tether-cutting reconnection region is above the polarity inversion line (greyscale in middle and right columns show the vertical component of the magnetic field at the photosphere). Row (b) shows the flare loop 20\ s later. Unlike single-slit spectrographs, \musen\ will be able to raster these FOVs with $<$ 20\ s cadence to capture such dynamic changes.}
    \label{fig:tethercutting}
\end{figure*}

While the flare model inspired by AR 12017 featured a trigger in the coronal region above the flux rope, other simulations suggest below-the-flux-rope reconnection may be responsible. Figure~\ref{fig:collision_zdir} shows one such example. This radiative simulation (\texttt{MURaM\_collision}) features a sunspot translating horizontally at the solar photosphere (and below), eventually colliding with a neighboring sunspot. This collision process creates a twisted magnetic flux rope above the PIL, akin to what is found in many flare-productive regions~\citep[e.g., ][]{ToriumiWang:2019,Chintzoglou:2019}. The figure shows maps of the total intensity, Doppler velocity and line width of the \ion{Fe}{15} 284 \AA\ and \ion{Fe}{21} 108 \AA\  lines at 40\ s and 20\ s before the soft x-ray peak of the simulated flare. At $t=-40$s, the total line width of \ion{Fe}{21} shows an enhancement (of up to 100 \kms) around the PIL (see the zoomed-in images in Fig.~\ref{fig:tethercutting}). Inspection of the MHD cubes reveals this occurs below the flux rope, which is consistent with the finding of~\citet{Harra:2013}, who reported non-thermal coronal line width enhancements from \hinodee\ measurements at the base of three ARs before they flared. In addition, the \ion{Fe}{21} Doppler map of the MHD models shows bidirectional flows in the region of enhanced line width. This is consistent with reconnection below the flux rope, i.e., tether-cutting reconnection.

A major difference between \MUSE\ and \hinodee\ is the almost of two-orders magnitude improved raster cadence of \MUSE. This allows AR-scale rasters to be available at 12\ s cadence, in contrast to the tens of minutes required for \hinodee. An \euvstn\ dense raster would still take several minutes to complete. In the {\tt MURaM\_collision} simulation, at $t=-20$ s ($20$ s after the identified trigger), a CME is already being initiated, accompanied by an EUV wave. \MUSE/SG has the cadence to capture this transient phenomenon while single-slit spectrographs would be too slow to keep up. Section~\ref{sec:ngspm_ii3} provides an extended discussion of how \musen/SG observations of EUV waves may be used to constrain and test CME models.

\begin{figure*}
\centering
\includegraphics[width=0.9\textwidth]{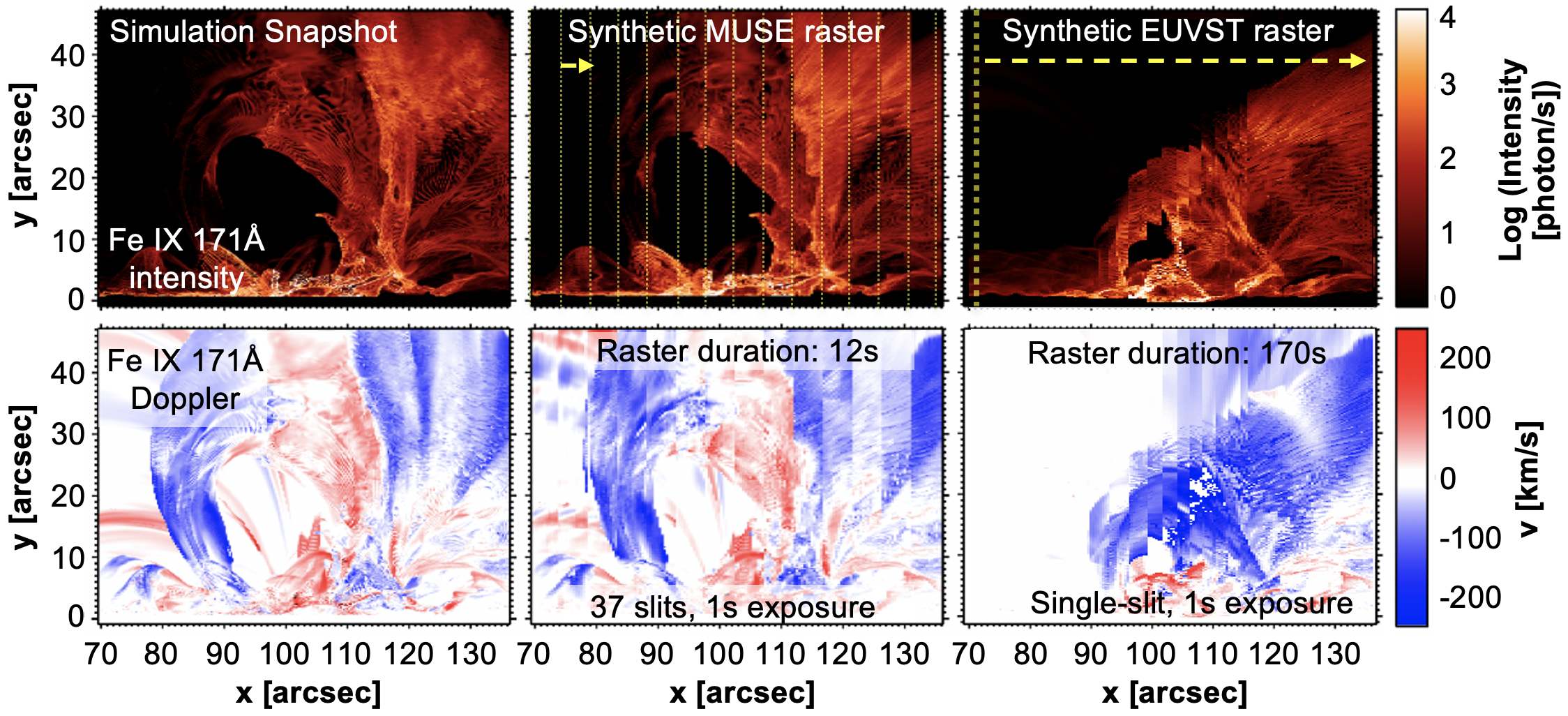} \\
\caption{\musen\ will capture the detailed evolution of solar eruptive events and their impact on the ambient corona. We show \musen\ \feixw\ synthetic observations from a radiative MHD simulation of a solar flare and CME ({\tt MURaM\_flare}, see Table~\ref{table_sims}, and Appendix~\ref{app:sim}). The \feix\ intensity and Doppler shift maps for a simulation snapshot (left column) are compared with corresponding line properties obtained via \musen\ (middle column) or \euvstn\ (right column) rasters. For both \musen\ and \euvstn\ we assume 0.4\arcsec\ steps, and 1~s exposures for each step. For \euvstn\ we assume a large dense raster covering the FOV shown ($\sim 65$\arcsec\ in the x direction perperndicular to the slit(s)). \musen\ high cadence large FOV observations would accurately capture the eruption properties. In the next figure (Fig.~\ref{fig:hgcr_erupt_euvst_2}) we show, for the same simulation, a time series of Doppler maps obtained with \musen\ and \euvstn\ rasters. 
}\label{fig:hgcr_erupt_euvst_1}
\end{figure*}

\begin{figure*}
\centering
\includegraphics[width=0.9\textwidth]{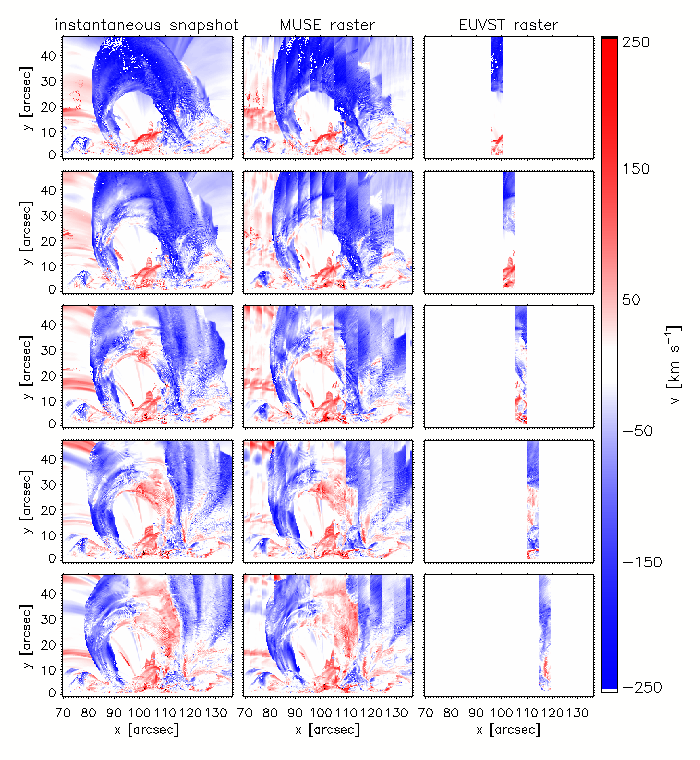} 
\caption{\musen\ high cadence spectroscopy over a large FOV is crucial to capture the detailed dynamics of solar flares and eruptive events. For the same radiative MHD simulation of a solar flare and CME ({\tt MURaM\_flare}, see Table~\ref{table_sims}, Appendix~\ref{app:sim})  shown in the previous figure (Fig.~\ref{fig:hgcr_erupt_euvst_1}) we show a time series, covering about 1 minute overall ($\sim 12$~s for each \musen\ raster), of \feix\ Doppler shift maps obtained with \musen\ rasters (middle column) and \euvstn\ rasters (right column). As in the previous figure for both \musen\ and \euvstn\ we assume 0.4\arcsec\ steps and 1~s exposures for each step, and dense rasters (i.e., $0.4$\arcsec\ raster steps in the x direction, perpendicular to the slit(s)). }\label{fig:hgcr_erupt_euvst_2}
\end{figure*}

While \EUVST\ rasters at AR-scale will miss the transient/TR coronal dynamics across the FOV (e.g., see Figs.~\ref{fig:hgcr_erupt_euvst_1} and~\ref{fig:hgcr_erupt_euvst_2}), sit-and-stare observations or rasters with fewer steps (and thus higher cadences) would capture the thermodynamic structure of the entire atmosphere along the LOS for a narrow field of view (FOV). In the absence of large-scale context data from \musen, it would be challenging to interpret the \EUVST\ rasters. But with datasets from the two observatories combined, we benefit from the spatio-temporal cadence of \musen, as well as the temperature coverage and density diagnostics of \EUVST. The photospheric and chromospheric slitjaw imaging (SJI) capability of \EUVST\ (in the 2833\AA\ continuum, 2852\AA\ Mg I and 2796\AA\ Mg II bands; none in coronal lines) will reveal the location and morphology of pre-eruption prominence material, and help locate the footpoints of reconnecting field, i.e., the flare ribbons. If \euvstn\ were rastering a region around the neutral line with the slit along the neutral line, it would be able to capture the low-T signatures of reconnection associated with tether-cutting, while \musen\ can capture the upper TR and coronal reconnection across the whole FOV. \EUVST/SJI will also facilitate alignment of the \EUVST\ $\times$ \MUSE\ dataset with \DKIST\ Visible Broadband Imager (VBI) observations. Photospheric and chromospheric magnetic field measurements will be provided by the \DKIST\ Visible Spectropolarimeter (ViSP) and the Diffraction-Limited Near-InfraRed Spectro-Polarimeter (DL-NIRSP), and for off-limb  regions, coronal field measurements by the Cryogenic Near-InfraRed Spectro-Polarimeter (Cryo-NIRSP). Due to integration times needed, Cryo-NIRSP would only provide before and after states of the coronal magnetic field, and would not provide dynamic changes during eruptions.

\begin{figure*}
    \centering
    \includegraphics[width=0.95\textwidth]{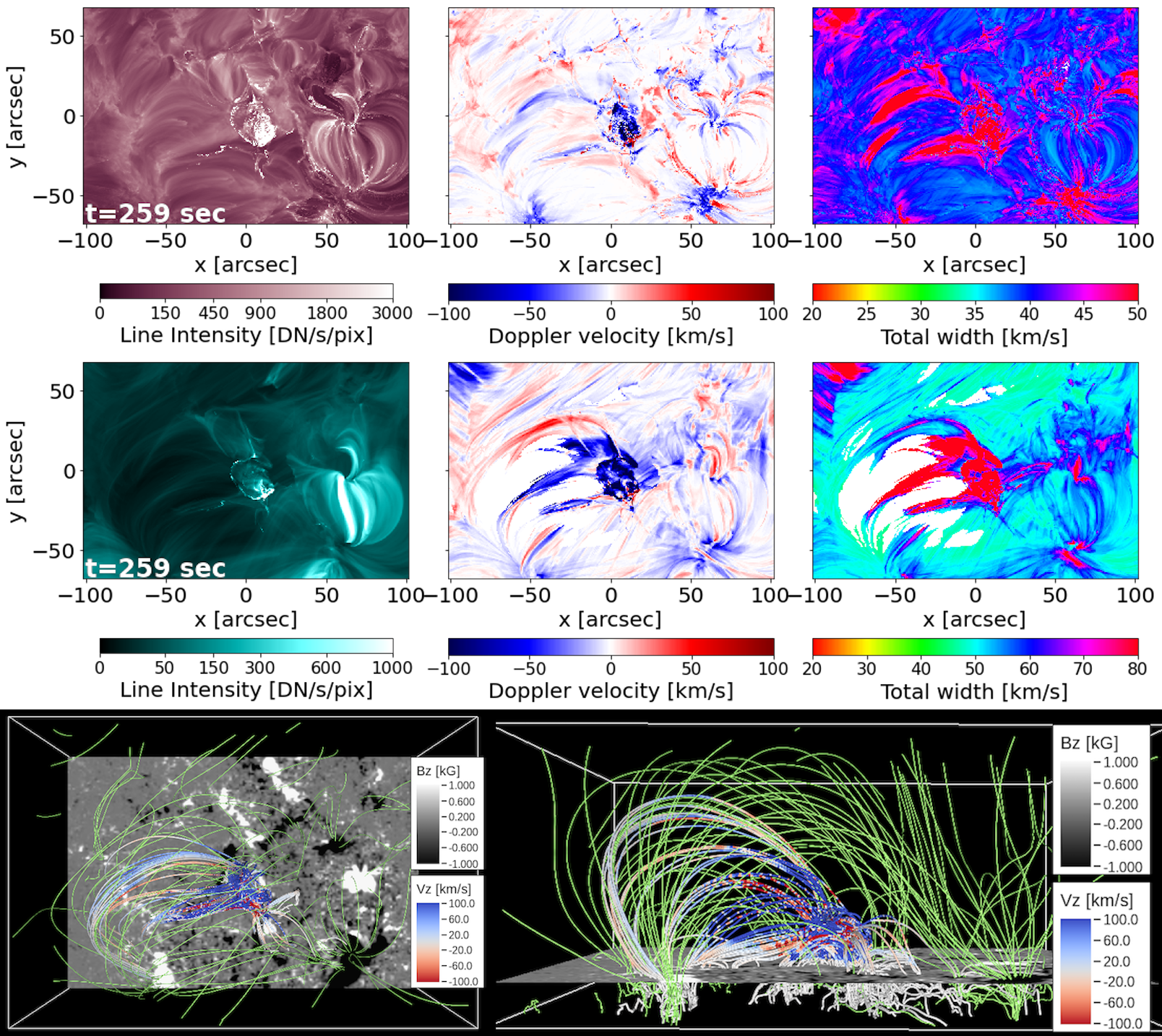}
    \caption{Top rows - Left: Zeroth (total intensity), first (Doppler shift) and second (line width) moments in the \ion{Fe}{15} 284 \AA\ and \ion{Fe}{19} 108 \AA\ lines for a simulation of a circular ribbon flare \citep[simulation {\tt MURaM\_circ\_rib}, see Table \ref{table_sims} and Appendix \ref{app:sim};][]{ChenRempelFan:2021}. During the onset of the circular ribbon flare \musen\ observables highlight the magnetic field structure consisting of a combination of locally closed and more distantly connected loops, which is in this case more complex than the classical fan-spine structure. Bottow row -  Field line traces for a top (left) and side (right) view. Field lines in green show the background field, field lines in red/blue color are selected to highlight fieldlines associated with the flare. They are randomly selected with a bias towards regions with a strong downward directed conductive heat flux. The color coding is based on the vertical flow velocity and shows close correspondence to the first moments observable with \musen\ in the top view. Rapid evolution requires a fast raster scan of the AR.}
    \label{fig:circular_ribbon_flare}
\end{figure*}

One particular flare topological configuration that would be particularly well suited for \musen\ observations is a circular ribbon flare with a fan-spine topology \citep{lau:1990}. They usually occur when new magnetic flux emerges in a region of dominant field, creating a parasitic polarity that connects locally creating a quasi-spherical domain or dome of close loops~\citep[e.g.,][]{Shibata:1994,MorenoInsertis:2008,WangLiu:2012}, with a coronal magnetic null and a spine line that goes through it. In this topology, field lines from different domains reconnect at the null as emergence progresses or as an embedded flux rope becomes unstable, producing observed intensity signatures at the quasi-circular ribbon that outlines the footpoints of the fan separatrix surface, and even more interestingly at distant brightenings that point at the location of the spine footpoint. These events have been observed with imaging instruments \citep[e.g.,][]{masson:2009,Sun:2013,Hernandez:2017,Xu:2017}, but are very challenging to be comprehensively observed with a single slit spectrograph when the distant flare signatures at the spine are typically 100-150\arcsec\ apart from the main ribbons. \musen\ is the ideal instrument to diagnose the still unknown spectral properties at the spine and circular ribbons and their spatial and temporal interdependency. The topology is uniquely favorable to characterize and understand the intensity, Doppler shift and non-thermal signatures of magnetic reconnection at and around a null point. Fig. \ref{fig:circular_ribbon_flare} shows a circular ribbon flare observed in simulation {\tt MURaM\_circ\_rib}. During the onset of the circular ribbon flare MUSE observables highlight the magnetic field structure consisting of a combination of locally closed and more distantly connected loops, which is in this case more complex than the classical fan-spine structure. The \ion{Fe}{15} 284 \AA\ maps show intensity and Doppler signatures at the circular ribbon flare ($[x,y]=[0,-10]$) and the distant footpoints ($[x,y]=[-25,-25]$\arcsec). The rapid evolution, that in the case of the Doppler signatures is at the scale of 1 minute or less, requires a fast (tens of seconds) raster scan of the AR.  \euvstn\ will be able to do a 50-60” dense raster of this compact example in $\sim$1-2.5min (0.5-1\ s exposures), but a typical 100\arcsec\ FOV would take 2-4 min, both insufficient to capture the dynamics of reconnection. An \euvstn\ sit-an-stare observation at the ribbons or fortuitously at the spine would provide a full temperature diagnostic at the appropriate cadence, at a single location. MUSE with its multiple slits, will be able to obtain diagnostics at both locations simultaneously, producing a raster of $170\arcsec \times 170\arcsec$ in 12\ s, making it the ideal instrument to firmly establish the spectral constraints and tests to current model predictions such as MURaM.

\subsection{II-3: Understand the evolution and propagation of CMEs and their effect on the surrounding corona}
\label{sec:ngspm_ii3}
The capability of forecasting hazardous solar eruptions depends critically on how well we understand the initial physical conditions of the source region(s) of CMEs, the physical environment through which the CME develops and propagates, and the physical processes involved when the CME interacts with the ambient corona. While \EUVST\ (with seamless temperature coverage) and \DKIST\ (with magnetic field information) can provide useful information when observing the CME-corona interaction region, the observations of \musen\ with a large field-of-view, and at higher temporal resolution are necessary for investigating this important aspect of solar eruptions. 

\subsubsection{Initial physical conditions of CME source regions}

\begin{figure*}[h]
\centering
\includegraphics[width=\textwidth]{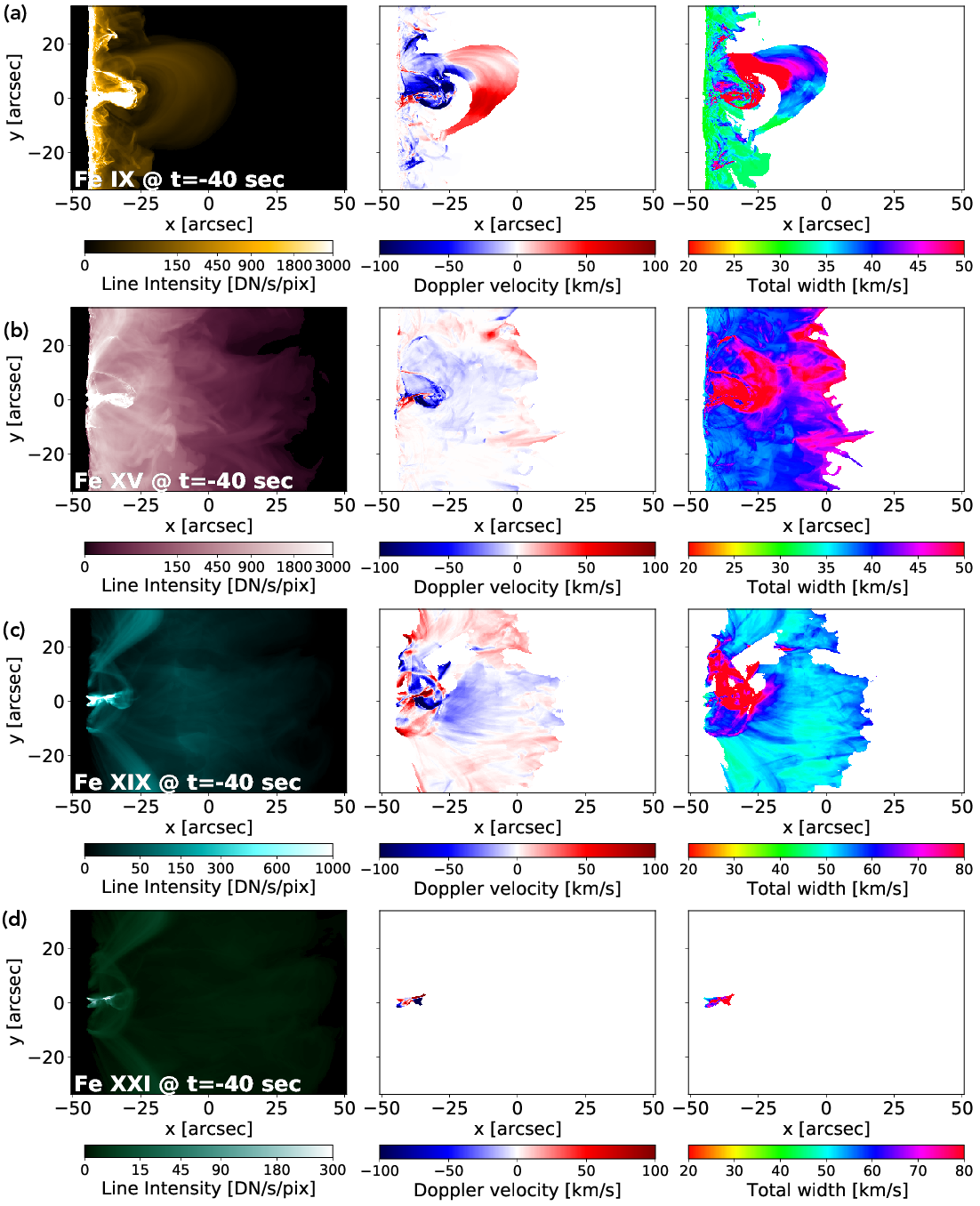}
\caption{\musen\ will capture the triggering and evolution of solar eruptive events and their impact on the ambient corona. Rows (a)-(d) show the synthetic \musen\ observables, namely the spectral line intensity (left column), Doppler velocity (middle) and line width (right column). (a): \ion{Fe}{9} 171~\AA; (b) \ion{Fe}{15} 284 \AA; (c) \ion{Fe}{19} 108 \AA; (d) \ion{Fe}{21} 108 \AA~line. These are synthesized from a radiative MHD simulation of a solar flare and CME (\texttt{MURaM\_collision}, see Table~\ref{table_sims}  and Appendix~\ref{app:sim}). See Fig.~\ref{fig:collision_zdir} for a top-down (disk center) view of this model. Hot plasma ($T \sim 12 $MK) is detectable in the \ion{Fe}{21} line showing magnetic reconnection under the flux rope. }\label{fig:collision_xdir}
\end{figure*}

\begin{figure*}[h]
\centering
\includegraphics[width=\textwidth]{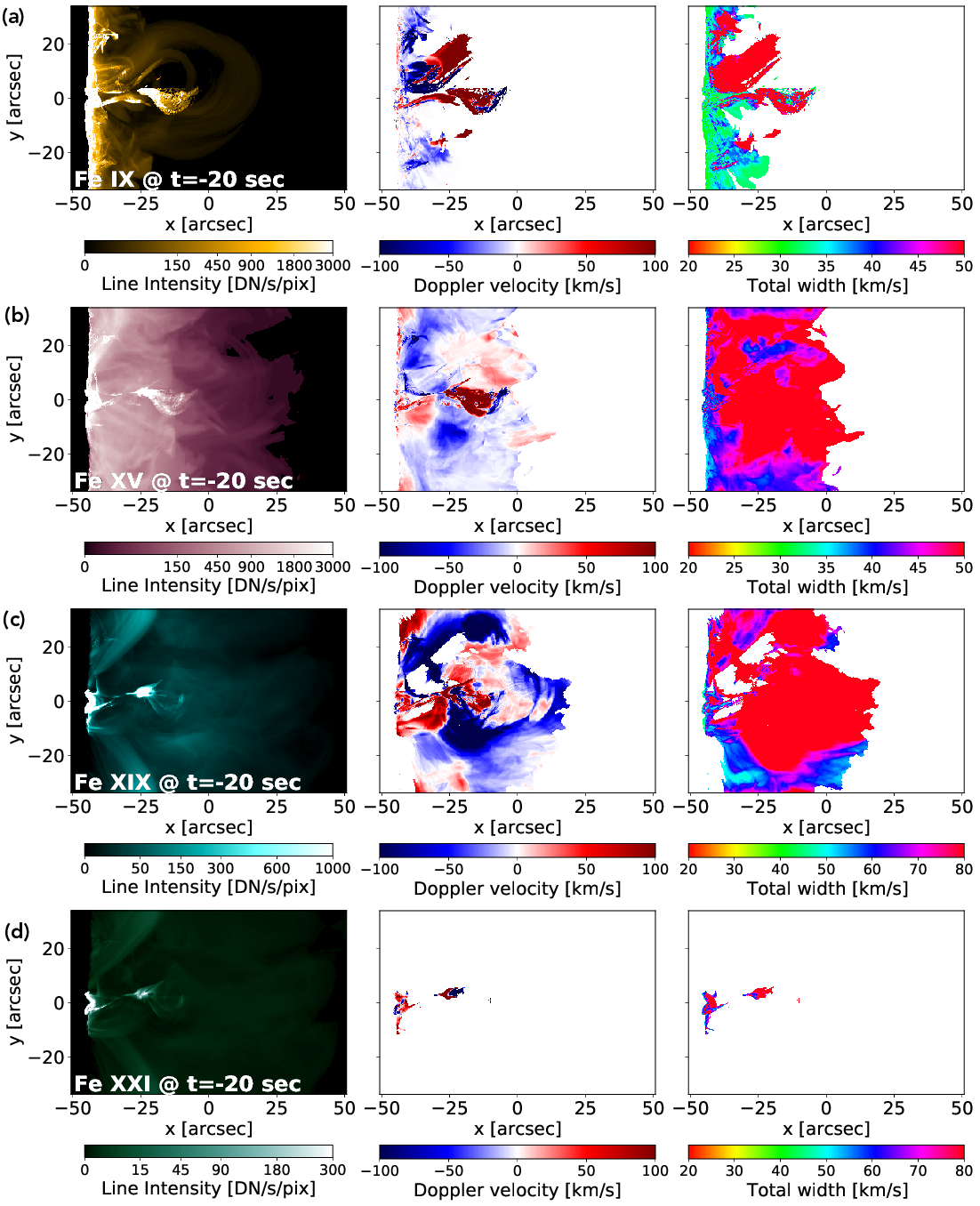}
\caption{Same as Fig.~\ref{fig:collision_xdir} but 20\ s later.}\label{fig:collision_xdir_later}
\end{figure*}

There are a number of ways \musen\ will improve models of CMEs. First of all,  \musen\ coronal imaging can be used to constrain 3D models of the coronal magnetic field (Section~\ref{sec:ngspm_ii}). This includes the identification of sigmoid-like loops (for on-disk observations) which are signatures of twisted magnetic fields (e.g., see intensity maps in Figs.~\ref{fig:multires} and \ref{fig:collision_zdir}),  the field structure of circular ribbon flares (see Fig.~\ref{fig:circular_ribbon_flare}), and coronal bubbles associated with hot flux ropes (for off-limb observations, see Fig.~\ref{fig:collision_xdir} and Fig.~\ref{fig:collision_xdir_later}). 

Secondly, \musen\ observations can help locate where and when eruption/flare triggers occur (e.g., \ion{Fe}{21} 108 \AA\ Doppler velocity and line width maps in Fig. \ref{fig:collision_zdir}; Section~\ref{sec:ngspm_ii2}). 

Thirdly, \musen\ will provide new constraints on the early phases CMEs to initialize data-constrained models of CMEs~\citep[e.g.,][]{Downs:2012,Shiota:2016,Toeroek:2018}. For example, take AWSoM+EEGGL \citep{vanderHolst:2014,Jin:2017}, a module delivered to NASA's Community Coordinated Modeling Center (CCMC) for the space weather community to run data-driven CME models. Currently, this model uses the observed CME speeds from coronagraphs as well as the photospheric magnetic field measurements from the CME source region to constrain the Gibson-Low flux rope parameters (e.g., location, size, magnetic strength, orientation, and helicity). However, the plasma properties within the flux rope are not currently constrained by observations.  \musen\ Doppler observation of the erupting flux rope could be used to specify the early velocity profile (e.g., the \ion{Fe}{15} Doppler map at $t=-20$s in Fig.~\ref{fig:collision_zdir}) of a magnetic flux rope, which is not explicitly specified in the current model. Such Doppler maps would be particularly important for constraining the early conditions of CMEs from on-disk source regions. Such CMEs are more likely to be Earth-directed and geoeffective.

For off-limb CME source regions, the spatial distribution of the flow pattern (e.g. as shown in Fig.~\ref{fig:collision_xdir}) will still be used to better constrain other flux rope parameters (e.g., orientation). Furthermore, the early stage kinematic CME information (e.g., speed, acceleration) obtained by \MUSE\ will provide important data for estimating the CME terminal speed and energetics, which are used to determine the magnetic flux of the GL flux rope in the EEGGL. Currently, this information is obtained from white-light coronagraphic observations when the CME is already at several solar radii. To obtain this information earlier and more accurately will be critical for improving the model capability for space weather forecast purposes. Last but not least, the \MUSE\ non-thermal line width observations will act as improved constraints of the wave heating parameters of global MHD models. The current AWSoM model uses constant heating parameters throughout the whole simulation domain~\citep{vanderHolst:2014}. The non-thermal broadening information from \MUSE\ will allow to apply spatially-resolved wave heating parameters for the CME source region. This can impact the properties of the solar wind solution, and the global magnetic topology of the ambient field. Both these effects are known to impact CME propagation and deflection~\citep{Lugaz:2011,Manchester:2017}.

\subsubsection{Interaction of Eruptions with the Ambient Corona}
CME propagation can be affected by the global coronal magnetic field as manifested, for example, in deflection and rotation (see review paper by \citealt{Manchester:2017}).  CMEs can also impact large-scale coronal structures, as seen, for example, in remote filament oscillations and sympathetic flares/eruptions~\citep{SchrijverTitle:2011,Jin:2016}. In order to understand the early evolution of CMEs and to obtain important information on coronal structures, we need to understand low-coronal signatures of CMEs (e.g., EUV waves, coronal dimmings, supra arcade downflows.) and to measure velocities projected in line-of-sight and on the plane-of-the-sky (POS).

To illustrate how \MUSE\ observations will advance our understanding of the interaction of eruptions with the ambient corona, Fig.~\ref{fig:collision_zdir} shows an example of the synthetic \MUSE\ observables from a radiative MHD simulation of a solar flare and CME (\texttt{MURaM\_collision}, see Table~\ref{table_sims}). At $t=-20$s, strong downward motions are evident outside the AR with a maximum value exceeding 100~km~s$^{-1}$. This phenomenon is due to the downward push of the CME during its expansion into the corona, which in rare cases has been observed by \hinodee\ with the ``sit-and-stare" mode with deep exposures of 45\ s \citep{Harra:2011, Veronig:2011} as well as in simulations (e.g., \citealt{Jin:2016}). Note that the flow pattern is highly structured and varies significantly on timescales of seconds, which will provide important diagnostics about the erupting flux rope (e.g. the coronal dimming of the flux rope footpoints; see \ion{Fe}{15} intensity map at $t=-20$s, at (x,y) = (-10,5)\arcsec ) and its interaction with the ambient corona. Therefore, while \EUVST\ (with better temperature coverage) can provide useful information when observing the CME-corona interaction region, \musen\ observations with a large field-of-view, and at higher temporal resolution are clearly required for investigating the non-local aspects of solar eruptions. \musen\ will capture the timing and location of local triggers  (e.g., tether-cutting, breakout reconnection) and possible remote triggers (e.g., EUV waves from other eruptions).

\subsection{II-4: Understand the processes of fast magnetic reconnection}
\label{sec:ngspm_ii4}

It is generally accepted that fast magnetic reconnection is necessary to power the acceleration of energetic particles, but the exact pathways by which this magnetic energy is converted into kinetic energy of particles, where the energy conversion happens, and how the energy is partitioned between thermal and non-thermal populations, is still under intense debate \citep{ 2011SSRv..159..357Z}. Although the EUV spectral lines observed by \musen\ will not directly probe plasma populations at super-hot (tens of MK and above) temperatures, it will provide important observations revealing the nature of fast reconnection (e.g., the plasmoid instability), and the plasma environment in which particle acceleration takes place (e.g., the structure of termination shock regions in reconnection outflows). \musen\ observations of flare ribbons can also be used as constraints on flare loop models investigating how loop atmospheres respond to injection of different particle/energy deposition mechanisms (see Section \ref{sec:ribbons}).

\subsubsection{Plasmoid Instability in Current Sheets}

\begin{figure}
    \centering
    \includegraphics[width=0.5\textwidth]{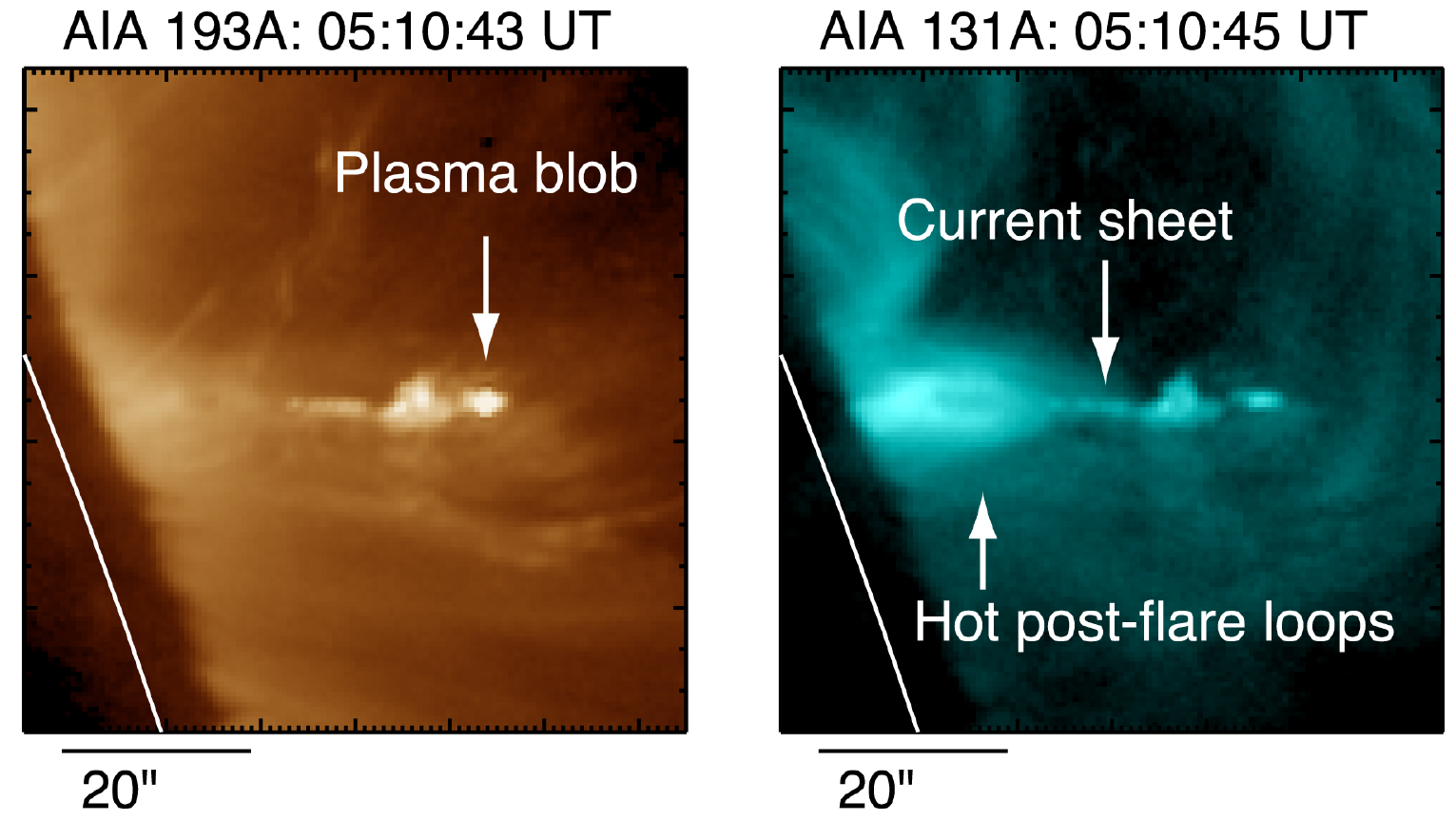}
    \caption{\sdo/\aia\ narrowband imaging data show bidirectionally moving (in the plane-of-sky) `plasma blobs' ejected from a current sheet~\citep{Takasao:2012,Takasao:Plasmoids}. The identified ejected plasmoids have widths of 2\arcsec-3\arcsec. \musen\ will provide imaging spectroscopic observations, testing model predictions that plasmoids are produced over a range of length-scales, and can coalesce and be ejected out of a current sheet.}
    \label{fig:takasao_plasmoids}
\end{figure}

\begin{figure*}
    \centering
    \includegraphics[width=1.\textwidth]{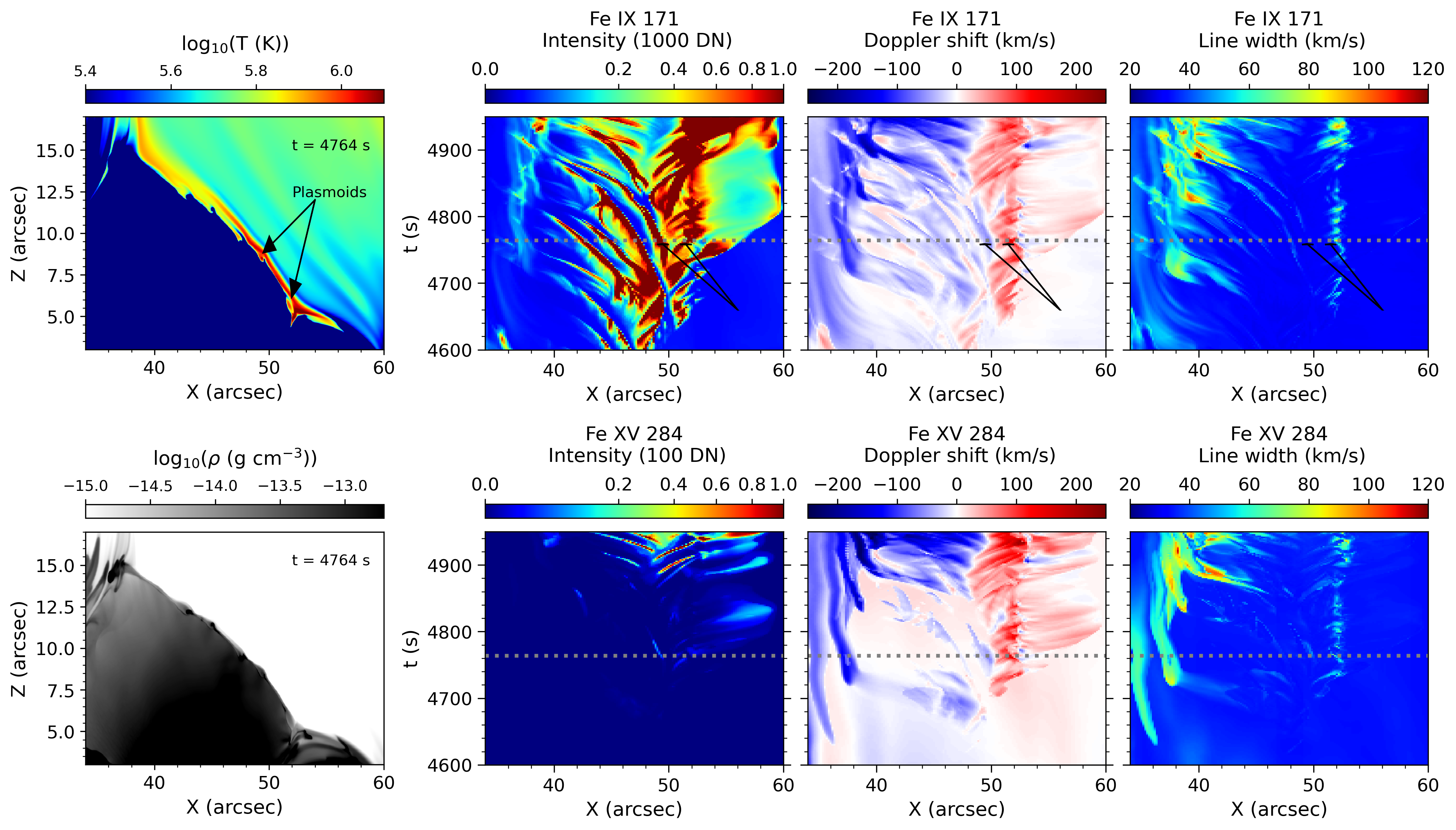}\\

    \caption{Formation and ejection of multiple plasmoids within an elongated current sheet in the Bifrost simulation {\tt B\_npdns03} (see Table~\ref{table_sims} and  Appendix~\ref{app:sim}). Left column shows temperature (top) and mass density (bottom) at time $t=4764$~s of the run. The panels on the right contain X-t maps of the MUSE \feixw\ (top) and \fexvw\ (bottom) moments, namely, intensity, Doppler shift, and line width. The horizontal dotted line in these panels correspond to the time shown in the temperature and density panels. An animation of the figure is also available showing the evolution of the temperature and density between $t=4600$~s and $t=4950$~s.}
    \label{fig:plasmsim}
\end{figure*}

In recent years, an emerging picture of magnetic reconnection suggests reconnection occurs in a dynamical fashion, unlike the earlier models of steady-state Sweet-Parker and Petschek-type reconnection scenarios. A robust result of numerical magnetic reconnection experiments at sufficiently high Lundquist numbers $S\gtrsim 10^4$ ($S = V_A L /\eta$, where $V_A$ is the Alfv\'en speed, $L$ the system-scale length, and $\eta$ the magnetic resistivity) is the formation of plasmoids~\citep{Loureiro:2007,Samtaney:2009,Bhattacharjee:2009,PucciVelli:2014,Shibayama:2015}.~\citet{ShibataTanuma:2001} have postulated that plasmoids exist in a spectrum of sizes exhibiting a fractal nature. These plasmoids are apparently a basic property of reconnection, seen in simulations where the initial conditions (magnetic field strength, thermodynamic variables) are symmetric or asymmetric about the current sheet (e.g., Fig.~\ref{fig:plasmsim}). While bidirectionally moving (in the plane-of-sky) plasma blobs resembling plasmoids have been identified in sequences of \sdo/\aia\ EUV images~\citep[][see Fig.~\ref{fig:takasao_plasmoids}]{Takasao:2012}, there exists no clear evidence that plasmoids are produced over a wide range of scales. Single-slit observations of non-thermal broadening of TR lines by \iris\ are consistent with the plasmoid instability~\citep{Innes:2015},  including the onset of fast reconnection mediated by plasmoids \citep{Guo:2020ApJ...901..148G}. However, the sit-and-stare observations used do not constrain the spatial structure and temporal evolution given the large plane of the sky motions \citep[see also][]{Rouppe-van-der-Voort:2017uo}. \iris\ slitjaw images have also revealed some features that could be plasmoids~\citep{Antolin:2021, Sukarmadji2021inprep}. With 0.4\arcsec\ resolution spectroscopic imaging data and 0.33\arcsec\ context imaging data, \MUSE\ will test whether plasmoids at sub-arcsec scales exist, and track their dynamical evolution.

The classical theory of fast Petschek reconnection involves standing slow-mode shocks across which magnetic energy is converted into kinetic energy. The fully-compressible MHD simulations of Harris sheet reconnection at $S\sim 10^4$ by~\citet{Shibayama:2015} predict the existence of dynamical (i.e., non-standing) slow-mode shocks between plasmoids. They also report that the existence of these Petschek-type slow-mode shocks is efficient in removing ejecta from localized reconnection regions, which enhances the reconnection inflow.  If the plasma ejecta of size $\sim3$\arcsec\ reported by~\citet{Takasao:2012} were indeed plasmoids, the accompanying dynamical slow-mode shocks can potentially be mapped by \MUSE, if they exist. Their discovery would provide support for the dynamical Petschek reconnection model. 

\euvstn, with a single slit and lacking coronal slitjaw imaging capability, will be unable to spectrally raster fast enough the elongated coronal currents to probe the highly dynamic evolution of the plasmoid instability. This observational gap will be filled by \musen's multi-slit spectral imaging capability, which will raster a FOV with comparable resolution to \EUVST\ at 30x to 100x the cadence. As a representative example of the capabilities of \musen , Figure \ref{fig:plasmsim} illustrates the formation of multiple plasmoids within an elongated current sheet (length $L>15\arcsec$) in the Bifrost simulation {\tt B\_npdns03}. This figure also shows the plasmoids being ejected out of the current sheet. In the image, the left column shows maps of the temperature $T$ (top) and mass density $\rho$ (bottom) at time $t=4764$ s of the run. At this time, asymmetric magnetic reconnection occurs between the chromosphere and the corona leading to a hot coronal jet. In addition, the current sheet becomes unstable to the tearing-mode instability \citep{Furth:1963aa} and several plasmoids are created and ejected due to the imbalance of the Lorentz force. The arrows indicate the location of two such plasmoids that are ejected from the current sheet. 
The trajectory of the different plasmoids are distinguishable as slanted lines in the intensity space-time plots (right-most three columns). They are initially detectable in \ion{Fe}{9} 171~\AA~maps, then eventually in the \ion{Fe}{15} 284~\AA\ maps as the reconnection proceeds because the newly created and ejected plasmoids are hotter. Moreover, inspecting these slanted lines, the X reconnection point can be identified as the location from which the plasma flows diverge in opposite directions. This is the case for the two plasmoids we have highlighted as examples: the left one moves upwards to greater coronal heights while the right one descend towards the lower atmosphere.  We can also discern the coalescence of plasmoids as they appear as almost perpendicular lines to the trajectories with changes in the Doppler shift values. In addition, in the line width panels, it is possible to know where the plasmoids impact after being ejected from the current sheet, e.g., between $x=38$\arcsec\ and $x=40$\arcsec\ and at $x=52$\arcsec. This way, \MUSE\ offers a unique capability to unravel the nature of the plasmoids as well as their coalescence and impact against preexisting magnetic field. Since plasmoids are considered possible environments where energetic electrons may be accelerated \citep[e.g.,][]{Drake:2006}, the ability of \musen\ to capture the evolution of plasmoids will provide important constraints for models on the possible injection time and location of non-thermal electrons (NTE). See Section \ref{sec:ribbons} for further discussion.

\EUVST\ rasters of the pre-flare conditions will be important for establishing the physical conditions of the entire stratified atmosphere in a narrow field-of-view before the onset of reconnection. To probe the dynamics of the plasmoid instability in thin current sheets requires sub-arcsecond resolution and imaging cadence much faster than one minute. \EUVST\ would achieve this only for very narrow rasters (e.g., 8\arcsec\ wide rasters with 0.4\arcsec\ step size, 1s slit dwell time for 20\ s cadence). Since it is not known {\em a priori} where current sheets will form, the likelihood of narrow \EUVST\ rasters capturing the events of interest will be low as demonstrated by the fact that there are very few spectroscopic observations of the plasma sheet region with single slit spectrometers \citep[e.g.,][]{Warren:2018ApJ...854..122W}. Furthermore, lack of coronal context imaging from \MUSE\ would make narrow rasters very difficult to interpret. This observational gap must be filled by \musen's multi-slit spectral imaging capability, which will raster a FOV with comparable resolution as \EUVST\ at up to 100x the cadence, capable to scan a whole reconnection site with less than $20$~s cadence. These fast spectral imaging will follow the multi-thermal dynamical development of plasmoids, the evolution of reconnection null points, the colliding plasmoids with the open field, and the reconnected retracting loops.

\DKIST, \EUVST\ and \MUSE\ coordinating as a distributed NSGPM will provide unprecedented observational constraints on reconnection physics from collisional to collisionless plasmas, and from weakly- to fully-ionized plasmas.  \DKIST\ Cryo-NIRSP coronagraphic spectropolarimetric measurements of the \ion{Fe}{14}~5303~\AA, \ion{Fe}{11}~7892~\AA, \ion{Fe}{13}~10746~\AA, \ion{Fe}{13}~10798~\AA, \ion{Si}{10}~14301~\AA, and \ion{Si}{9}~39343~\AA~lines will provide diagnostics of the line-of-sight component of the coronal magnetic field, as well as orientation of the plane-of-sky components~\citep{SchadDima:2020}. Spectropolarimetric observations by ViSP will map the photospheric and chromospheric magnetic field, directly showing locations of current sheets in the lower atmosphere. High cadence imaging from \DKIST\ VBI would also reveal whether current sheets operating in the fully collisional, weakly-ionized regime produce plasmoids as predicted from multi-fluid numerical experiments~\citep{Leake:2012}. Finally, the distributed NSGPM can study reconnection between magnetic fields loaded with plasma of different temperatures (e.g. chromospheric and coronal), an example of which is shown in Fig.~\ref{fig:plasmsim} \citep[see also][]{Rouppe-van-der-Voort:2017uo}.

\subsubsection{Fine structure of Termination Shock(s) and Reconnection Outflows}

There is increasing evidence that the sunward directed reconnection outflow impinging onto the flare arcade leads to the formation of a fast-mode termination shock~\citep{Chen:2015,Chen:2019,Luo:2021}. Recent work on modeling the September 2017 X8.2 flare and comparisons with EUV and radio observations has suggested that a termination shock is a plausible region for particle acceleration~\citep[as opposed to near the reconnection region, see][]{Chen:2020}. 

One spectroscopic signature predicted by the models are the deflecting flows downstream of the shock, which would be observed as large ($\approx$~200 \kms\ for \ion{Fe}{21}) blue and redshifts in the spectra of high-temperature EUV/UV lines \citep[e.g.,][see also the \ion{Fe}{21} Doppler map in Fig.~\ref{fig:collision_zdir} at $t=-20$s]{guo2017}. However, observations have remained very rare and elusive \citep[e.g.,][]{imada2013,polito2018b}, mostly due to the difficulty of observing the reconnection region at the right time, in the correct location and with the best orientation of the instrument with a single-slit spectrometer.

Some MHD models of the dynamical evolution of reconnection outflows impinging on arcade loops~\citep{Takasao:2015,Takasao:2016,Takahashi:2017,Kong:2019} predict the termination shock region to consist not of a single fast-mode shock, but multiple interacting fast-mode shocks. The magnetic field near this region has an upward concave geometry suitable for trapping particles (so-called~\emph{magnetic bottle geometry}), which allows particles to be accelerated to higher energies. This is a viable model to explain coronal loop-top x-ray sources with photon energies $E_{\rm ph} \gtrsim 25$ keV. However, there is currently a lack of direct evidence for the multi-part structure of termination shock regions. 

\begin{figure*}
\includegraphics[width=\textwidth]{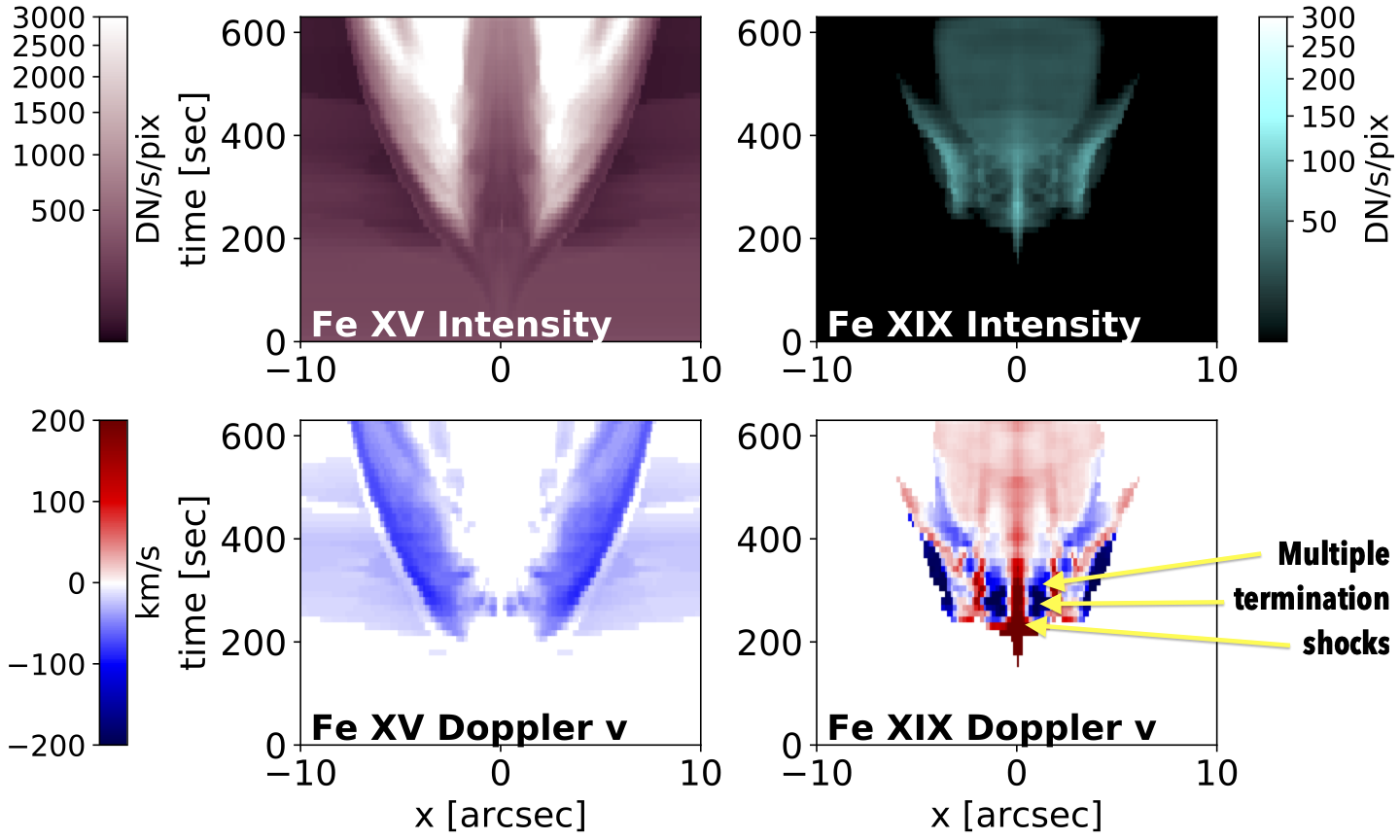}
\caption{Synthesized \musen\ observables (top-down view) from the flare arcade reconnection model {\tt Termination\_shocks} (see Table~\ref{table_sims} and Appendix~\ref{app:sim}) of \citet{Takasao:2015}. Shown are distance-time diagrams of the line intensity and Doppler shift of the \ion{Fe}{15} (284 \AA, 2 MK, left) and the \ion{Fe}{19} (108 \AA, 10 MK, left) lines. Interacting fast mode shocks in the sunward reconnection outflow appear as oppositely directly Doppler shifts in the 108~\AA~ line (near $x=0$). \musen\ has the spatiotemporal resolution to detect this type of~\emph{magnetic tuning fork} structure in flares.} \label{fig:tuning_fork}
\end{figure*}

Since \MUSE/SG will capture FOVs spanning $170\arcsec \times170\arcsec$ at 12\ s raster cadence (and faster for sit-n-stare and step sizes larger than 0.4\arcsec), it will capture the evolution of flare termination shock regions with much greater chances of success than single-slit instruments. Figure~\ref{fig:tuning_fork} shows how the multi-part termination shock structure in the simulation of~\citet{Takasao:2015} would appear as \musen\ observables (at a spatio-temporal sampling rate comparable to \musen's capability) for a top-down (i.e., disk center) view. The coronal current sheet in the simulation is located at $x=0$. The \fexixw\ line shows alternating patterns of blue- and red-shifts of approximately $\pm 100$ \kms, a signature of the multi-part termination shock structure. These regions are also accompanied by enhanced total line width of a $\sim 100$~\kms\ (see top left panel of Fig.~\ref{fig:takasao_snapshot} in Appendix~\ref{app:sim}). Detection of these signatures in \musen\ observations of loops would support models of the multi-shock nature of termination shock regions. Comparison of such dynamic models of the evolution of the termination shock region with \musen\ observables will constrain their magnetic geometries and evaluate their importance as sites for particle acceleration.

It has been proposed that such outflows are unlikely to be laminar \citep[][]{1993ApJ...418..912L}, and are instead likely to develop a turbulent structure that, cascading down to kinetic scales, is capable of bulk acceleration of electrons \citep[e.g.,][]{2010A&A...519A.114B,2014SoPh..289..881M}. Indeed, high cadence imaging observations by \aia\ are highly suggestive of the presence of turbulence in reconnection outflows \citep[e.g.,][]{2018ApJ...866...64C}. \cite{2017PhRvL.118o5101K} inferred a timescale for electron energization in such a region on the order of 1-10s. \musen/CI is capable of providing TR and coronal images in the \heiiw\ and \fexii/\ion{Fe}{24} 195 \AA\ bands at 0.33\arcsec\ resolution at cadences down to 8s/4s (dual/single channel). Furthermore, \musen/SG sit-and-stare rasters can run at a cadence as fast as 0.5~s when targeting flares. The combined capability will characterize the intermittency of the reconnection outflows, providing evidence for dynamical reconnection. While \euvstn\ could provide DEMs and density diagnostics with sit-and-stare or narrow raster sampling a region of the outflow, with \musen\ characterising turbulence throughout a larger volume.

\subsubsection{Supra-arcades, plasmoids and their relations to QPPs}

The origins of supra-arcade downflows (SADs), supra-arcade downflow loops~\citep[SADLs; see][]{Savage:2011}, Quasi-Periodic Pulsations \citep[QPPs; see][]{Nakariakov:2009SSRv..149..119N}, and their relationship with each other are still under debate. QPPs might be signatures of repeated/bursty reconnection, intermittent collision of plasmoids/SADs, MHD sausage mode oscillations, and more. They likely carry information about the energy release process in flares. \musen\ spectroscopic rasters will provide an unprecedented opportunity to study SADs, SADLs and QPPs in unprecedented detail.

The magnetic tuning fork structure in Fig.~\ref{fig:tuning_fork} has been proposed as the source of quasi-periodic pulsations (QPPs) emanating from flare loops~\citep{Takasao:2016}. The interacting fast mode shocks generate oscillations that radiate away from the termination shock region. An alternative explanation proposed as the driver of QPPs is the intermittent collision of plasmoids ejected from the current sheet colliding with flare arcade loops \citep{Samanta:2021}. MHD waves (fast sausage modes) are yet another possible explanation for QPPs. 

\begin{figure*}
\includegraphics[width=\textwidth]{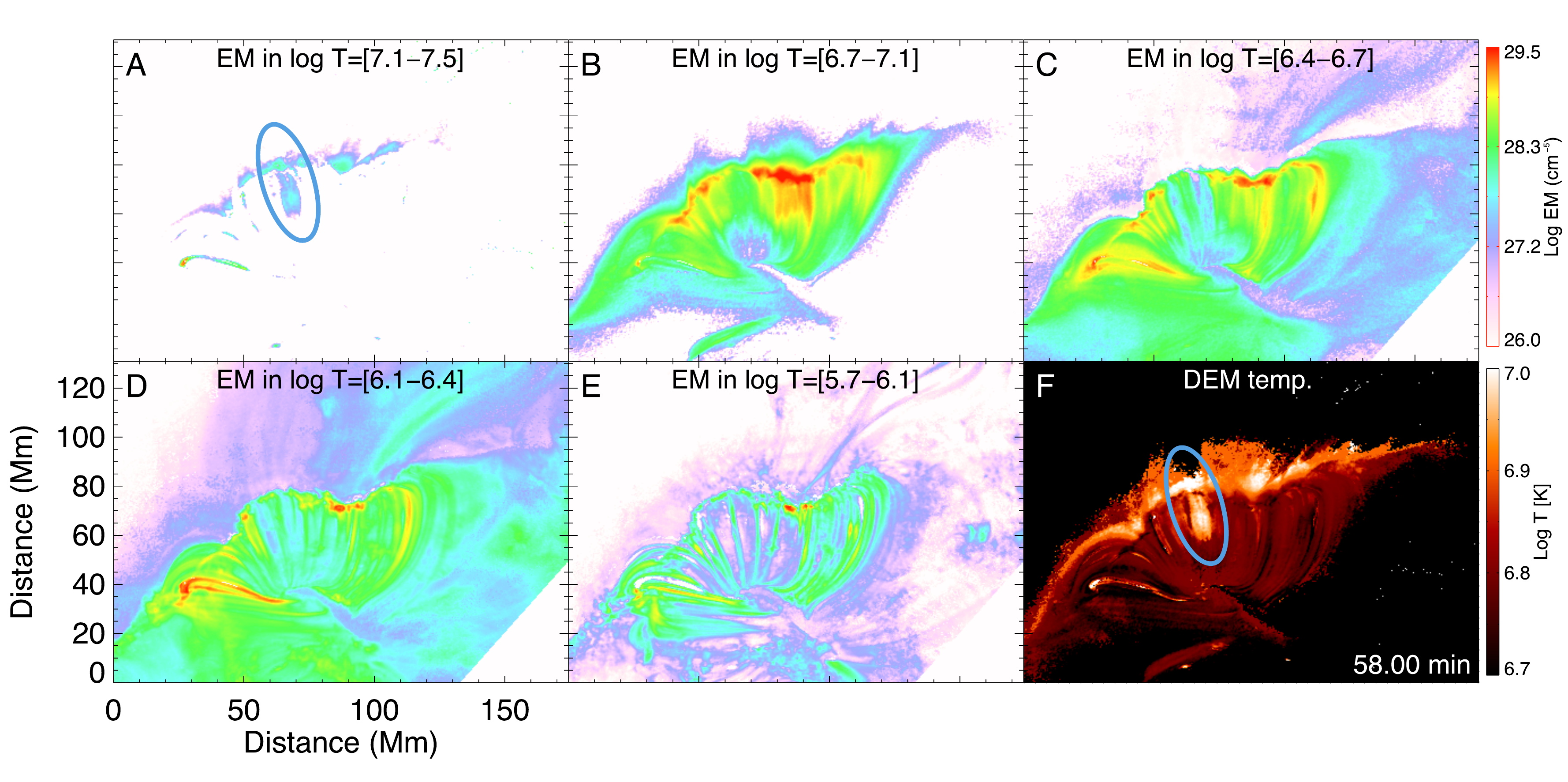}
\caption{\aia\ observations of flare loops and supra-arcades show dynamical structures with spatial coherence parallel, and across neighbouring flare loops. For example, this differential emission measure (DEM) map of such a system \citep[figure from][]{Samanta:2021} shows emission measure enhancements (see oval in panel A) along loops with width down to the \aia\ resolution ($\sim 1$\arcsec), as well as cross-loop extensions spanning several Mm. Spectroscopic imaging data by MUSE at 0.4\arcsec\ resolution will probe whether these enhancements have signatures of termination shocks predicted by MHD models (e.g., see Fig.~\ref{fig:tuning_fork}).}\label{fig:samanta}
\end{figure*}

Using narrowband EUV imaging data from \sdo/\aia\, \citet[][see Fig.~\ref{fig:samanta}]{Samanta:2021} reported the detection of episodic temperature and density enhancements in a flare arcade following the apparent collision of SADLs with the arcade loops. The authors propose that individual QPP pulses are driven by the collision of retracting SADLs with the underlying arcade. SADs and SADLs have typical speeds of hundreds of \kms\ and are spatially and temporally intermittent. Single-slit spectroscopic rasters with cadences of a few minutes are insufficient to track their evolution. Furthermore, flare arcade loops are not typically straight in the plane-of-sky. So it is very difficult to catch the evolution of plasma along flare loops when operating a single-slit experiment in sit-and-stare mode. \musen's multi-slit approach addresses the need to capture the dynamics of SADs, SADLs and QPPs at sufficiently high spatio-temporal cadence to test models of their physical origin.

Magneto-acoustic waves, in particular fast sausage modes, are another possible interpretation for QPPs \citep{Li:2020SSRv..216..136L}. \citet{Tian:2016} attempted to detect oscillations of the width of the \ion{Fe}{21} line observed by IRIS. However, the cadence was not fast enough to observe the expected line width oscillation, or perhaps the location of the slit was not ideal. The multi-slit coverage of \musen\ will be able to capture such a line width oscillation, if it exists.

\subsubsection{Heating and Magnetic Evolution at Flare Ribbons}\label{sec:ribbons}
The coronal magnetic reconnection that facilitates energy release in flares leads to intense heating of the lower solar atmosphere, up to temperatures normally considered `coronal' \citep[e.g.,][]{2013ApJ...767...83G,2013ApJ...771..104F}. This results in the appearance of flare ribbons in the EUV, UV, and optical wavelengths~\citep[e.g.,][]{Isobe:2007,2011SSRv..159...19F,Yadav:2021}. Studying these ribbons helps bridge the gap between the reconnection and the eventual dissipation of the energy that is released. It is particularly important to observe flare ribbons at subarcsecond resolutions since ground-based H$\alpha$ observations of coronal rain in flare indicate loop widths as low as $\sim 100$ km~\citep{Jing:2016}. In the standard 2D flare picture~\citep[][]{Hirayama:1974SoPh...34..323H}, ribbons occur at the interface between distinct magnetic volumes, forming a topological discontinuity. In the 3D extension to the flare model~\citep[e.g.,][]{Janvier:2015}, conjugate footpoints of ribbons not only separate as new flux is reconnected, they also have a displacement along the direction of the PIL due to slipping reconnection. In both 2D and 3D cases, the topological change in the field results in reconnected field lines that can relax to a lower energy state. The Lorentz force work due to the field relaxation provides power to heat loops. In order to properly track the evolution of these loops (and thus the energy sources) in 3D requires high-cadence spectral imaging observations over AR-scale FOVs.
 
Combining flare ribbon observations with measurements of the magnetic field and its variation during flares gives direct information on the overall rate of flux transfer associated with magnetic reconnection \citep[e.g.,][]{2001SoPh..204...69F,2002ApJ...565.1335Q}. Using 1600~\AA~imaging data from \sdo/\aia\ and \trace\ for example, correlation studies of intensity and magnetic reconnection rate have been made on the scale of ARs, but for individual bright features within ribbons, e.g., \citet{2007ApJ...654..665T} where, for example, a 2-second cadence from \trace\ was used to demonstrate that parts of the ribbon where a high reconnection rate is measured are associated with the most energetic sources - i.e., the hard X-ray emitting regions \citep[][]{,2009A&A...493..241F}. These studies were possible only because of the high time resolution available in optical, UV and hard x-ray imaging observations.

\begin{figure*}[htb]
\begin{center}
	\vbox{
	\subfloat{\includegraphics[width = 0.6\textwidth, clip = true, trim = 0.cm 0.cm 0.cm 0.cm]{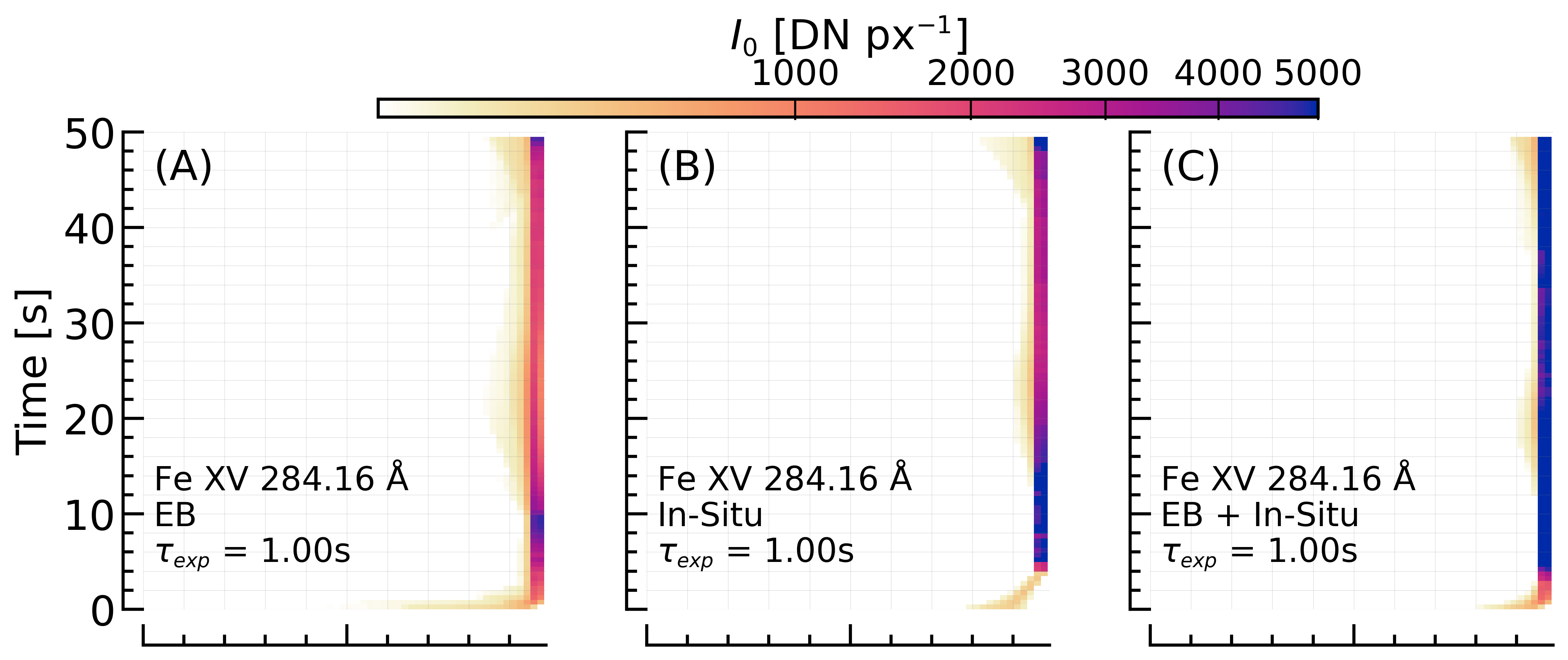}}	         
	}
	\vbox{
	\subfloat{\includegraphics[width = 0.6\textwidth, clip = true, trim = 0.cm 0.cm 0.cm 0.cm]{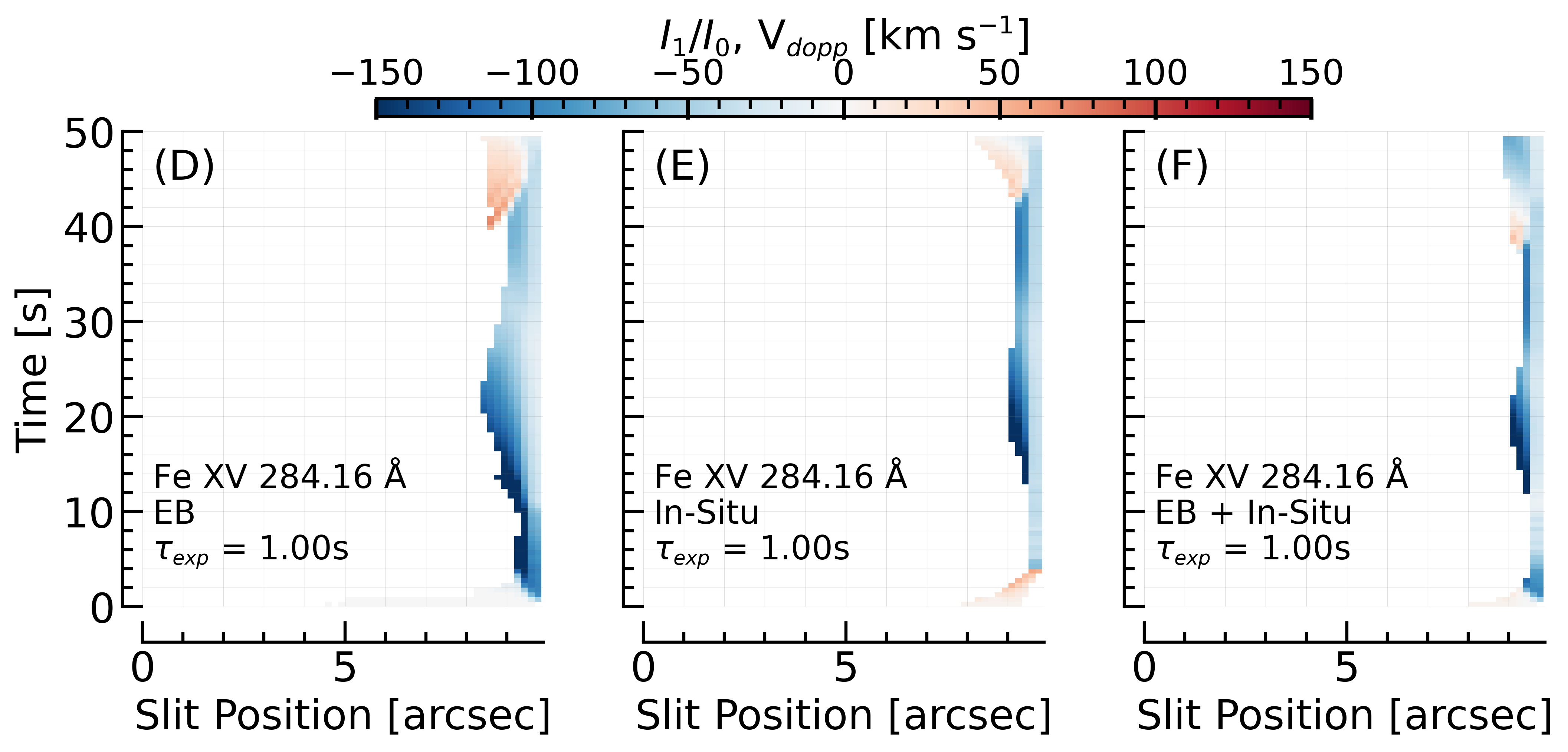}}
	}
	\vspace{0.05in}
	\vbox{
	\subfloat{\includegraphics[width = 0.6\textwidth, clip = true, trim = 0.cm 0.cm 0.cm 0.cm]{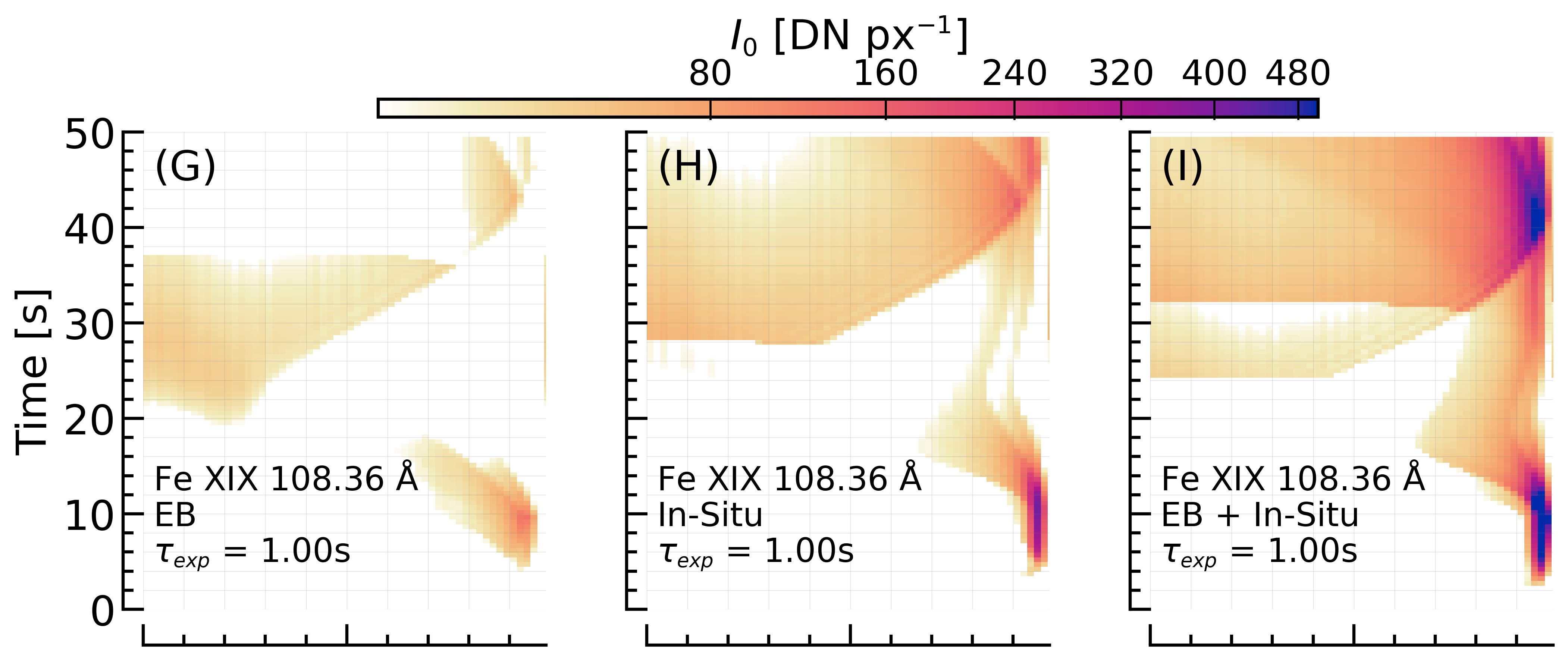}}
	}
	\vbox{
	\subfloat{\includegraphics[width = 0.6\textwidth, clip = true, trim = 0.cm 0.cm 0.cm 0.cm]{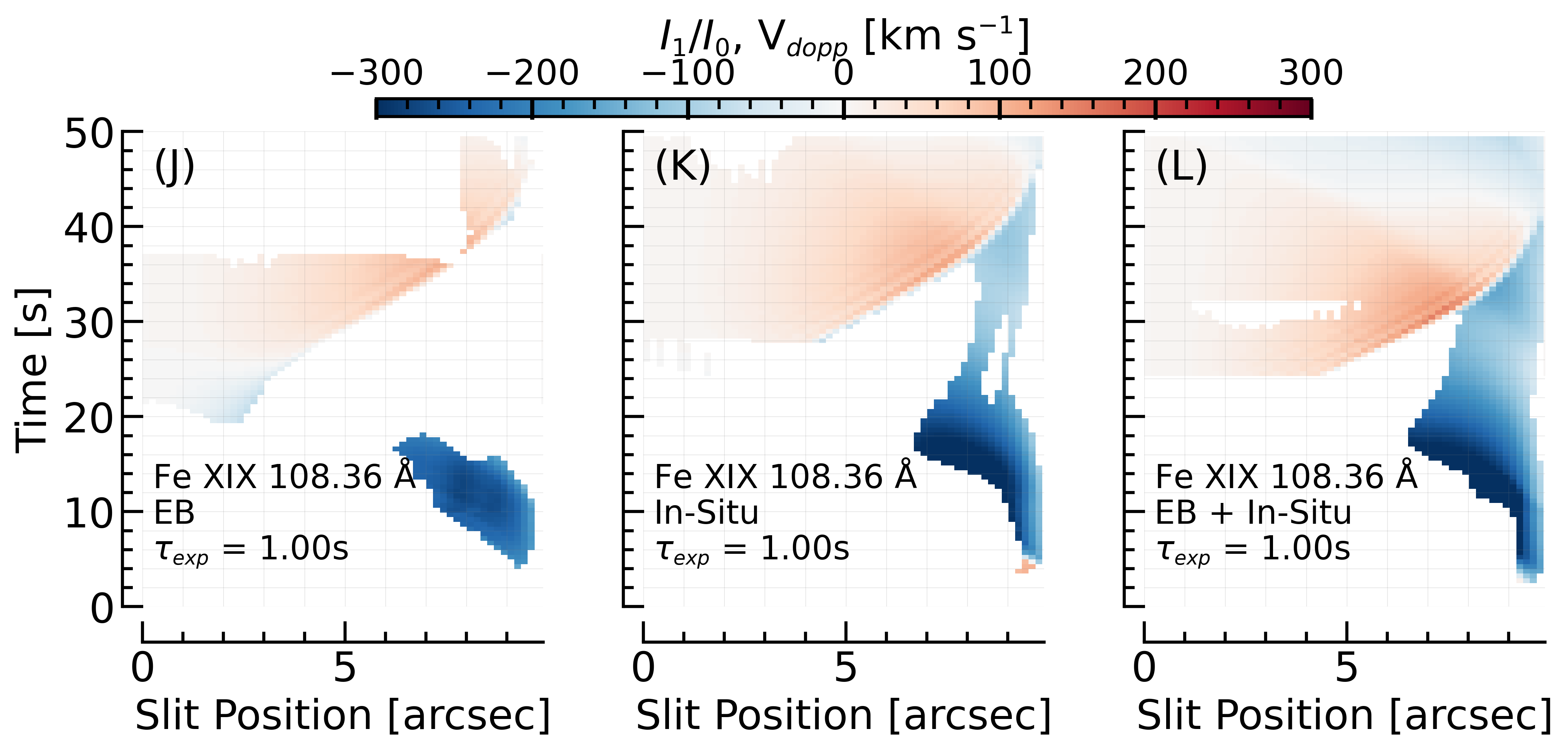}}
	}
	\caption{
	The timings and locations of flare responses in each line varies depending on the flare heating, making it important to have comprehensive spatial coverage on short timescales.
	We show first and second moment (see Appendix~\ref{app:moments}) of synthetic \musen\ lines (\fexvw\ and \fexixw\ in top and bottom two rows respectively), from three RADYN flare simulations with different flare heating models (see Appendix~\ref{app:sim} for details on the models shown here): an electron beam driven flare ({\tt RADYN\_warm\_EB}, left column), a flare driven by \textsl{in-situ} coronal heating ({\tt RADYN\_warm\_TC}, middle column), and a tandem EB plus \textsl{in-situ} scenario ({\tt RADYN\_warm\_EB\_TC}, right column). 
	}
	\label{fig:radyndiffheating}
\end{center}
\end{figure*}

Different energy transport and heating mechanisms result in distinctive thermal and dynamical properties at the flare ribbons. For example, evaporative upflows arising from heating by high-energy electrons are predicted to have a different behavior as a function of time from those due to conductive heating, as shown in simulations discussed below (Fig.~\ref{fig:radyndiffheating}). High time resolution is critical:  hard X-ray timescales for impulsive energy input are at least as short as 10~s, and the simulations suggest that transient phenomena at flare onset can be a distinguishing feature of different heating models. Previous \iris\ sit-and-stare observations at 1.7~s cadence also captured the rapid onset of transition-region flows and line broadening preceding the flare heating by some 10~s, posing a challenge for our understanding of lower atmosphere energization \citep[][]{2018SciA....4.2794J}. The large FOV of \musen\ will give simultaneous access to different parts of the flare ribbons on a spatial scale large enough to examine whether different energy transport mechanisms dominate at different times and locations in the flare (e.g. non-thermal electrons in at the strongest footpoints during the impulsive phase versus widespread conductively-driven evaporation later on) and with a temporal and spatial resolution sufficient to capture ribbon variability. Coupled with magnetic field measurements over the \dkistn/VTF FOV, and \euvstn\ spectroscopy providing additional plasma diagnostics over a narrower FOV, rapid progress on flare energy transport and its relationship to magnetic restructuring can be expected.

We demonstrate via flare radiation hydrodynamic modeling how \musen\ observations of hot flare plasma at high spatio-temporal resolution will shed light on the partition of energy following fast reconnection.

\begin{figure}
\includegraphics[width=0.45\textwidth]{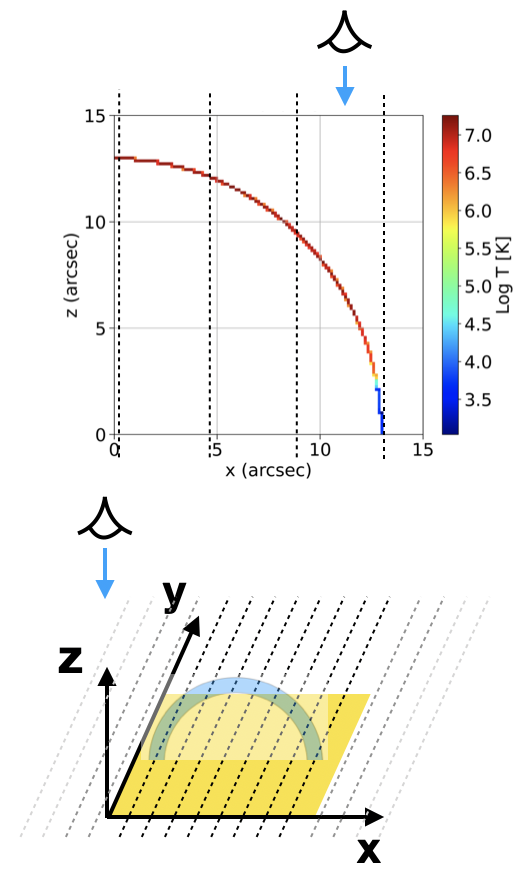}
\caption{\label{fig:radyn_mapping} Sketch to illustrate the mapping of 1D RADYN loop models to a 2D semicircular loop (lying in the $x-z$ plane; the top panel shows the simulated plasma temperature along half-loop for a snapshot of one of the RADYN models). We are assuming that  the line-of-sight coincides with the $z$ direction (i.e., the \musen\ slits are in the $x-y$ plane, as shown in the bottom  panel). \musen\ will allow to observe at different loop locations simultaneously the atmospheric response of the loop plasma to the energy flare release,  uniquely constraining its properties, as shown in Figures~\ref{fig:preflarediffs}, and~\ref{fig:radyndiffheating}.}
\end{figure}

\begin{figure*}[htb]
\begin{center}
	\vbox{
	\subfloat{\includegraphics[width = 0.7\textwidth, clip = true, trim = 0.cm 0.cm 0.cm 0.cm]{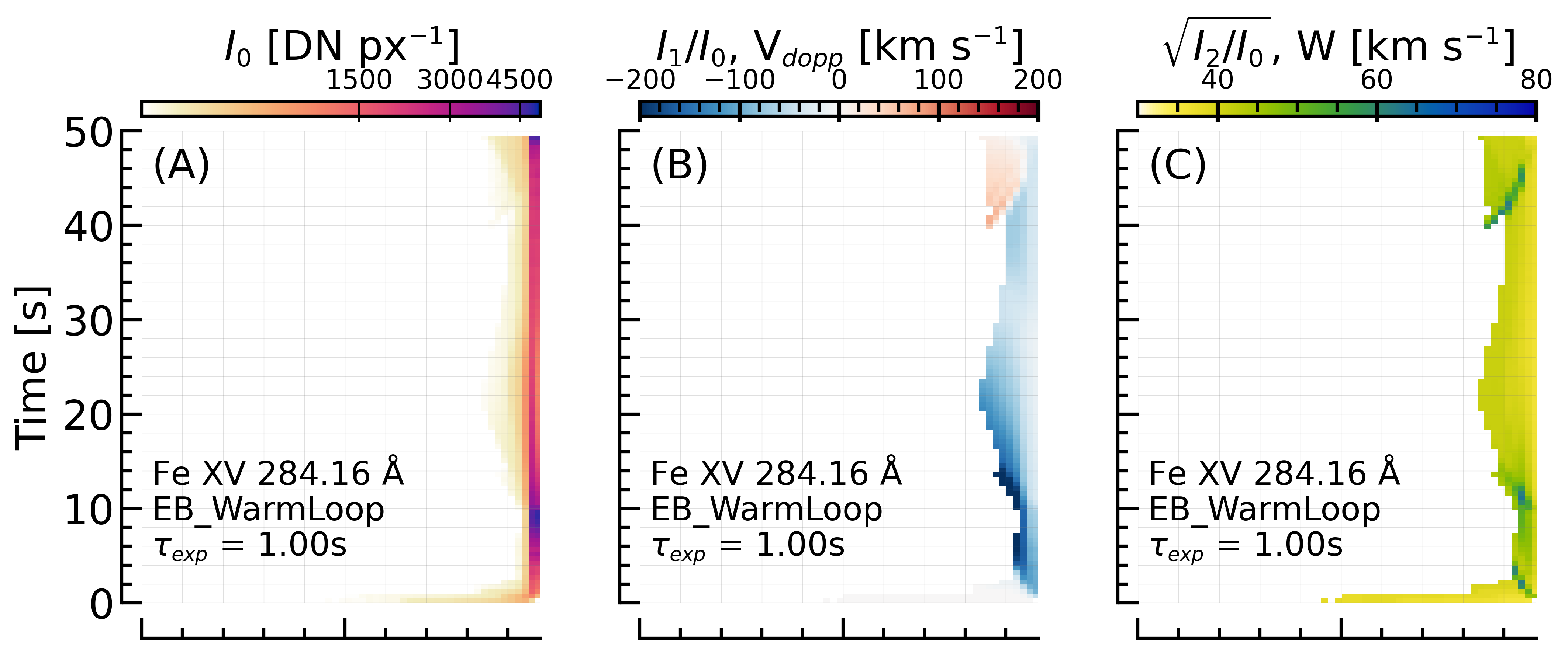}}	         
	}
	\vbox{
	\subfloat{\includegraphics[width = 0.7\textwidth, clip = true, trim = 0.cm 0.cm 0.cm 0.cm]{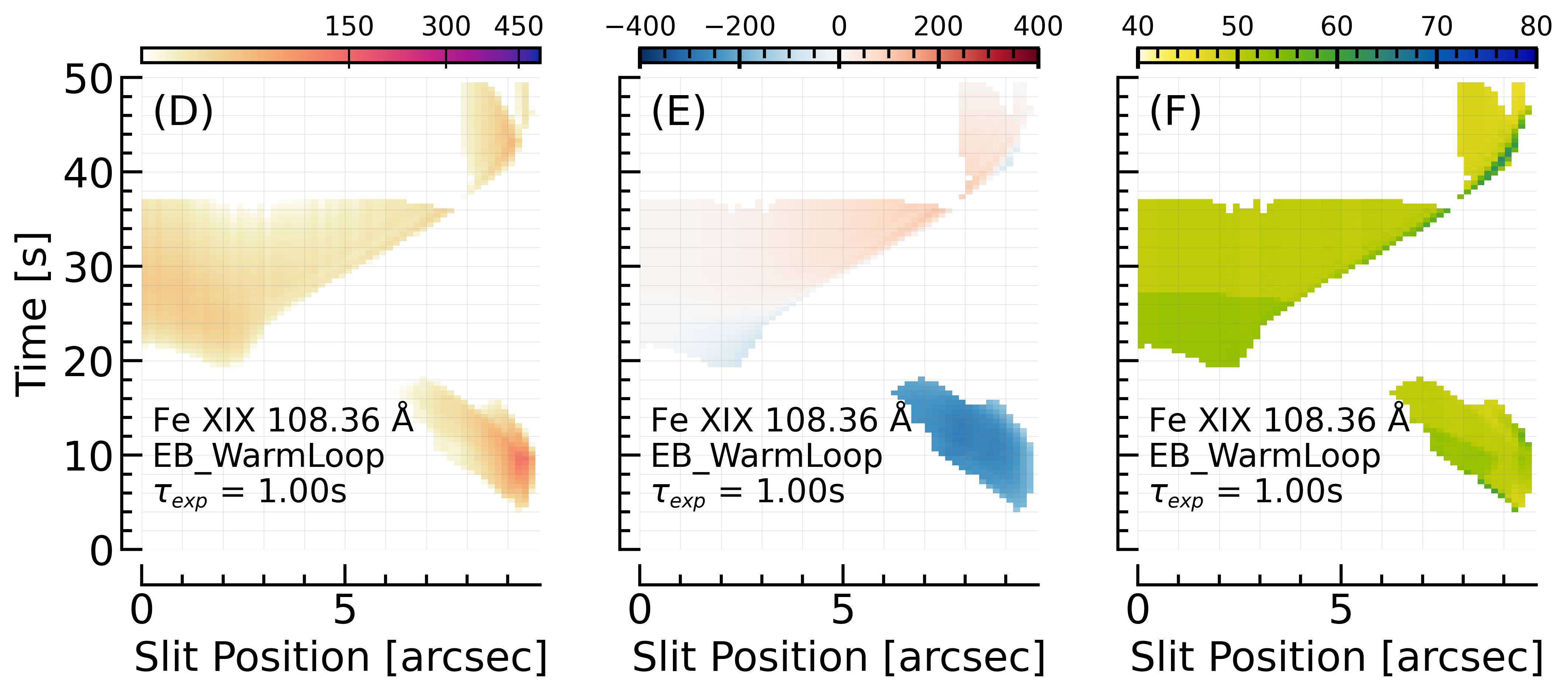}}
	}
	\vspace{0.1in}
	\vbox{
	\subfloat{\includegraphics[width = 0.7\textwidth, clip = true, trim = 0.cm 0.cm 0.cm 0.cm]{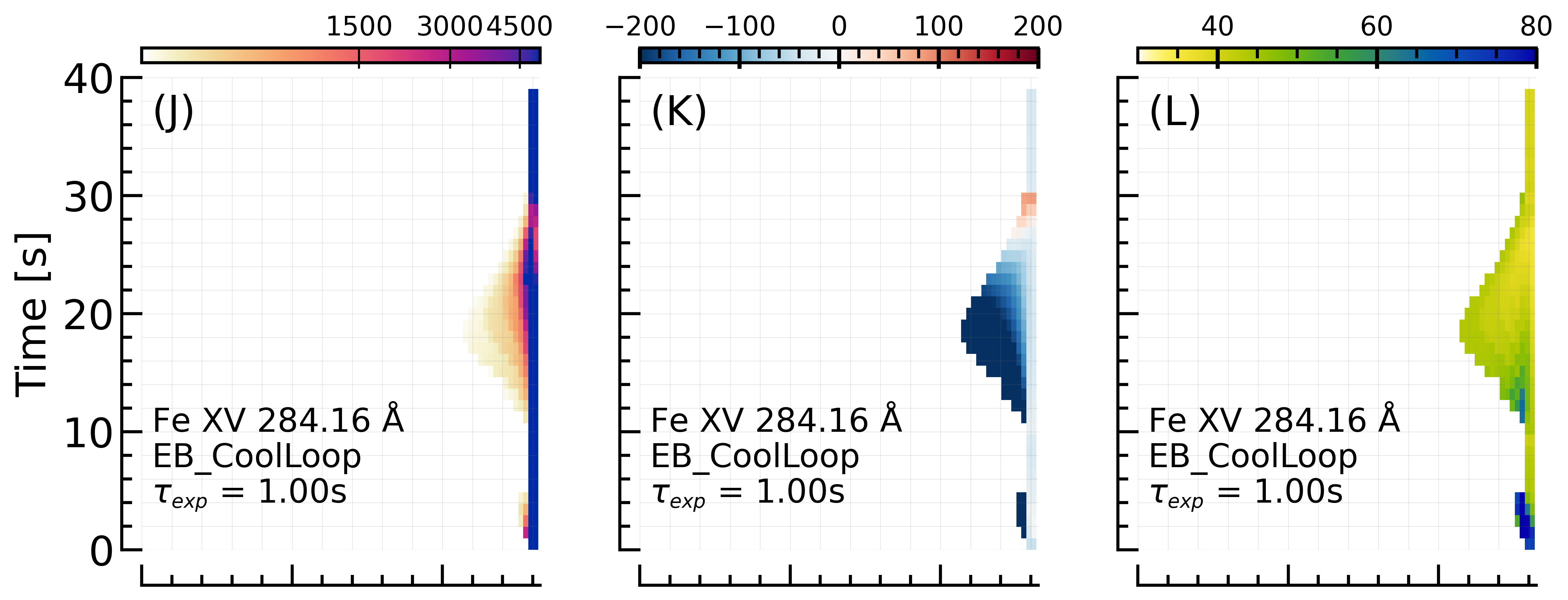}}
	}
	\vbox{
	\subfloat{\includegraphics[width = 0.7\textwidth, clip = true, trim = 0.cm 0.cm 0.cm 0.cm]{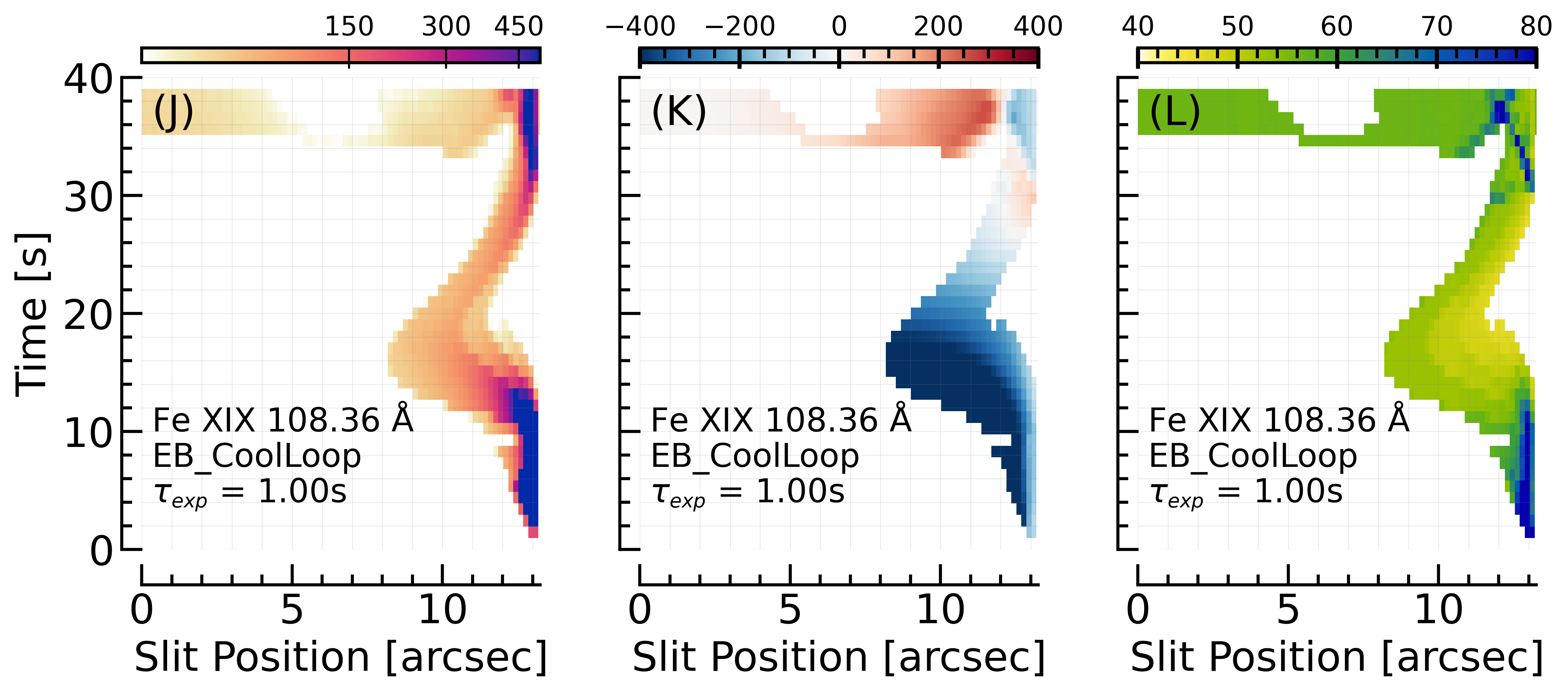}}
	}
	\caption{
	\musen\ spectral observations of the whole loop both before and during the flare will provide tight constraints on the properties of the heating and energy transport (thermal conduction, non-thermal particles) during the flare.
	Shown are the moments (see Appendix~\ref{app:moments}) of \musen\ \fexv\ and \fexix\ lines synthesized from two RADYN flare simulations with the same injected electron beam properties, but with different pre-flare conditions (see text and Appendix~\ref{app:sim} for details): model {\tt RADYN\_warm\_EB} (panels A-F), and model {\tt RADYN\_cool\_EB} (panels J-L).
	}
	\label{fig:preflarediffs}
\end{center}
\end{figure*}

Field-aligned radiative hydrodynamic models such as RADYN (see Sec.~\ref{sec:sims}, and Appendix~\ref{app:sim}) allow us to study the response of the plasma to the flare heating as a function of many parameters including the initial physical conditions of the atmosphere and the details of the heating properties. Comparisons between the models and spectroscopic observations have been shown to provide crucial diagnostics of the heating mechanisms at play in flares  \citep[e.g.,][to cite just a few recent results]{Reep2015,Polito2016,Polito2019,Kerr2020,Kerr2021}. We illustrate this by comparing the \musen\ synthetic spectral observables for a few field-aligned modeling experiments.
In the first experiment, we use two different electron beam simulations, with the same non-thermal particle distribution injected into two different pre-flare atmospheres, a loop with an initially cool and low density corona (model {\tt RADYN\_cool\_EB}), and a loop with initially hot and dense corona (model {\tt RADYN\_warm\_EB}; see Appendix~\ref{app:sim} for details).
In order to synthesize \musen\ observables, the RADYN 1D simulations are mapped to a 2D semicircular loop as illustrated in Figure~\ref{fig:radyn_mapping}, which also shows the assumed line-of-sight.
Synthetic \musen\ \fexvw\ and \fexixw\ observables for the two experiments are shown in Figure~\ref{fig:preflarediffs}.  Figure~\ref{fig:preflarediffs} shows that the plasma response to the flare heating can vary significantly depending on the initial physical conditions (temperature and densities) of the pre-flare loop atmosphere. 
When the NTE are released in the initially emptier and cooler loop ({\tt RADYN\_cool\_EB}) the footpoint brightenings are characterized by brighter and broader lines, and larger flows compared to the denser and hotter initial atmosphere ({\tt RADYN\_warm\_EB}).  
This shows the importance of providing constraints on the initial physical conditions of the loop when diagnosing different heating models from the observations. \musen, thanks to its multi-slit design, will allow to capture simultaneous spectral images of the loops prior to and after the flare.

We also experimented with \textsl{in-situ} heating in the corona, with the thermal conductive flux subsequently heating the transition region and chromosphere (model {\tt RADYN\_warm\_TC}). Finally, a tandem experiment was performed containing both \textsl{in-situ} heating plus electron beam energy deposition (with the same parameters as the individual heating scenarios; model {\tt RADYN\_warm\_EB\_TC}). 
In Figure~\ref{fig:radyndiffheating}, we compare the \musen\ \fexvw\ and \fexixw\ synthetic spectra as a function of time for the three heating models (i.e., electron beam (EB), \textsl{in-situ}, and \textsl{in-situ} + EB), given the same initial physical conditions (warm and dense loop). The synthetic spectra exhibit several distinctive differences between the models. For instance, panels B and E show that the \ion{Fe}{15}~line is brighter at the loop footpoints in the \textsl{in-situ} and hybrid scenarios than in the EB model. In addition, in the  \textsl{in-situ} model the spectra are redshifted at the footpoint by up to 50 km~s$^{-1}$ in the first few seconds of the simulation. The redshift is more modest (only 5-10 km~s$^{-1}$) or absent in the hybrid and EB cases respectively. Another interesting feature is the fact that the hotter \ion{Fe}{19} is brightest in the \textsl{in-situ} and hybrid models, which are more effective at heating the corona to higher temperatures than the EB only case, where the electrons deposit their energy mostly in the lower atmosphere.  
While the loops shown here have an average length of $\approx$~20--30~Mm, longer flare loops (up to $\approx$~100~Mm or even longer) are also observed \citep[e.g.,][]{Polito2019}. For those cases, the large FOV coverage of \musen\ is even more crucial to fully capture all the dynamics across the loop arcade.

As demonstrated above, field-aligned models provide a valuable and flexible tool to investigate the plasma response to the flare heating for a large range of parameter space with significantly reduced computational cost as compared to 3D MHD models. On the other hand, they inherently lack crucial information about line-of-sight effects that can be obtained using 3D models. To overcome this challenge, we have recently developed the {\tt  RADYN\_Arcade} model \citep{Kerr2020}, where the RADYN field-aligned atmospheres are grafted onto observed AR loops (see also Appendix~\ref{app:sim}). Using {\tt  RADYN\_Arcade} we demonstrate below that \musen\ can be used to study non-thermal line broadening in flares, in particular how the spatial distribution of non-thermal line broadening is an important metric in understanding flares. 

\citet{Reep:2020} recently showed, using 1D hydrodynamic loop modeling, that electron beam heating is insufficient to produce coronal rain, which is a common feature of flares in the late phase. This suggest other heating mechanisms may be important. It has been suggested that  Alfv\'enic waves propagating downward from the reconnection site to the lower atmosphere may play a role in the heating of chromosphere and transition region during flares \citep{1982SoPh...80...99E,2008ApJ...675.1645F}. These waves may also play a role in local acceleration of particles in the chromosphere or in the generation of turbulence \citep{2008ApJ...675.1645F}. Recent numerical experiments have demonstrated that waves of sufficient frequency ($f \gtrsim 1$~Hz) can penetrate the transition region and efficiently damp energy in the chromosphere \citep{2013ApJ...765...81R}. Simulations of flares driven by high-frequency Alfv\'en waves have been successful in heating the lower atmosphere, generating mass flows, and have suggested that the differences in energy deposition profiles may result in observable signatures \citep{2018ApJ...853..101R,2016ApJ...827..101K,2016ApJ...818L..20R}. However, such waves have yet to be directly observed due to the absence of sub-second observations with high spatial resolution. Thus the energy flux carried by waves in these  experiments is largely unconstrained. 

One potential way to constrain the Poynting flux carried by Alfv\'en waves in flares would be to measure the non-thermal width of spectral lines at various temperatures (i.e., formed at different heights), times and spatial locations within the flare arcade. The non-thermal width of a spectral line broadened by a passing Alfv\'en wave is directly related to the amplitude of that wave \citep[][]{1991SoPh..131...41M,2009A&A...501L..15B}. 

\begin{figure*}[htb]
\begin{center}
	\vbox{
	\subfloat{\includegraphics[width = 0.80\textwidth, clip = true, trim = 0.cm 0.cm 0.cm 0.cm]{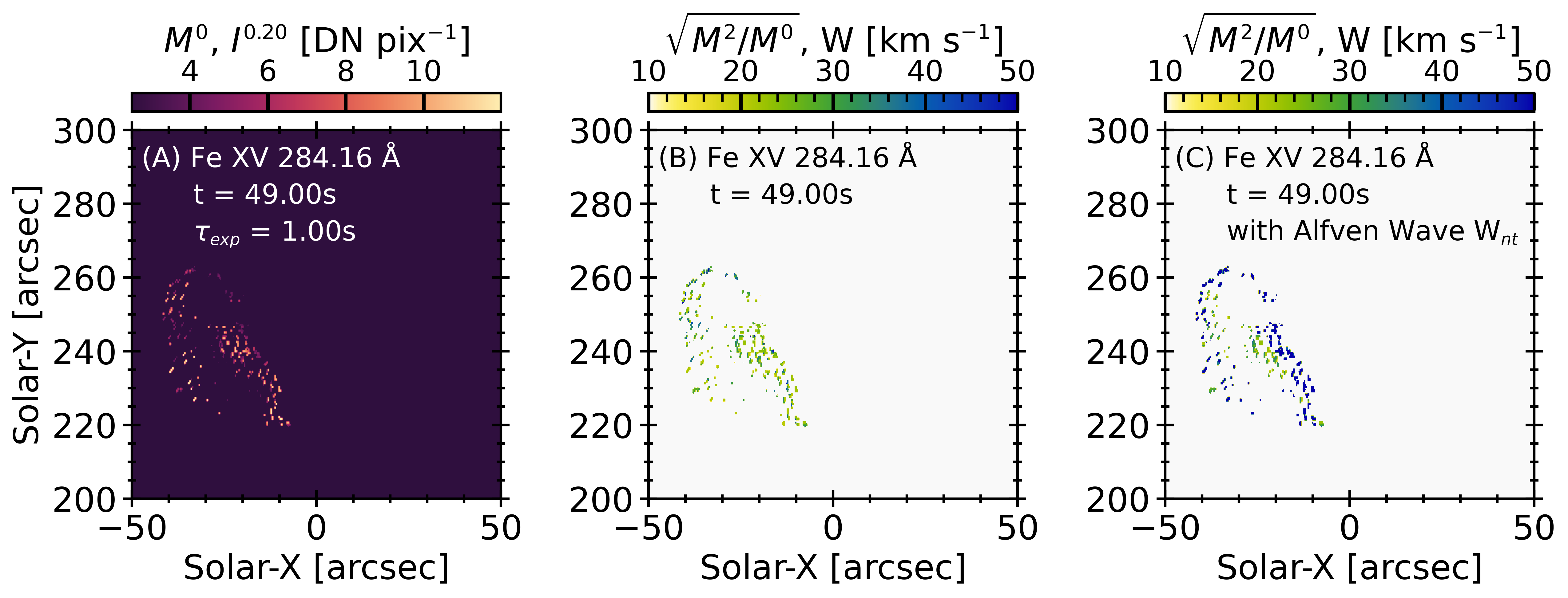}}	         }
	\vbox{
	\subfloat{\includegraphics[width = 0.80\textwidth, clip = true, trim = 0.cm 0.cm 0.cm 0.cm]{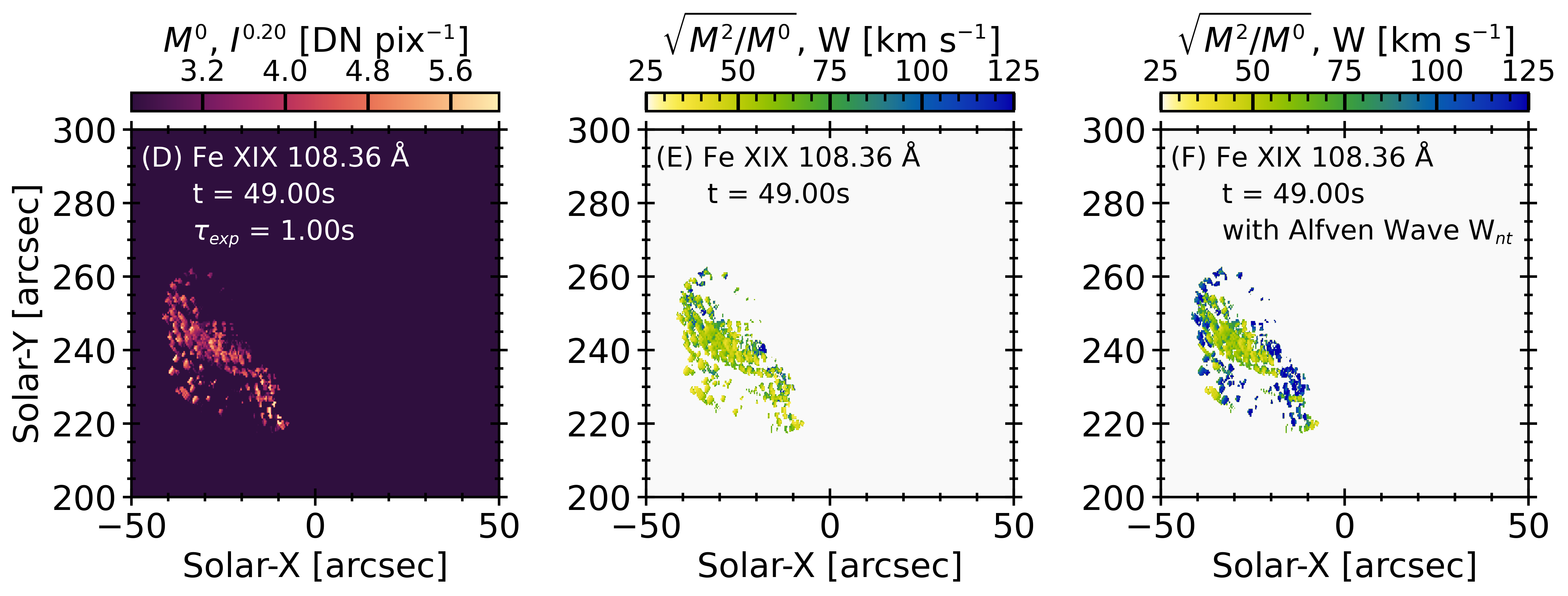}}
	}
 \vbox{
 	\subfloat{\includegraphics[width = 0.25\textwidth, clip = true, trim = 0.cm 0.cm 0.cm 0.cm]{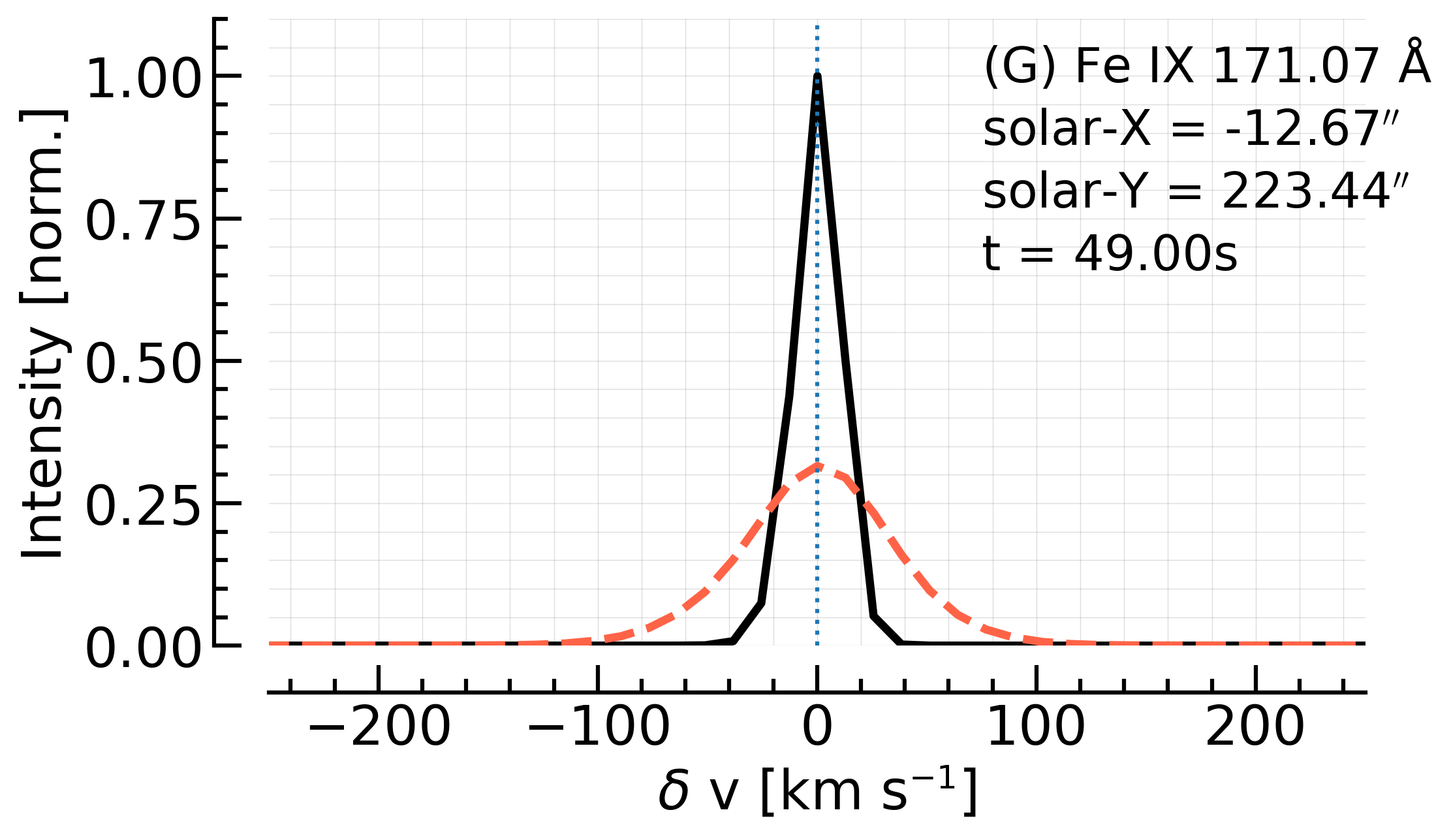}}
 	\subfloat{\includegraphics[width = 0.25\textwidth, clip = true, trim = 0.cm 0.cm 0.cm 0.cm]{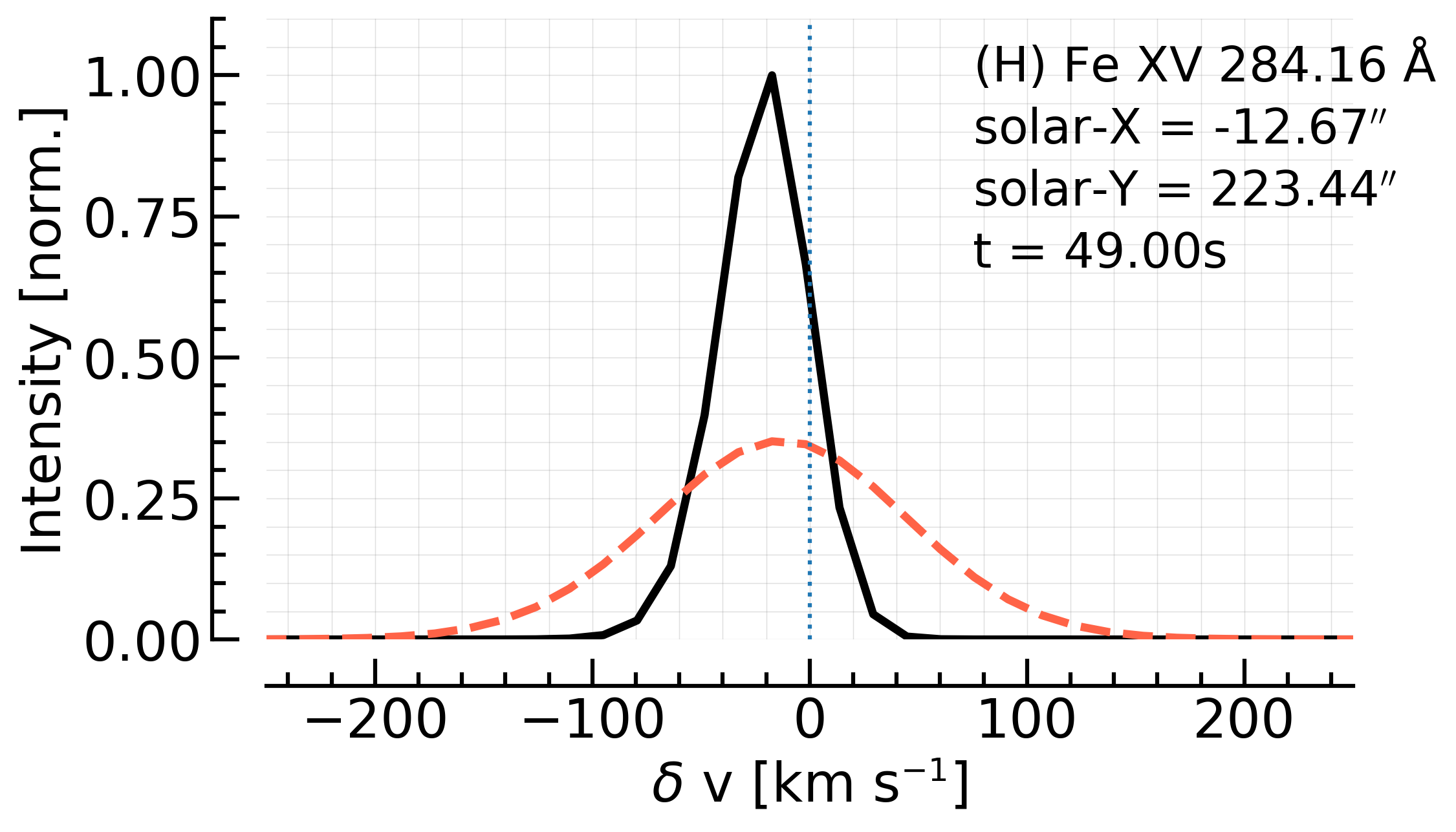}}
    \subfloat{\includegraphics[width = 0.25\textwidth, clip = true, trim = 0.cm 0.cm 0.cm 0.cm]{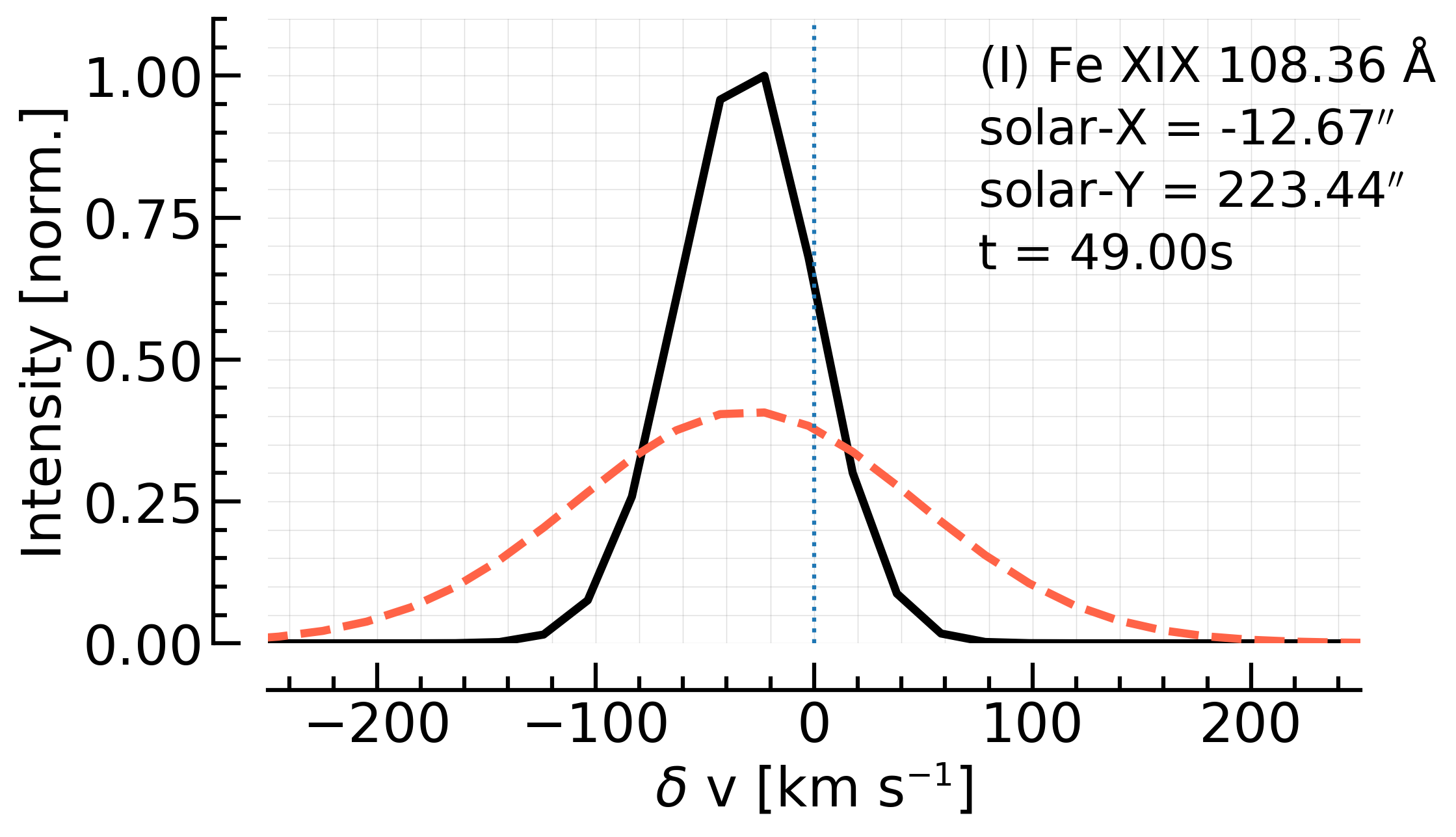}} 
 }
	\caption{Illustration of the impact of non-thermal broadening due to presence of Alfv\'en waves during the flare. The 2nd moment of the Fe XV 284.4~\AA\ line (top row) and Fe XIX 108.3~\AA\ line is shown in the case with and without Alfv\'en wave broadening. The bottom row shows the difference in line width without Alfv\'en waves (black lines) and with Alfv\'en waves (red dashed lines) from a single pixel near a loop footpoint. By simultaneously observing conjugate footpoints, loops and loop apexes MUSE can study the spatial distribution of non-thermal broadening which can be compared to models of flares that include Alfv\'enic waves.}
		\label{fig:arcade_exs_aw}

\end{center}
\end{figure*}

Figure~\ref{fig:arcade_exs_aw} shows that including broadening due to an Alfv\'en wave can result in line widths that exceed those in which only thermal motions and line-of-sight superposition of field aligned flows (e.g., along different loops) are considered. In the original RADYN 1D model we impose a time-independent magnetic field stratification that varies with loop pressure, and calculate the Alfv\'en speed over time from the density evolution. From that we can calculate the non-thermal broadening along the loop as a function of time, assuming a Poynting flux of $10^{10}$~erg~s$^{-1}$~cm$^{-2}$. This non-thermal broadening is included for the duration of the heating phase of each loop, and is zero in the cooling phase. We then include this as an additional line broadening term when synthesizing the spectral lines in the {\tt  RADYN\_Arcade} model, using the magnetic field geometry to project the broadening appropriately. 

Each spectral line will experience a different degree of broadening, and at different locations along the loop. The duration of the enhanced non-thermal width would be the time taken for the wave(s) to propagate past that location. Tracking the non-thermal line broadening over time and space with the high resolution afforded by \musen\ could allow constraints on the energy carried by MHD waves during flares.

To summarize, the large FOV of \musen\ is necessary to follow ribbon evolution over a significant area to (i) calculate flux transfer rate over significant portion of rapidly-moving ribbons; (ii) identify conjugate footpoints from their correlated time evolution; (iii)  disambiguate spatial and temporal ribbon evolution, e.g., to test theoretical relationships 
between intensity and plasma flow speeds as a function of time (e.g., to constrain properties of non-thermal electrons). 
With spectroscopy at high cadence we can (i) track in detail the onset of mass flows as a flare starts, and their relationship to flare ribbon brightening and field evolution; (ii) examine the development of non-thermal line broadening, possibly due to  turbulence, MHD waves, or non-thermal electrons, before and during ribbon brightening, and their spatio-temporal relationship to changes in the magnetic field.

The large amount of energy transported from the corona to the chromosphere and back in flares means that those atmospheric layers are strongly coupled. Any theories and models must be able to explain the response of the full flaring atmosphere. \musen\ will provide essential spectroscopic images of the flaring corona with a large spatial coverage, required to fully appreciate \euvstn\ and \dkistn\ observations. Observations from \euvstn\ and \dkistn\ and other GBOs will provide information about the flaring chromosphere, the source of evaporated material. 

Fast, but narrow, rasters with \euvstn\ will capture smaller segments of the flare ribbons, with an emphasis on detailed plasma diagnostics to complement those from \musen. Density diagnostics of heated ribbon plasma can be carried out, for example using the density pairs \ion{C}{3} $\lambda$~977/1176 \AA\ and \ion{Fe}{10} $\lambda$~175/177 \AA , and the sequence of Fe lines between \ion{Fe}{9} and \ion{Fe}{14} will be used to provide DEM measurements \citep[e.g.,][]{Baker:2015ApJ...802..104B,Baker:2018ib}. Non-thermal line widths and Doppler shifts from lines formed at different temperatures will also provide information about gradients and stratification of flows and turbulence throughout the atmosphere. 

\DKIST/VBI observations  will provide photospheric and chromospheric imaging (in \ion{Ca}{2}~K,  G band, blue continuum and $H\beta$), to track the overall ribbon spatial and temporal evolution. 
Blue continuum and G-band “quasi-white-light” images will reveal the locations where the flare kernels are the brightest, hence the locations of strongest energy deposition.
VBI will also allow to follow in detail the dynamics of the ribbons by identifying/tracking features within them, to obtain local reconnection electric field and flux transfer rate  \citep[a.k.a.\ ``reconnection rate”][]{Kazachenko:2017}.
In addition, \DKIST/VTF will provide photospheric and chromospheric spectropolarimetry in \ion{Fe}{1} 630.2\ nm and \ion{Ca}{2} 854.2\ nm respectively, which will track the evolution and re-orientation of the magnetic field as the the flare ribbons expand and travel across the AR while the coronal reconnection occurs. 
This will permit the identification of sites of strong discontinuities in the magnetic field, indicating possible current sheet locations. 

\subsection{II-5: Understand the formation mechanism of sunspots, in particular delta sunspots}
\label{sec:ngspm_ii5}

Regardless of whether an AR has sunspots that are delta-type (and prone to flares), the spots do not form as newborn monolithic structures. Rather, interaction with subsurface convective flows leads to a fragmented emergence, and subsequent reorganization~\citep[opposite polarity patches counter-streaming, e.g.,][]{Pariat:2004} into larger structures that eventually become sunspots~\citep[][]{Cheung:2010,Rempel:2014,Chen:2017,ToriumiHotta:2019}. The counter-streaming polarities drive reconnection events in the photospheric/chromospheric layers that can reach TR temperatures with broad profiles observed by \iris~\citep[e.g.,][]{Peter:2014,Young:2018}.  They are considered signatures of reconnection of sea-serpent field lines, which are field lines emerging into the solar atmosphere, modulated by granulation flows. This reconnection allows the emerging magnetic field to remove its mass burden~\citep[][]{Cheung:2010}. This process allows new coronal loops to form, eventually connecting the two opposite polarity nascent sunspots. Rapid cadence \musen\ rasters of the corona above such emerging flux regions will reveal how the newly formed loops are structured. In addition, the large FOV and high-cadence raster scans will greatly increase the probability (compared to single-slit spectrographs) of capturing the initial stages of newly emerging flux and the resulting coronal response. Studying such regions with \musen\ will reveal how mass is unloaded from emerging flux regions, test whether magnetic buoyancy instabilities are at play in transporting magnetic flux to coronal heights~\citep[see review by][]{CheungIsobe:2014}, and sunspots of opposite polarities are eventually connected.

Although radiative MHD numerical simulations are able to model entire sunspots with umbral and dynamic penumbral filaments~\citep[][]{Rempel:2009}, there are still many open questions with regards to how the penumbra is formed and sustained. The current simulations with fully-fledged penumbral filaments start from initial conditions consisting of a monolithic funnel of magnetic field introduced into an originally weak-field convecting environment. \cite{Rempel:2011} report that artificially increasing the field inclination at the top boundary (located at the top of the photosphere) forced the model sunspot to have an extended penumbra. When the artificial forcing is switched off and the top boundary field is matched to a potential field, the model penumbra diminishes. However, a recent study concluded that model sunspots without artificial forcing at the top boundary had photospheric magnetic distributions that are more consistent with observations~\citep{Jurcak:2020}. Yet, those models without artificial forcing also lacked extensive Evershed flows. In a study of where penumbral filaments begin to form around the periphery of nascent sunspots, \cite{Murabito:2018} reported roughly half to the sunspots have penumbrae forming in the region between the opposite polarity spots, while the remainder had penumbra form at the outer edges (i.e., opposite side from emerging flux region).

Sub-arcsecond \musen\ intensity images from the SG and CI will be important for constraining the geometry and 3D structure of the coronal field before, during and after the formation or penumbrae (e.g. either as inputs to extrapolation models, or as validation of models, see Section~\ref{sec:ngspm_ii}). Furthermore, spectroscopic observables like the Doppler velocity and line width can be used to track how umbral and penumbral waves (from \dkistn/GBOs and \euvstn) propagate into the corona~\citep[e.g.,][]{Zhao:2016}. For an extended discussion of wave models of coronal loops and predicted \musen\ observables, we refer the reader to the companion paper on coronal heating~\citep{DePontieu:MUSE_corheating}.

A particular class of sunspots that form the focus of NGSPM SO II-5 is the class of delta sunspots. The characteristic size of delta spots span the spectrum of AR sizes (up to 100 Mm) and the associated collisional polarity inversion lines (cPILs) range from 10 to 100 Mm \citep{Liu:2021}. Photospheric and subsurface flows pressing strong (umbral/penumbral) fields of opposite polarity into a compact region drive a plethora of dynamic phenomena in the overlying atmosphere. In particular, delta sunspots are known to be particularly flare productive~\citep[e.g., see][and references therein]{ToriumiWang:2019}. 

\euvstn\ would be able to provide AR-scale rasters of delta-spot regions at a cadence of minutes (assuming 1s slit dwell time and 0.4\arcsec\ step size for dense rasters). Continuous coverage of the AR over hours and days would allow for monitoring of the detailed thermal structure of the AR from the photosphere to corona. However, delta spot regions often produce multiple flares and eruptions as flux emerges, and as shearing and flux cancellation along the PIL ensue. As illustrated in Figs.~\ref{fig:hgcr_erupt_euvst_1} and~\ref{fig:hgcr_erupt_euvst_2}, single slit spectrographs miss out of the important dynamics of flares and eruptions. This gap is filled by the high cadence ($<20$ s), 0.4\arcsec\ resolution rasters from \musen. Within smaller FOVs of what \musen\ would cover, \euvstn\ would provide seamless temperature coverage from 10,000 K to 15 MK, as well as density diagnostics.

Delta spot regions often spawn homologous eruptions~\citep[e.g.,][]{CheungDeRosa:2012,Li:2013,Panesar:2016,Polito:2017,2017ApJ...839...67S,Mitra:2020}. To piece together a quantitative picture of the family of homologous flares/eruptions, it is important to capture the evolution and structure of all the homologous events in their entirely via \musen 's multi-slit approach. This would then permit one to study how the eruptions vary throughout the evolution of the AR. As for the trigger mechanisms, Section~\ref{sec:ngspm_ii2} discusses in detail how \musen\ spectral moment maps can be used to identify the location and timing of triggers. 

\section{Conclusions} \label{sec:conclusions}

In summary, no spectroscopic observations have ever been captured at the spatio-temporal scales characteristic of the corona. The highest resolution coronal data to date are based on imaging, which is blind to many of the processes that drive coronal energetics and dynamics. As shown by \iris, \hinode\ and GBOs for the low solar atmosphere, we need high-resolution spectroscopic measurements with simultaneous imaging to understand the dominant processes. \euvstn\ will provide sub-arcsecond resolution spectra with seamless temperature coverage from the photosphere to the flaring corona, but due to its single-slit design and the lack of context/slitjaw coronal imaging, it will not be able raster AR-scale regions with sufficient cadence to freeze coronal dynamics (see, e.g.,  Figs.~\ref{fig:hgcr_erupt_euvst_1} and \ref{fig:hgcr_erupt_euvst_2}), especially during flares and eruptions. \musen\ provides the essential spectroscopic imaging capability at high resolution and cadence to capture EUV waves, CME triggering events and plasmoids (if they exist) on a routine basis. These observations are necessary to discriminate models of flares/eruptions, to detect signatures of turbulent structure in current sheets and reconnection outflows, and to constrain physics-based space weather models. 

As case studies in Section~\ref{sec:casestudies} demonstrate, coordinated observations between \musen, \euvstn\ and \dkistn\ (and other GBOs with sub-arcsecond spectropolarimetric capabilities) can be considered a distributed implementation of the NGSPM mission concept. This distributed mission can tackle the NSGPM science objectives to understand how energy accumulation occurs in ARs, how flares and eruptions are triggered (e.g., Figs.~\ref{fig:hgcr_trigger} and~\ref{fig:collision_zdir}), how CMEs evolve from their source region to interact with the ambient corona (Figs.~\ref{fig:collision_zdir}, \ref{fig:collision_xdir} and \ref{fig:collision_xdir_later}), how to characterize the consequences of fast magnetic reconnection (e.g., multi-shock structure in the termination shock region, see Fig.~\ref{fig:tuning_fork}; electron beam footpoint heating (e.g. in flare ribbons), see Fig. \ref{fig:radyndiffheating}), and how sunspots (especially delta-spots) form. Table~\ref{tab:observables_flares} lists the relevant physical phenomena and predicted \musen\ diagnostics. The combined capabilities of the NSGPM will also provide unprecedented insight into the source regions of disturbances and waves in the heliosphere as measured by the Parker Solar Probe, Solar Orbiter, and planetary missions with instruments monitoring space weather conditions in the solar system.

\begin{table*}[!htbp]
\centering%
\caption{\label{tab:observables_flares} \musen\ diagnostics for eruptive and flaring events for case studies addressing NGSPM Science Objectives II-1 to II-5 (see Table \ref{table:NGSPM}).}
\begin{tabular}{|p{\dimexpr0.04\textwidth-1\tabcolsep-\arrayrulewidth\relax}|
                p{\dimexpr0.21\textwidth-1\tabcolsep-\arrayrulewidth\relax}|
                p{\dimexpr0.54\textwidth-1\tabcolsep-\arrayrulewidth\relax}|
                p{\dimexpr0.09\textwidth-1\tabcolsep-\arrayrulewidth\relax}|
                p{\dimexpr0.09\textwidth-1\tabcolsep-\arrayrulewidth\relax}|
              }
\hline
SO & Physical Phenomena &  Predicted Diagnostic / Observational Constraint & $\lambda$\tablenotemark{*} [\AA] & Figures \\
\hline
II-1 & - Quasi-separatrix layer & - $\sim 10$-$20$~\kms\ blueshifts in fan loops at edge of AR & 284 & \ref{fig:fan_loops} \\ 
\textcolor{white}{c}  &  - Emerging flux  & - Loops connecting opposite polarities with alternating blue and redshifts of tens of \kms & 171, 284 &  \ref{fig:emergence} \\ 
\textcolor{white}{c}  &  - Sheared field near PILs & - Loop morphology constraints (e.g. sigmoidal loops) & 195, 284, 108  & \ref{fig:multires}, \ref{fig:collision_zdir}, \ref{fig:tethercutting} \\ 

\hline 
II-2 & - Triggers of flares and eruptions & To distinguish between flare/CME initiation models: location and timing of transient bidirectional flows, brightenings and enhanced non-thermal width identifying the reconnection site & 284, 108 & \ref{fig:hgcr_trigger}-\ref{fig:tethercutting},  \ref{fig:collision_xdir}-\ref{fig:plasmsim} \\ \hline
II-3 & -  Pre-eruption coronal field structure & - Sigmoidal loops, coronal cavity/bubble, hot flux ropes & 171, 195, 284, 108 & \ref{fig:multires}, \ref{fig:collision_zdir}, \ref{fig:tethercutting}, \ref{fig:collision_xdir}, \ref{fig:collision_xdir_later} \\ 
\textcolor{white}{c} & - CME-ambient corona interaction (including sympathetic eruptions \& flares) & - EUV wavefront and ambient coronal response & 195, 284, 108 & \ref{fig:collision_zdir}, \ref{fig:collision_xdir_later} , \ref{fig:circular_ribbon_flare} \\ 
\textcolor{white}{c} & \textcolor{white}{c} & - Redshifted loops in ambient corona from CME expansion & 284, 108 & \ref{fig:collision_zdir} \\ 

\textcolor{white}{c} & -  CME source region & - Doppler shift maps of early CME rise to constrain flux-rope driven CME models & 171, 284, 108 & \ref{fig:collision_zdir}, \ref{fig:collision_xdir_later} \\

\textcolor{white}{c} & - Coronal Heating & - Spectral line intensities and line widths as constraints on ad hoc heating terms in global MHD models with Alfv\'en wave heating & 171, 284, 108 & \textcolor{white}{c} \\
\hline
II-4 & - Plasmoid instability & - Plasma blobs in intensity images, spectral rasters  & 171, 195, 284, 108 & \ref{fig:takasao_plasmoids}, \ref{fig:plasmsim} \\ 
\textcolor{white}{c} & \textcolor{white}{c} & - Trajectory and thermal evolution of plasmoids & 171, 195, 284, 108 &\ref{fig:plasmsim} \\ 
\textcolor{white}{c} & \textcolor{white}{c} & -  Coalescence of plasmoids from Doppler maps & 284, 108 & \ref{fig:plasmsim}, \ref{fig:radyndiffheating} \\ 
\textcolor{white}{c} & \textcolor{white}{c} & - Enhanced line broadening & 284, 108 & \ref{fig:plasmsim}, \ref{fig:preflarediffs},  \ref{fig:radyndiffheating}  \\ 
\textcolor{white}{c} & - Multi-part termination shock structure & -  Alternating, pulsating patterns of blue and redshifts & 108 & \ref{fig:tuning_fork} \\ 
\textcolor{white}{c} & - Plasmoid impact on arcade loops & - Reconnection outflows impinging on arcade loops  & 108 & \ref{fig:samanta}  \\
\textcolor{white}{c} & - Loop heating mechanisms & - Spatio-temporal evolution of spectral moments along loop (including footpoints) constrains heating properties (\textsl{in-situ}, NTE) & 284, 108 &\ref{fig:radyndiffheating} \\ 
\textcolor{white}{c} & & - Line widths provide constraints on Alfv\'en wave flux & 284, 108   &\ref{fig:arcade_exs_aw} \\ 

\hline
II-5 & - Formation of delta spots & - Sigmoidal loops & 195, 284, 108  &\ref{fig:collision_zdir}, \ref{fig:circular_ribbon_flare} \\

\textcolor{white}{c} & - Penumbra formation & - Coronal loop morphology & 171, 195, 284 &\textcolor{white}{c}  \\
\textcolor{white}{c} & - Umbral/penumbral waves & - Tracking of waves into the corona from Doppler and line width maps. & 171, 284 &\textcolor{white}{c}  \\ 
\hline
\end{tabular}
\begin{flushleft}
\tablenotetext{}{PIL: Polarity inversion line. NTE: Non-thermal electrons.}
\tablenotetext{*}{For 304 and 195 imaging is desired. For 171, 284, and 108, intensity, Doppler shift, and line broadening are typically desired. Note that when referring to the 108\AA\ spectral band we typically refer to both the \fexix\ 108.35\AA\ and \fexxi\ 108.12\AA\ line.}
\end{flushleft}
\end{table*}

To understand the complex dynamics in the turbulent solar atmosphere requires not only high cadence, high resolution observations, and spectral diagnostics, but also advanced numerical simulations. A core component of the \musen\ science investigation involves the use of numerical models, for the following reasons. First of all, \musen\ (and, in general, the NGSPM) observations will allow us to test, discriminate, and improve existing models.  Secondly, they guide the team to develop the science requirements which flow down to the instrument requirements. It is through this process that we identify the need for high-cadence imaging, and show how single-slit spectrographs like \euvstn\ (which, complementarily, have better temperature coverage, and diagnostics of density and chemical composition) would miss important dynamics and the relevant context. The simulations used in this and a companion paper are just a subset of what can be done to compare models with \musen\ observations. The purpose of the \musen\ mission, and of this paper, is not to show that existing models are correct. Rather, \musen\ will provide the necessary measurements at the appropriate cadence and resolution to test our understanding of solar eruptive activity, to discriminate between models, and to build a solid foundation for physics-based models of flares and CMEs. The combined capabilities of the NSGPM will also provide unprecedented insight into the the source regions of disturbances and waves in the heliosphere as measured by the Parker Solar Probe, Solar Orbiter, and planetary missions with instruments monitoring space weather conditions in the solar system.

{\bf Acknowledgements:} We would like to thank Amy Winebarger, Stuart Bale, Harry Warren, Toshifumi Shimuzu, Lars Frogner, Charles Kankelborg, Jenna Samra, Sami Solanki, Hardi Peter, Jorrit Leenaarts, Paolo Pagano, Fabio Reale, Peter Young, and Wei Liu for their valuable contributions and discussions of the results.  We are grateful to Jim Lemen for his technical leadership in the MUSE investigation. We gratefully acknowledge support by NASA contract 80GSFC21C0011 (MUSE Phase A). Some of this work was also supported by NASA contract NNG09FA40C (IRIS), and NASA grants 19-HTMS19\_2-0025, 80NSSC18K1285, 80NSSC21K0737, and 80NSSC19K0855. DNS acknowledges funding from the Synergy Grant number 810218 (ERC-2018-SyG) of the European Research Council, and the project PGC2018-095832-B-I00 of the the Spanish Ministry of Science, Innovation and Universities. PA acknowledges funding from the STFC Ernest Rutherford Fellowship (No.\ ST/R004285/2). VP acknowledges support from NASA's HGI grant\# 80NSSC20K0716. LF acknowledges support from STFC Consolidated Grant ST/T000422/1. GC, MR and MCMC acknowledge support from NASA grant 80NSSC19K0855 “Investigating the Physical Processes Leading to Major Solar Activity”. MCMC acknowledges support from NASA's \sdo/\aia\ contract (NNG04EA00C) to LMSAL. \aia\ is an instrument on board \sdo, a mission for NASA's Living With a Star program. GSK acknowledges support from NASA's Early Career Investigator Program (Grant\# 80NSSC21K0460), and Heliophysics Supporting Research program (Grant\# 80NSSC19K0859). The simulations have been run on clusters from the Notur project, and the Pleiades cluster through the computing project s1061, s8305 and s2169 from the High End Computing (HEC) division of NASA. This material is based upon work supported by the National Center for Atmospheric Research, which is a major facility sponsored by the National Science Foundation under Cooperative Agreement No.\ 1852977. We would like to acknowledge high-performance computing support from Cheyenne (doi:10.5065/D6RX99HX) provided by NCAR's Computational and Information Systems Laboratory, sponsored by the National Science Foundation. SD is supported by a grant from the Swedish Civil Contingencies Agency (MSB) and the Knut and Alice Wallenberg foundation (2016.0019). Simulation {\tt MURaM\_emergence} was performed on resources provided by the Swedish National Infrastructure for Computing (SNIC) and the European Union’s Horizon 2020 research and innovation program under grant agreement No. 824135 (SOLARNET). The synthesis and analysis of the various numerical models have been performed on the Google Cloud Platform, allowing sharing the synthetic data and models, developing common tools, and access to instances with various specifications and Graphics Processing Units (GPUs). This project has been supported by a grant (project lunar-campaign-29341) by Google Could to the University of Oslo.  

To analyze the data we have used IDL, Python and PyTorch~\citep{PyTorch}. This research is also supported by the Research Council of Norway through its Centres of Excellence scheme, project number 262622, and through grants of computing time from the Programme for Supercomputing. \iris\ is a NASA small explorer mission developed and operated by LMSAL with mission operations executed at NASA Ames Research Center and major contributions to downlink communications funded by ESA and the Norwegian Space Centre.

\appendix
\section{Numerical simulations}\label{app:sim}
Throughout the paper we presented synthetic observables from several numerical models, listed in Table~\ref{table_sims} of Section~\ref{sec:sims}, and for which here we provide a description.

{\bf MURAM:}
The MURaM code aims to address the most relevant physical processes in the outer solar atmosphere, i.e., photosphere, chromosphere, TR, and lower corona. MURaM can cover a spatial range from deep in the convection zone to a coronal scale height ($\sim 50$~Mm) or more.  The simulations presented here are based on the coronal extension of the MURaM code as described in \cite{Rempel:2017zl}, and includes: single fluid MHD, 3D grey radiative transfer, a tabulated LTE equation of state, Spitzer heat conduction and CHIANTI based optically thin radiative loss in the corona. As for the Bifrost experiments, the Poynting flux that heats the chromospheric and coronal parts of the simulation domain is generated through magnetoconvection in the photosphere and convection zone. 
Here we analyze several different simulations (see Table~\ref{table_sims}):
\begin{itemize}
    \item Model {\tt MURaM\_flare}  This is a 3D radiative MHD simulation of a solar flare inspired by the evolution of AR 12017. The region spawned dozens of flares (Cs, Ms, and one X-class flare) as a parasitic bipole emerged near the pre-existing leading sunspot. The MHD simulation mimics this process, and created a C4 flare. See \cite{CheungRempel:2019} for details.
    \item Model {\tt MURaM\_circ\_rib}: This model is based on the flux emergence setup of \citet{ChenRempelFan:2017} that couples a global dynamo simulation \citep{FanFang2014} with MURaM. This work was extended into the corona, resulting in a $197$ Mm wide domain reaching $113$ Mm above the photosphere as will be detailed in \citet{ChenRempelFan:2021}. We did not specifically set up the simulation to produce a circular ribbon flare, the conditions just arose as part of a complex flux emergence process. The analysis shown in Figure \ref{fig:circular_ribbon_flare} is restricted to a subdomain of $147\times 98\times 66$ Mm$^3$. 
    \item  Model {\tt MURaM\_emergence} 
    is aimed at simulating the plasma dynamics in an emerging flux region (see Figure~\ref{fig:emergence}). The simulation domain has an extent of $40 \times 40  \times 22$~Mm, with $8$~Mm protruding below the photosphere.  The resulting model is generated in phases, similarly to previous runs. The initial magnetic field of $200$~G is added to well developed non-magnetic convection simulation to form extended magnetic field concentrations at meso- to super-granular spatial scales. The computational domain was then extended to include the upper solar atmosphere and the magnetic field from the pre-existing simulation was used for potential field extrapolation into the rest of the domain. The new simulation was then run until a relaxed state is achieved \citep{Sanja_AGU}. 
    In this model, the additional bipolar flux system is advected through the bottom boundary over an ellipsoidal flux-emergence region with the major axes $(a, b) =  (3,  1) $~Mm and $B_0=8000$~G field strength \citep{CheungRempel:2019}.  The emergence resulted in a flare after $4.6$ hours of solar time. Earlier, low activity, phases of this simulation are analyzed in the companion paper to this, which studies coronal heating \citep{DePontieu:MUSE_corheating}.
\end{itemize}

{\bf Bifrost:}
 The Bifrost models cover a domain that ranges from the convection zone up to the corona, include self-consistent magneto-convection, and self-consistently produce a chromosphere and hot corona, through the Joule dissipation of electrical currents that arise as a result of footpoint braiding in the photosphere and convection zone \citep{Hansteen2015}. Bifrost \citep{Gudiksen2011} solves the MHD equations, including thermal conduction along the magnetic field, non-LTE and non-gray radiative transfer with scattering \citep{Skartlien2000,Hayek:2010ac}, parameterized radiative losses \citep{Carlsson:2012uq} in the upper chromosphere, TR, and corona, and characteristic boundary conditions on the upper and lower boundaries. Optically thin radiative losses in the corona are based on CHIANTI emissivities \citep[e.g.,][]{Chianti2021}

The main free parameter for these simulations is the seed magnetic field (and its spatial distribution and strength) which can produce drastically different atmospheres 
\citep{Hansteen2010}. 

The Bifrost model we use here ({\tt B\_npdns03}, see Table~\ref{table_sims}) \ref{fig:plasmsim}~is a 2D simulation aimed at studying coronal bright points and their conspicuous emission in extreme-utraviolet and X-rays.  The initial condition was created imposing a potential nullpoint configuration at 8 Mm of height in the corona over a preexisting statistically stationary 2D snapshot that mimics a coronal hole and encompasses from the uppermost layers of the solar interior up to the corona. The physical domain is $0.0$~Mm $\leq x \leq 64.0$~Mm and $-2.8$~Mm $\leq z \leq 67.0$~Mm, where $z=0$~Mm corresponds to the solar surface. This domain is solved with $4096\times4096$ grid cells using a uniform numerical grid, in both the horizontal and vertical directions, with $\Delta x \approx 15.6 $~km and $\Delta z \approx 17.0 $~km, respectively. 

\textbf{RADYN:} The RADYN 1D field-aligned radiation hydrodynamic (RHD) code \citep{Carlsson:1992kl,1997ApJ...481..500C,1995ApJ...440L..29C,2005ApJ...630..573A,2015ApJ...809..104A} 
solves the coupled equations of hydrodynamics, charge conservation, NLTE radiation transport and non-equilibrium atomic level populations, on an adaptive grid \citep{1987JCoPh..69..175D} for a half of a symmetric semi-circular loop spanning from the sub-photosphere through corona.
RADYN has been widely used to model impulsively heated loops, and, through comparisons with observations, it provides crucial diagnostics of heating properties and energy transport in a variety of events from nanoflares (e.g., \citealt{Testa2014,Polito2018,Testa2020}; see also the companion paper \citealt{DePontieu:MUSE_corheating}) to large  flares \citep[e.g.,][]{2015SoPh..290.3487K, 2017ApJ...836...12K, 2016ApJ...827...38R, 2016ApJ...827..101K, 2019ApJ...871...23K, Kerr2020,Kerr2021, 2017A&A...605A.125S,2018ApJ...862...59B}.
Flares are typically simulated by injecting a distribution of non-thermal electrons at the apex of the loop, in a power-law form so that there is some instantaneous energy flux ($F$) carried by electrons with a spectral index ($\delta$) above some low-energy cutoff ($E_{c}$). Electrons lose their energy primarily via Coulomb collisions when they impact the denser lower atmosphere, depositing energy in the chromosphere/transition region. 
Particles are transported by solving the Fokker-Planck equations with no need to make any cold or warm target assumption. Most of the simulations presented here use the \cite{2015ApJ...809..104A} version of RADYN, but some use an updated treatment of the Fokker-Planck transport solver \citep{2020ApJ...902...16A}.
As well as flare energy injection via non-thermal electrons we present some results where flare energy is deposited directly into the corona, which is then conducted to the rest of the atmosphere.
Here we use RADYN in two ways, (1) performing field-aligned loop modeling as well as (2) arcade modeling that takes into account superposition of loops and line of sight effects of forward modelled optically thin radiation \citep[RADYN\_Arcade,][]{Kerr2020}. In the latter, field-aligned models are grafted onto observed AR loops from \cite{2018arXiv180700763A}.
\begin{itemize}
    \item {\bf 1D loop modeling:} We run several RADYN experiments using different initial conditions and loop lengths. Here we list the models for which results are shown in  the paper (Figures~\ref{fig:preflarediffs}, \ref{fig:radyndiffheating},   \ref{fig:arcade_exs_aw}):
    \begin{itemize}
        \item {\tt RADYN\_cool\_EB}: is a loop model with semi-length 15~Mm, and initial cool and low density corona  ($T \sim 1$~MK and $n_e \sim 5 \times 10^8$~cm$^{-3}$), heated by a non-thermal electron (NTE) beam characterized by a power-law distribution with the following parameters: $F=1.2\times10^{11}$~erg~cm$^{-2}$~s$^{-1}$ injected for $t=10$~s (giving a total flux of $F=1.2\times10^{12}$~erg~cm$^{-2}$), $\delta = 5$, and $E_{c} = 20$~keV \citep{Polito2019}.
        \item {\tt RADYN\_warm\_EB}: is a loop model with semi-length 10~Mm, and initial hotter and denser corona ($T \sim 3$~MK and $n_e \sim 5 \times 10^9$~cm$^{-3}$), heated by NTE with the same parameters as model {\tt RADYN\_cool\_EB}, i.e., $F=1\times10^{11}$~erg~cm$^{-2}$~s$^{-1}$ injected for $t=10$~s (giving a total flux of $F=1\times10^{12}$~erg~cm$^{-2}$), $\delta = 5$, and $E_{c} = 20$~keV.
        \item {\tt RADYN\_warm\_TC}: is a loop model with semi-length 10~Mm, and initial hotter and denser corona ($T \sim 3$~MK and $n_e \sim 5 \times 10^9$~cm$^{-3}$), heated \textsl{in-situ}, with total energy of $1.4\times10^{11}$~erg~cm$^{-2}$, deposited for 10~s over the top half of the loop.
        \item {\tt RADYN\_warm\_EB\_TC}: is a loop model with semi-length 10~Mm, and initial hotter and denser corona ($T \sim 3$~MK and $n_e \sim 5 \times 10^9$~cm$^{-3}$), heated by the combination of NTE -- with the same parameters as for model {\tt RADYN\_warm\_EB} -- and \textsl{in-situ} heating -- with the same parameters as for model {\tt RADYN\_warm\_TC}.
    \end{itemize}
    \item {\bf Arcade modeling:} For the {\tt RADYN\_Arcade} model, we use the same simulation as presented in \cite{Kerr2020}. This implementation is described in detail in \cite{Kerr2020}, but in short: (1) the 3D magnetic structure in an AR was obtained via extrapolation, (2) a subset of observed loops were selected to form the modelled flare arcade, (3) at a specified time each loop was filled with VDEM \citep[velocity differential emission measure; following definition of][]{Cheung:SDC} from the appropriate time in a 1D RADYN flare simulation, (4) within each pixel of the loop the resulting emission in each of the \musen\ lines was synthesised, (5) the voxels through which that loop passed were projected onto a 2D x-y observational plane, and the spectra added to that pixel with the Doppler shift applied taking into account viewing angles, (6) if some other voxel previously or subsequently projects to the same observational pixel then emission is summed within that pixel (so that superposition of loops along the line of sight is accounted for). The field-aligned model grafted onto each observed loop was an electron-beam driven flare, with time-varying flux $(1-6)\times 10^{10}$~erg~s$^{-1}$~cm$^{-2}$, $\delta = 7.2$, $E_{c} = 25.3$~keV. The heating duration was 25~s. Each loop was activated at different times so that any one snapshot contains loops that are newly activated, some that at experiencing their impulsive phase, and some that are experiencing their gradual phase. The \goes\ class of the flare was M2.0.
\end{itemize}

\begin{figure}
    \centering
    \includegraphics[width=\textwidth]{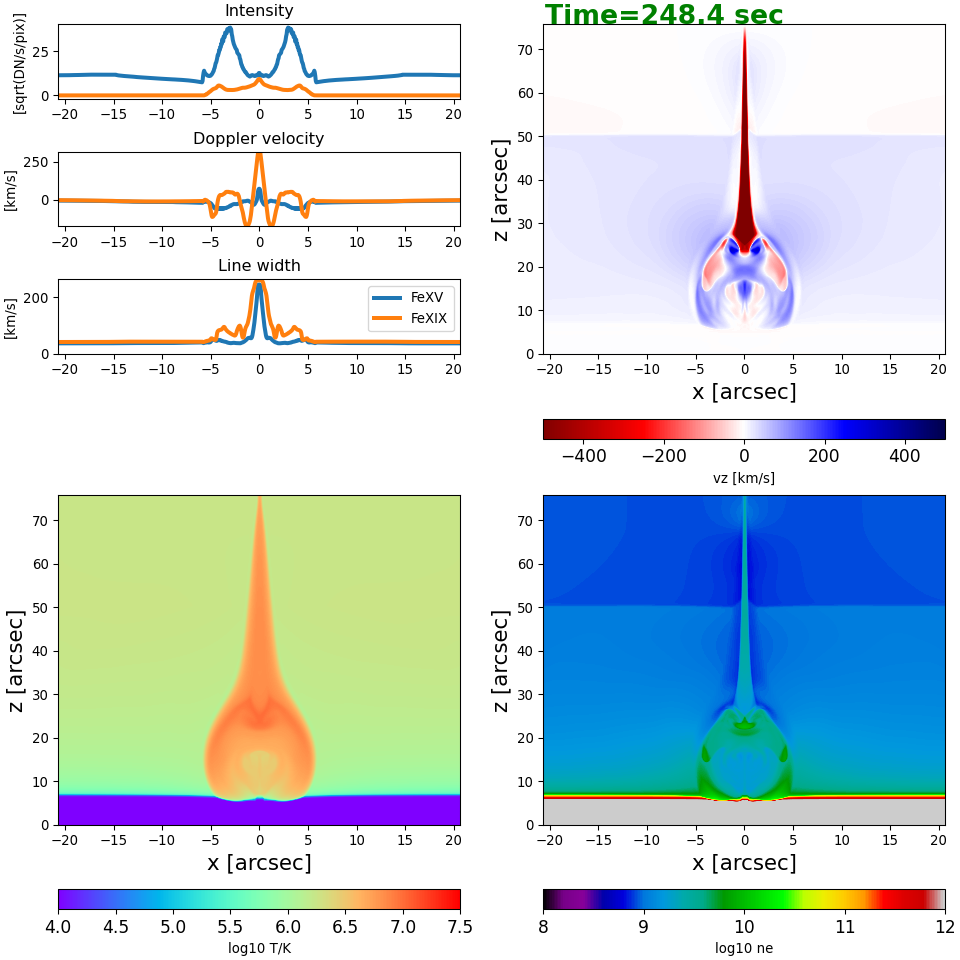}
    \caption{Snapshot of the {\tt Termination\_shocks} model \citep[see Table~\ref{table_sims}  and Appendix~\ref{app:sim};][]{Takasao:2015} at $t=248$~s. Top right: vertical component of plasma velocity ($v_z$, negative/red is downflow) showing the sunward reconnection outflow. Bottom right: distribution of the free electron number density $n_e$. Bottom left: temperature distribution. Top left: 0th (line intensity), 1st (Doppler velocity; positive is redshift) and 2nd (line width) moments of \ion{Fe}{15} 284 \AA\ and \ion{Fe}{19} 108 \AA\ lines, when observed along the $-z$ direction (i.e., top-down) view). See Fig.~\ref{fig:tuning_fork} for time evolution of the intensity and Doppler shifts.}
    \label{fig:takasao_snapshot}
\end{figure}

{\bf Termination shock region -- model \texttt{Termination\_shocks}:}
This two-dimensional MHD model is detailed in~\citet{Takasao:2015} and~\citet{Takasao:2016}. It begins with a plane-parallel stratified atmosphere with an initial density and temperature stratification with a sharp contrast representing the transition region between the chromosphere and the corona (see Fig.~\ref{fig:takasao_snapshot}). The atmosphere is threaded with purely vertical magnetic field of uniform amplitude, but of opposite orientation on either side of $x=0$ (i.e., left and right sides of the domain have opposite polarity field), which means the initial condition has a current sheet at $x=0$. Reconnection is triggered at the top boundary at $x=0$, which continues over the course of the simulation (duration is $10.5$ min). The downward directed reconnection outflow develops multiple fast mode shocks. For distance-time plot displayed in Fig.~\ref{fig:tuning_fork}, the line-of-sight integration is carried out in the vertical ($z$) direction. 

\section{Synthesis of \musen\ spectral observables from models} \label{app:synthesis}

The \musen\ observables shown in the paper are synthesized from different models using the \musen\ response functions. The \musen\ spectral response functions provide the detector response across all 1024 spectral pixels for all three channels, per unit emission measure ($10^{27}$~cm$^{-5}$), at a specified slit (1–37), temperature, and Doppler velocity, as described in detail in \cite{BDP:MUSE}. MHD models are first sampled as velocity-DEM distributions~\citep{Cheung:SDC} and then folded with the \musen\ response functions to generate synthetic spectra. Here  we focus on the main lines: \fexix\ and \fexxi\ 108\AA, \feixw, and \fexvw.  
The response functions are computed using the latest CHIANTI database version, and include instrumental effects (such as instrumental line broadening) and thermal broadening of the lines.
The response functions calculate the predicted spectra in units of [DN~s$^{-1}$~pix$^{-1}$], where the pixel is \musen\ spectral raster pixel which has a sampling rate of 0.4\arcsec\ in the rastering (perpendicular to slits) direction, and a sampling of 0.167\arcsec\ along slits. For this paper, even when the synthetic images shown have a different sampling than a MUSE raster (e.g., MURaM simulations sampled at 192 km, equivalent to 0.265\arcsec), we have opted to continue using the same response functions (and thus effective area) to demonstrate the count rate is sufficient for the relevant features. 

\section{Definition of spectral moments} \label{app:moments}

In the paper we show 0th ($I_0$, a.k.a.\ line intensity), 1st ($I_1$, a.k.a.\ Doppler velocity) and 2nd ($I_2$, a.k.a.\ total line width), moments of the \musen\ synthetic observables from models, and they are defined, respectively, as follows:

\begin{eqnarray}
I_0 = \sum_j F_j ~[{\rm DN/pix/s}]\\
I_1 = \frac{\sum_j F_j \times v_j}{I_0}~[\text{km s}^{-1}] \\ 
I_2 = \sqrt{\frac{\sum_j F_j \times (v_j-I_1)^2}{I_0}}~[\text{km s}^{-1}]
\end{eqnarray}
\noindent where $j$ is the index of the spectral bin, $F_j$ is the intensity in spectral bin $j$ in units of DN~s$^{-1}$~pix$^{-1}$, and $v_j$ is the Doppler velocity in km~s$^{-1}$.

\bibliography{sample631,refs}{}
\bibliographystyle{aasjournal}



\end{document}

%% file: author_list.tex
\author[0000-0002-8370-952X]{Mark C. M. Cheung}
\email{cheung@lmsal.com}
\affil{Lockheed Martin Solar \& Astrophysics Laboratory, 3251 Hanover St, Palo Alto, CA 94304, USA}

\author[0000-0002-0333-5717]{Juan Mart\'inez-Sykora}
\affil{Lockheed Martin Solar \& Astrophysics Laboratory, 3251 Hanover St, Palo Alto, CA 94304, USA}
\affil{Bay Area Environmental Research Institute, NASA Research Park, Moffett Field, CA 94035, USA.}
\affil{Rosseland Centre for Solar Physics, University of Oslo, P.O. Box 1029 Blindern, N-0315 Oslo, Norway}
\affil{Institute of Theoretical Astrophysics, University of Oslo, P.O. Box 1029 Blindern, N-0315 Oslo, Norway}

\author[0000-0002-0405-0668]{Paola Testa}
\affil{Harvard-Smithsonian Center for Astrophysics, 60 Garden St, Cambridge, MA 02193, USA}

\author[0000-0002-8370-952X]{Bart De Pontieu}
\affil{Lockheed Martin Solar \& Astrophysics Laboratory, 3251 Hanover St, Palo Alto, CA 94304, USA}
\affil{Rosseland Centre for Solar Physics, University of Oslo, P.O. Box 1029 Blindern, N-0315 Oslo, Norway}
\affil{Institute of Theoretical Astrophysics, University of Oslo,
P.O. Box 1029 Blindern, N-0315 Oslo, Norway}

\author[0000-0002-1253-8882]{Georgios Chintzoglou}
\affil{Lockheed Martin Solar \& Astrophysics Laboratory, 3251 Hanover St, Palo Alto, CA 94304, USA}
\affil{University Corporation for Atmospheric Research, Boulder, CO 80307-3000, USA}

\author[0000-0001-5850-3119]{Matthias Rempel}
\affiliation{High Altitude Observatory, NCAR, P.O. Box 3000, Boulder, CO 80307, USA}

\author[0000-0002-4980-7126]{Vanessa Polito}
\affil{Bay Area Environmental Research Institute, NASA Research Park, Moffett Field, CA 94035, USA.}
\affil{Lockheed Martin Solar \& Astrophysics Laboratory, 3251 Hanover St, Palo Alto, CA 94304, USA}

\author[0000-0001-5316-914X]{Graham S. Kerr}
\affil{Department of Physics, The Catholic University of America, 620 Michigan Avenue NW, Washington DC 20064, USA}
\affil{NASA Goddard Space Flight Center, Heliophysics Sciences Division, Code 671, 8800 Greenbelt Road, Greenbelt, MD 20771, USA}

\author[0000-0002-6903-6832]{Katharine K. Reeves} 
\affil{Harvard-Smithsonian Center for Astrophysics, 60 Garden St, Cambridge, MA 02193, USA}

\author[0000-0001-9315-7899]{Lyndsay Fletcher} 
\affil{SUPA School of Physics \& Astronomy, University of Glasgow, Glasgow, G12 8QQ, UK}
\affil{Rosseland Centre for Solar Physics, University of Oslo, P.O. Box 1029 Blindern, N-0315 Oslo, Norway}

\author[0000-0002-9672-3873]{Meng Jin} 
\affil{Lockheed Martin Solar \& Astrophysics Laboratory, 3251 Hanover St, Palo Alto, CA 94304, USA}
\affil{SETI Institute, 189 North Bernardo Avenue, Suite 200, Mountain View, CA 94043, USA}

\author[0000-0002-7788-6482]{Daniel N\'obrega-Siverio}
\affil{Instituto de Astrof\'isica de Canarias, E-38205 La Laguna, Tenerife, Spain}
\affil{Universidad de La Laguna, Dept. Astrof\'isica, E-38206 La Laguna, Tenerife, Spain}
\affil{Rosseland Centre for Solar Physics, University of Oslo, P.O. Box 1029 Blindern, N-0315 Oslo, Norway}
\affil{Institute of Theoretical Astrophysics, University of Oslo,
P.O. Box 1029 Blindern, N-0315 Oslo, Norway}

\author[0000-0002-2344-3993]{Sanja Danilovic}
\affiliation{Institute for Solar Physics, Department of Astronomy, Stockholm University, AlbaNova University Centre, 106 91 Stockholm, Sweden}

\author[0000-0003-1529-4681]{Patrick Antolin}
\affil{Department of Mathematics, Physics \& Electrical Engineering, Northumbria University, Newcastle Upon Tyne, NE1 8ST, UK}

\author[0000-0003-4227-6809]{Joel Allred}
\affil{NASA Goddard Space Flight Center, Heliophysics Sciences Division, Code 671, 8800 Greenbelt Road, Greenbelt, MD 20771, USA}

\author[0000-0003-0975-6659]{Viggo Hansteen}
\affil{Lockheed Martin Solar \& Astrophysics Laboratory, 3251 Hanover St, Palo Alto, CA 94304, USA}
\affil{Bay Area Environmental Research Institute, NASA Research Park, Moffett Field, CA 94035, USA.}
\affil{Rosseland Centre for Solar Physics, University of Oslo, P.O. Box 1029 Blindern, N-0315 Oslo, Norway}
\affil{Institute of Theoretical Astrophysics, University of Oslo,
P.O. Box 1029 Blindern, N-0315 Oslo, Norway}

\author[0000-0001-5503-0491]{Ignacio Ugarte-Urra}
\affil{Space Science Division, Naval Research Laboratory, Washington, DC 20375, USA}

\author[0000-0001-7416-2895	]{Edward DeLuca}
\affil{Harvard-Smithsonian Center for Astrophysics, 60 Garden St, Cambridge, MA 02193, USA}

\author[0000-0003-2102-0070]{Dana Longcope}
\affil{Department of Physics, Montana State University, Bozeman, MT 59717, USA}

\author[0000-0003-3882-3945]{Shinsuke Takasao}
\affil{Department of Earth and Space Science, Graduate School of Science, Osaka University, Toyonaka, Osaka 560-0043, Japan}

\author[0000-0002-6338-0691]{Marc DeRosa}
\affil{Lockheed Martin Solar \& Astrophysics Laboratory, 3251 Hanover St, Palo Alto, CA 94304, USA}

\author{Paul Boerner}
\affil{Lockheed Martin Solar \& Astrophysics Laboratory, 3251 Hanover St, Palo Alto, CA 94304, USA}

\author[0000-0001-5459-2628]{Sarah Jaeggli}
\affil{National Solar Observatory, 22 Ohi'a Ku, Makawao, HI 96768, USA}

\author[0000-0001-6119-0221]{Nariaki Nitta}
\affil{Lockheed Martin Solar \& Astrophysics Laboratory, 3251 Hanover St, Palo Alto, CA 94304, USA}

\author[0000-0002-9288-6210]{Adrian Daw}
\affil{NASA Goddard Space Flight Center, Heliophysics Sciences Division, Code 671, 8800 Greenbelt Road, Greenbelt, MD 20771, USA}

\author[0000-0001-9218-3139]{Mats Carlsson}
\affil{Rosseland Centre for Solar Physics, University of Oslo, P.O. Box 1029 Blindern, N-0315 Oslo, Norway}
\affil{Institute of Theoretical Astrophysics, University of Oslo,
P.O. Box 1029 Blindern, N-0315 Oslo, Norway}

\author[0000-0001-9638-3082]{Leon Golub}
\affil{Harvard-Smithsonian Center for Astrophysics, 60 Garden St, Cambridge, MA 02193, USA}

\author{the \musen\ team}